\documentclass[reprint,superscriptaddress,amsmath,amssymb,twocolumn]{revtex4-2}

\usepackage{amsmath}
\usepackage{mathtools}
\usepackage{amssymb}
\usepackage{graphicx}
\usepackage[colorlinks, linkcolor=blue, citecolor=blue, urlcolor=blue, breaklinks=true]{hyperref}
\usepackage{booktabs}
\usepackage[dvipsnames]{xcolor}
\usepackage{ragged2e}
\usepackage{orcidlink}
\usepackage[T1]{fontenc}
\usepackage{mathrsfs}
\usepackage{centernot}

\DeclareMathSymbol{:}{\mathpunct}{operators}{"3A}
\newcommand{\figpanel}[2]{\hyperref[#1]{\ref*{#1}(#2)}}
\newcommand{\Tr}{\operatorname{Tr}}
\newcommand{\cG}{\mathcal G}
\newcommand{\Vpi}{\mathcal V_{\pi}}
\newcommand{\Cclass}{\mathcal C}

\begin{document}

\date{\today}

\title{Thermodynamic Irreversibility from Inaccessible Endogenous Quantum Histories}

\author{Borhan Ahmadi\orcidlink{0000-0002-2787-9321}}
\email{borhan.ahmadi@ug.edu.pl}
\affiliation{International Centre for Theory of Quantum Technologies, University of Gda\'nsk, ul. prof. Marii Janion 4, 80-309 Gda\'nsk, Poland}

\begin{abstract}
Microscopic unitary dynamics preserves all fine information, yet an isolated finite system can show a robust thermodynamic window in which the entropy associated with a restricted record rises. We ask why information hidden from the current record usually fails to rebuild a low-entropy macrostate. For a fixed projective record, every finite step separates exactly into the evolution predicted from the record alone and an exact correction carried by unresolved microscopic structure. The record-only contribution starts only at second order in time, so all instantaneous change of the record comes from hidden currents between macrostates. We derive the exact entropy rate, separate entropy-spreading from return-oriented currents, and resolve those currents into energy-gap amplitudes. Transitions with the same gap add coherently, revealing how the Hamiltonian and the microscopic state organize hidden information for return. In interacting mixing dynamics the current power is spread over many frequencies; free and deliberately commensurate controls progressively concentrate it, and the engineered dynamics reconstructs a low-entropy macrostate. An independent distribution-level test separates hidden dynamical activity from finite-step return, and an exact classical measure-preserving counterpart shows which parts of the construction are not uniquely quantum. Within the specified record and observation window, irreversibility is therefore not loss of microscopic information, but the failure of that information to organize currents that restore macroscopic order.
\end{abstract}

\maketitle

\section{Introduction}

The foundational tension of statistical mechanics is not that a thermodynamic entropy can be assigned to a restricted description, but that this entropy acquires a preferred direction while the underlying dynamics preserves complete information. Boltzmann connected the macroscopic arrow to reduced distributions and overwhelmingly large macrostate multiplicities, whereas Gibbsian evolution leaves the exact phase-space density unchanged~\cite{boltzmann1872weitere,boltzmann1877beziehung,gibbs1902elementary}. Loschmidt's reversal argument and Zermelo's recurrence objection then made the difficulty precise: reversible microscopic laws permit specially organized entropy-lowering trajectories, and finite measure-preserving systems cannot sustain strict monotonicity for all mathematical times~\cite{loschmidt1876waermegleichgewicht,zermelo1896satz}. Planck encountered the same obstruction in radiation theory, where an additional restriction on microscopic phase organization was needed before reversible field equations produced an irreversible law~\cite{planck1900strahlung}. These developments established a distinction that remains central today: defining a thermodynamic entropy and explaining its dynamical increase are different problems~\cite{lebowitz1993arrow,uffink2007compendium,brown2009htheorem}.

Thermodynamics sharpens this distinction by fixing the system boundary before entropy generation is assigned. Heat and work are boundary modes of energy transfer: heat is energy transferred because of a temperature difference and carries entropy across the boundary, whereas work transfers energy without entropy; entropy generation measures irreversibility produced inside the chosen system~\cite{kondepudi1998modern,degroot1984nonequilibrium,lebellac2004equilibrium,cengel2024thermodynamics}. Thus, for an isolated system there is no entropy transfer and the entropy balance reduces locally in time to \(\dot S=\dot S_{\rm gen}\geq0\) during an irreversible thermodynamic evolution, with the accumulated entropy obtained only afterward by integration. An earlier analysis by some of us emphasized the complementary microscopic point that unitary dynamics cannot generate von Neumann entropy, \(\dot S_{\rm vN}=0\), and therefore cannot account for thermodynamic entropy generation if the latter is identified with the fine-state entropy~\cite{ahmadi2020noentropy}. The question addressed here is the resulting one: which entropy can grow under exact unitary dynamics, and what \textit{endogenous} microscopic motion generates that growth?

Reversal protocols reveal what this second problem must address. Maxwell and Szilard showed that microscopic discrimination and conditional control can turn information hidden from ordinary thermodynamic operations into macroscopic order or work~\cite{maxwell1871theory,szilard1929entropy}. Spin echoes provide the dynamical analogue: a vanished macroscopic signal can return because the microscopic phases were dispersed rather than destroyed~\cite{hahn1950echo,ridderbos1998spin,anastopoulos2011spin}. The lesson is not that thermodynamic entropy is ill defined. It is that coarse graining alone cannot explain irreversibility, because the information omitted from the record may remain physically present and, when suitably organized, become macroscopically effective again.

Several mature formalisms isolate pieces of this mechanism. Ehrenfest coarse graining and Jaynesian maximum entropy construct representative states from incomplete macroscopic data. Zwanzig and Mori show how unresolved variables re-enter projected dynamics through memory and source terms, while modern extensions make the resulting non-Markovian structure explicit~\cite{ehrenfest1959conceptual,jaynes1957information,zwanzig1960ensemble,zwanzig1961memory,mori1965transport,grabert1982projection,aristoff2023coarse,espanol2026memory}. Ulam's transfer-operator construction gives a stochastic cell-transition matrix for densities taken uniform within each cell~\cite{ulam1960collection,froyland1999ulam}; classical observational entropy and nondestructive coarse entropies retain the exact Hamiltonian evolution while assigning entropy to a restricted record~\cite{safranek2020classical,ding2025hamiltonian}. These lines explain representation, memory, and cell-to-cell transport. They do not yet isolate, for the actual present microscopic state, the finite-step term through which hidden structure can reconstruct macroscopic return.

The quantum setting has an equally developed entropy lineage. Von Neumann's macrospaces and observational entropy associate a projective record with the entropy \(-k_{\rm B}\sum_Rp_R\ln(p_R/\Omega_R)\)~\cite{vonneumann1929ergodic,safranek2019thermodynamics,safranek2019thermalization}. The corresponding macro-uniform state is also the Bayesian/Petz-recovered coarse state for a uniform microscopic prior~\cite{buscemi2023coarse,bai2024priors,nagasawa2025macroscopicity,schindler2025unification}. When the microscopic state already equals this representative, its observational entropy cannot decrease under unitary evolution, and generic unitaries increase it strictly~\cite{strasberg2021firstsecond,nagasawa2024generic}. Recent closed-system and restricted-information laws establish endpoint or most-time constraints, thermodynamic-limit macroscopic laws, finite-resolution thermodynamics, and bounds associated with hidden microscopic resources~\cite{meier2025secondlaw,chiba2026secondlaw,rubino2026coarse,xsqg-xvgc,rignonbret2026algebraic}. These results do not resolve the accumulation of entropy along the actual autonomous trajectory. We therefore do not claim the representative, its entropy, or its macrostate monotonicity as new. The unresolved dynamical question is narrower: how does microscopic structure absent from the present record generate the instantaneous motion of that record, and when does this motion become organized as thermodynamic return?

We answer this question for one isolated quantum system viewed through a fixed projective record. A thermodynamic preparation can be compatible with many fine initial states, but once the Hamiltonian and one fine initial state are specified, the microscopic history is unique. We call that history endogenous because no bath, measurement, stochastic intervention, or external information-erasing mechanism is introduced to generate the relaxation.

Our construction keeps the exact microscopic state evolving while using the established macro-uniform state only as the state specified by the current record. Over any finite interval, the future record separates into the part predicted from that record alone and the exact contribution of the microscopic structure that the record does not resolve. The key short-time result is that the record-only part produces no first-order motion: the entire instantaneous change is carried by hidden currents between macrostates. We derive how those currents generate or oppose the growth of thermodynamic entropy and then resolve them spectrally, showing how equal energy gaps can synchronize microscopic contributions into return. A complementary relative-majorization test determines whether the full macrodistribution has left the class of multiplicity-preserving contractions.

We test this picture in interacting hard-core and Bose--Hubbard chains, an interacting two-dimensional lattice, and number-conserving Floquet dynamics. During their primary relaxation windows, hidden microscopic information remains dynamically active while the entropy rate stays positive and no distributional return is detected at the tested resolution. We then keep the preparation and thermodynamic record fixed but deliberately reorganize the microscopic dynamics: an interacting mixing Hamiltonian spreads the current over many frequencies, a free chain is intermediate, and a commensurate control concentrates the current into a few synchronized gap channels and reconstructs a low-entropy macrostate. The same finite-step structure has an exact finite-measure classical counterpart, while the record entropy independently recovers the dilute classical free-expansion endpoint. Heat and work require an additional thermodynamic boundary and control specification and are therefore left to a companion study. Our claims are restricted to the stated records, controls, finite systems, and observation windows; we do not assert global monotonicity or a universal spectral criterion.

\section{Thermodynamic records and finite-step propagation}

A thermodynamic record is a restriction on the description of the same isolated system, not a second physical subsystem. Let the relevant Hilbert space, or an invariant sector fixed by exact conservation laws, be partitioned into mutually orthogonal macrospaces \(P_R\), with \(\sum_RP_R=I_{\mathcal H}\). Their dimensions \(\Omega_R=\Tr P_R\) count the microscopic directions compatible with record value \(R\), while \(p_R=\Tr(P_R\rho)\) gives the current probability of that record.

The state that preserves these probabilities and nothing more is
\(\cG[\rho]=\sum_R(p_R/\Omega_R)P_R\). This is the established observational coarse state: it lives on the same Hilbert space as \(\rho\), but is uniform within each unresolved macrospace. Its entropy is
\begin{equation}
S_R(\rho)
=
-k_{\rm B}\Tr\!\left[\cG[\rho]\ln\cG[\rho]\right]
=
k_{\rm B}\left[\ln\Omega_{\rm tot}-D(p\Vert\pi)\right],
\label{eq:accessible-entropy}
\end{equation}
where \(\Omega_{\rm tot}=\sum_R\Omega_R\), \(\pi_R=\Omega_R/\Omega_{\rm tot}\), and \(D(p\Vert\pi)=\sum_Rp_R\ln(p_R/\pi_R)\). The same uniform prior that defines conventional observational entropy induces this multiplicity distribution~\cite{bai2024priors}. The accompanying observational deficit,
\begin{equation}
I_R(\rho)
\equiv
D\!\left(\rho\Vert\cG[\rho]\right)
=
\frac{S_R(\rho)-S_{\rm vN}(\rho)}{k_{\rm B}}
\geq0,
\label{eq:observational-deficit}
\end{equation}
measures how much microscopic information is hidden from the present record. It does not measure irreversibility by itself. Two microscopic states can carry the same amount of record-hidden information and nevertheless evolve very differently under different Hamiltonians. The record fixes the thermodynamic description; the microscopic physics determines how the exact state moves through it. That dynamics first enters through the unitary propagation below and, more explicitly, through the Hamiltonian matrix elements that generate the instantaneous currents in Sec.~\ref{sec:instantaneous-currents}. The representative and its information-theoretic status are reviewed in Note~\ref{supp:note-coarse-entropy} of the Supplementary Material (SM)~\cite{SuppMat}.

Write the exact current state as \(\rho_t=\cG[\rho_t]+\chi_t\), with \(\Tr(P_R\chi_t)=0\) for every \(R\). This is the standard retained--unresolved projection split specialized to the thermodynamic record; the new question is what the unresolved part does to that record later. Evolving both pieces with the same unitary gives the exact future record
\begin{equation}
\begin{aligned}
p(t+\tau)
&=
K^{(U)}(\tau)p(t)+r(t,\tau),
\\
K^{(U)}_{RR'}(\tau)
&=
\frac{1}{\Omega_{R'}}
\Tr\!\left[P_RU_\tau P_{R'}U_\tau^\dagger\right],
\end{aligned}
\label{eq:finite-step-propagation}
\end{equation}
where \(r_R(t,\tau)=\Tr[P_RU_\tau\chi_tU_\tau^\dagger]\). Here \(t\) labels the current point on the microscopic trajectory, whereas \(\tau>0\) is the forward propagation interval from \(t\) to \(t+\tau\); all limits \(\tau\to0^+\) below are taken at fixed \(t\). The term \(K^{(U)}p(t)\) is not an approximation to the microscopic evolution. It is the future macrodistribution obtained if the exact present state were replaced by the macro-uniform state with the same current record and then evolved by the same unitary. The vector \(r\) is the exact difference between that record-only prediction and the true future record. It therefore isolates the dynamical effect of microscopic information absent from the current macroprobabilities without introducing stochastic dynamics, a Markov closure, or information loss.

This interpretation also fixes the relation to projection methods. With \(\cG\) viewed as an idempotent state-space projection, \(K^{(U)}\) is the coordinate representation of the full retained-to-retained finite-time block, whereas \(r\) is the retained response initiated by the unresolved component already present at the beginning of the step. Retained--unresolved--retained excursions generated during the interval are already resummed inside \(K^{(U)}\); the closest Mori--Zwanzig analogue of \(r\) is therefore the finite-time response to the inhomogeneous source, not the memory kernel as a whole~\cite{zwanzig1960ensemble,zwanzig1961memory,mori1965transport,grabert1982projection}. The block structure and the corresponding non-semigroup composition defect are derived in Note~\ref{supp:note-projection-classical} of the SM~\cite{SuppMat}.

Unitarity makes \(K^{(U)}\) stochastic and preserves the multiplicity reference, \(K^{(U)}\pi=\pi\), while \(\sum_Rr_R=0\). The record-only reference evolution therefore satisfies the established finite-step contraction
\begin{equation}
\Delta_\tau S_R
=
\Delta_\tau S_{\rm mix}
+
\Delta_\tau S_{\rm corr},
\quad
\Delta_\tau S_{\rm mix}\geq0,
\label{eq:finite-step-entropy-balance}
\end{equation}
where \(\Delta_\tau S_{\rm mix}=S_R[K^{(U)}p(t)]-S_R[p(t)]\) and \(\Delta_\tau S_{\rm corr}=S_R[p(t+\tau)]-S_R[K^{(U)}p(t)]\). The inequality is established structure~\cite{strasberg2021firstsecond,nagasawa2024generic}. We use it only to separate a finite-step record-only contribution from the effect of hidden microscopic structure. It is not the local Second-Law statement for the isolated dynamics~\cite{kondepudi1998modern,degroot1984nonequilibrium,lebellac2004equilibrium,cengel2024thermodynamics}.

For the isolated system considered here, once \(S_R\) is taken as the thermodynamic entropy associated with the specified record, there is no entropy transfer across the outer boundary. Within a thermodynamic relaxation window, the local Second-Law condition is therefore
\[
\dot S_R(t)=\dot S_{\rm gen}(t)\geq0,
\]
and a positive finite change \(\Delta S_R\) is the accumulated consequence of that rate. The finite-step inequality in Eq.~\eqref{eq:finite-step-entropy-balance} does not by itself establish this local arrow. In fact, the record-only contribution has no first-order entropy rate. Expanding at \(\tau=0\) gives
\begin{equation}
\begin{aligned}
\left.\partial_\tau K^{(U)}(\tau)\right|_{\tau=0}
&=0,
\quad
\dot S_{\rm mix}(t)
\equiv
\lim_{\tau\rightarrow0}
\frac{\Delta_\tau S_{\rm mix}}{\tau}
=0,
\\
\lim_{\tau\rightarrow0}
\frac{r(t,\tau)}{\tau}
&=\dot p(t).
\end{aligned}
\label{eq:instantaneous-limit}
\end{equation}
The record-only contribution begins only at second order, \(K^{(U)}(\tau)=I+O(\tau^2)\), whereas the hidden-state residual is linear, \(r(t,\tau)=\tau\dot p(t)+O(\tau^2)\). Hence every nontrivial instantaneous motion of the accessible record originates in \(\chi_t\). At differentiable times, \(\Delta_\tau S_{\rm corr}/\tau\to\dot S_R(t)\). The boundary cases of the probability simplex are treated in Note~\ref{supp:short-time-baseline} of the SM~\cite{SuppMat}.

The finite-step structure is not uniquely quantum. For invertible measure-preserving classical evolution \(\Phi^\tau\) on a finite invariant measure space and a measurable partition \(\{C_R\}\),
\begin{equation}
K^{(\Phi)}_{RR'}(\tau)
=
\frac{\mu[C_{R'}\cap(\Phi^\tau)^{-1}C_R]}{\mu(C_{R'})},
\quad
K^{(\Phi)}\pi^{\rm cl}=\pi^{\rm cl},
\label{eq:classical-cell-map}
\end{equation}
with \(\pi_R^{\rm cl}=\mu(C_R)/\mu(\Gamma)\). This is the established Ulam cell-transition matrix~\cite{ulam1960collection,froyland1999ulam}. The additional step is to retain the exact nonuniform within-cell component, which completes the closed Ulam iteration as \(p(t+\tau)=K^{(\Phi)}p(t)+r^{\rm cl}(t,\tau)\). The construction requires a finite invariant measure---for example a bounded invariant region, a finite-measure energy surface, or an invariant probability measure---and does not normalize unrestricted Liouville measure on an unbounded phase space. The proof and its thermodynamic consequences are given in Note~\ref{supp:note-projection-classical} of the SM~\cite{SuppMat}.
\begin{figure}[t]
\centering
\includegraphics[width=1\columnwidth]{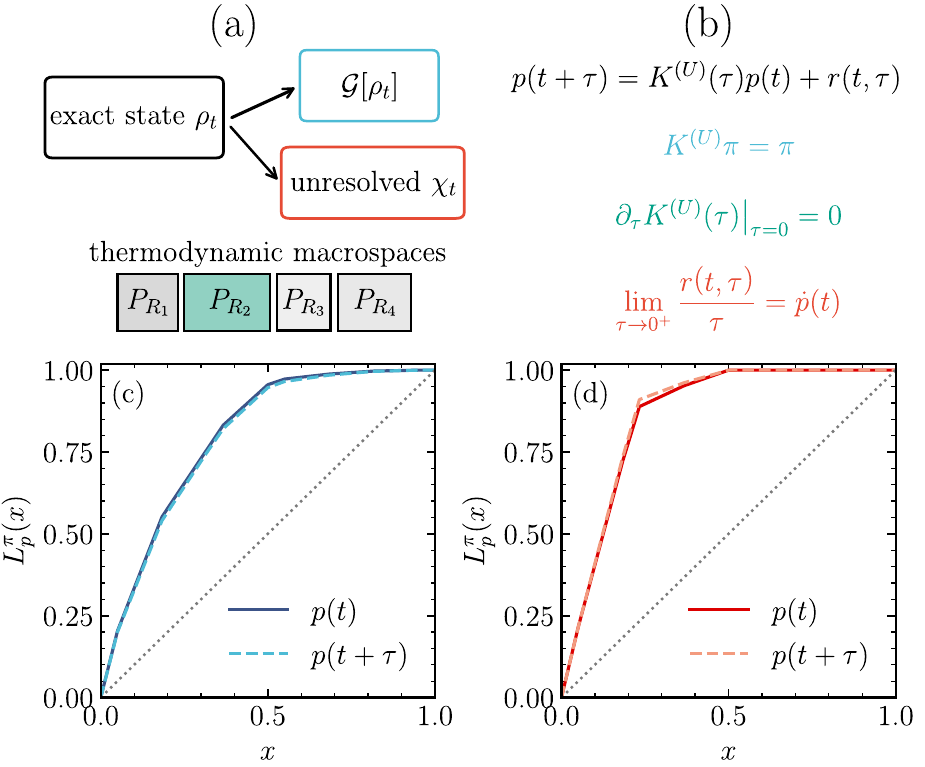}
\caption{
Closed-system thermodynamic coarse graining and finite-step return structure.
(a) The exact state and its thermodynamic representative describe the same isolated system. The representative \(\cG[\rho_t]\) retains the macroprobabilities, while \(\chi_t=\rho_t-\cG[\rho_t]\) contains structure invisible to the current record.
(b) Every finite step obeys \(p(t+\tau)=K^{(U)}(\tau)p(t)+r(t,\tau)\). The record-only contribution preserves \(\pi\) but is stationary to first order, whereas \(r(t,\tau)/\tau\to\dot p(t)\).
(c) During primary mixing, the later generalized Lorenz curve lies below the earlier one, so \(p(t)\succ_\pi p(t+\tau)\).
(d) During organized return, the curves cross and the relative-majorization defect becomes nonzero.\justifying}
\label{fig:mechanism}
\end{figure}

\section{Instantaneous hidden currents and exact spectral organization}
\label{sec:instantaneous-currents}

Equation~\eqref{eq:instantaneous-limit} turns the hidden component from a static information deficit into the source of the actual thermodynamic motion. For \(R\neq R'\), define the current from macrospace \(R'\) into \(R\) and the corresponding thermodynamic affinity by
\begin{equation}
\begin{aligned}
J_{RR'}(t)
&=
2\operatorname{Im}
\Tr\!\left(P_RHP_{R'}\chi_t\right),
\quad
\dot p_R=\sum_{R'}J_{RR'},
\\
A_{RR'}(t)
&=
\ln\!\left[
\frac{p_{R'}(t)/\Omega_{R'}}{p_R(t)/\Omega_R}
\right].
\end{aligned}
\label{eq:instantaneous-current}
\end{equation}
The macro-uniform representative carries no inter-macrospace current. Equation~\eqref{eq:instantaneous-current} also shows explicitly where the microscopic physics enters. The record fixes the macrostates and their multiplicities, but the Hamiltonian selects which macrostates are coupled and with what strength through \(P_RHP_{R'}\), while the actual state supplies the hidden coherence and correlation blocks through \(P_{R'}\chi_tP_R\). Neither the record entropy nor the amount of hidden information alone determines the arrow. The same record can therefore support ordinary mixing under one Hamiltonian and organized return under another.

The affinity compares probability per available microscopic direction, so its sign identifies whether a current spreads probability toward less occupied microscopic directions or concentrates it back toward a more ordered record. At differentiable times the exact entropy balance is
\begin{equation}
\dot S_R
=
\frac{k_{\rm B}}{2}
\sum_{R,R'}J_{RR'}A_{RR'}
=
\sigma_{\rm spread}-\sigma_{\rm ret},
\label{eq:instantaneous-current-balance}
\end{equation}
where \(\sigma_{\rm spread}\) and \(\sigma_{\rm ret}\) collect the positive and negative unordered-pair contributions. This is the local irreversibility statement for the isolated system within the chosen thermodynamic window. Entropy is generated when the spreading contribution exceeds the return-oriented one; individual entropy-lowering currents may remain active without reversing the total entropy accumulation. Thus irreversibility is not equivalent to eliminating hidden microscopic motion. It depends on how the actual Hamiltonian organizes that motion relative to the thermodynamic affinities.

The microscopic support of return is correspondingly sharp. Trace-norm duality gives
\begin{equation}
\begin{aligned}
\sigma_{\rm ret}(t)
\leq{}&
2k_{\rm B}
\sum_{\substack{R<R'\\J_{RR'}A_{RR'}<0}}
|A_{RR'}|
\left\|P_RHP_{R'}\right\|_\infty
\\
&\quad\times
\left\|P_{R'}\chi_tP_R\right\|_1.
\end{aligned}
\label{eq:microscopic-return-bound}
\end{equation}
Only hidden inter-macrospace blocks that generate currents against the thermodynamic affinity contribute instantaneously to \(\sigma_{\rm ret}\). Structure confined within one macrospace can matter later, but only after the Hamiltonian converts it into these current-carrying blocks. This is the point regarding the statement that information is hidden becomes a physical statement about irreversibility: the couplings, phases, correlations, and spectrum of the actual Hamiltonian decide whether that information remains dispersed or becomes organized into return.

To see this organization directly, we resolve the exact current network spectrally. Write \(H=\sum_\alpha E_\alpha\Pi_\alpha\) in terms of projectors onto distinct energies and, for each unordered coupled macrospace pair \(e\), define \(\widehat I_e=i(P_{R'}HP_R-P_RHP_{R'})\). For the prepared state \(\rho_0\), let
\begin{equation}
\begin{aligned}
c_{e,\alpha\beta}
=
\Tr\!\left(\widehat I_e\Pi_\alpha\rho_0\Pi_\beta\right),\quad
C_{e,g}
=
\sum_{\substack{\alpha\neq\beta\\E_\alpha-E_\beta=g}}
c_{e,\alpha\beta}.
\end{aligned}
\label{eq:spectral-current-amplitudes}
\end{equation}
Then each macrocurrent has the exact decomposition
\begin{equation}
J_e(t)
=
J_e^\omega
+
\sum_{g\neq0}C_{e,g}e^{-igt},
\quad
J_e^\omega
=
\Tr(\widehat I_e\omega),
\label{eq:spectral-current-decomposition}
\end{equation}
where \(\omega=\sum_\alpha\Pi_\alpha\rho_0\Pi_\alpha\). Complete energy eigenspace projectors make the amplitudes basis independent within degenerate subspaces. The key physical point is simple: microscopic transitions with the same energy gap oscillate at the same frequency, so their amplitudes add before their contribution to the current power is formed.

This motivates a separation between the transition content available to the current operators and the amount that survives after equal-gap amplitudes are combined,
\begin{equation}
\begin{aligned}
P_{\rm pair}
&=
\sum_e\sum_{\alpha\neq\beta}|c_{e,\alpha\beta}|^2,
\\
P_{\rm spec}
&=
\sum_e\sum_{g\neq0}|C_{e,g}|^2,
\\
\Gamma_{\rm spec}
&=
\frac{P_{\rm spec}}{P_{\rm pair}},
\quad P_{\rm pair}>0.
\end{aligned}
\label{eq:spectral-coherence-factor}
\end{equation}
If \(D_G\) is the largest number of distinct-energy pairs sharing one nonzero gap, Cauchy--Schwarz yields
\begin{equation}
\begin{aligned}
\overline{\|\mathbf J-\mathbf J^\omega\|_2^2}
&=
P_{\rm spec}
=
\Gamma_{\rm spec}P_{\rm pair},
\\
P_{\rm spec}
&\leq
D_GP_{\rm pair},
\quad
0\leq\Gamma_{\rm spec}\leq D_G.
\end{aligned}
\label{eq:spectral-current-theorem}
\end{equation}
Nondegenerate nonzero gaps force \(\Gamma_{\rm spec}=1\): there are then no equal-gap amplitudes to combine. Degeneracy only makes coherent enhancement possible; the prepared state and the current operators determine whether the corresponding amplitudes actually reinforce or cancel one another. The Fourier decomposition itself is standard. What matters here is that it is applied to the exact thermodynamic currents, allowing the microscopic spectrum to be connected directly to entropy-spreading and return-oriented motion.

Current power is not yet thermodynamic return. The affinities decide whether a given current spreads or concentrates the record, and at every time
\begin{equation}
\sigma_{\rm ret}(t)
\leq
k_{\rm B}
\|\mathbf A(t)\|_2
\|\mathbf J(t)\|_2.
\label{eq:spectral-current-return-envelope}
\end{equation}
For a finite observation window the current power is given exactly by a sinc kernel coupling distinct gaps; in the infinite-time average only equal-gap groups survive. The finite-window identity, the time-reversal condition \(\mathbf J^\omega=0\), and the reconstruction tests are developed in Note~\ref{supp:note-spectral-return} of the SM~\cite{SuppMat}.

\section{Relative majorization and history-sensitive return}

The instantaneous entropy rate tells us whether entropy is being generated locally in time, but one scalar cannot describe the full ordering of the macrodistribution over a finite interval. Relative majorization provides that stronger, separate test. For distributions \(p\) and \(q\) with full-support reference \(\pi\), write \(p\succ_\pi q\) when \(q=Mp\) for a stochastic map satisfying \(M\pi=\pi\)~\cite{renes2016relative,buscemi2017lorenz}. Equivalently, the generalized Lorenz curve of \(p\) lies nowhere below that of \(q\). We quantify departure from this order by
\begin{equation}
\begin{aligned}
\Vpi(p\rightarrow q)
&=
\sup_{0\leq x\leq1}
\left[L_q^\pi(x)-L_p^\pi(x)\right]_+,
\\
\Vpi[p(t)\rightarrow p(t+\tau)]
&=0
\Longrightarrow
S_R(t+\tau)\geq S_R(t).
\end{aligned}
\label{eq:return-defect}
\end{equation}
The implication is deliberately one-sided. A Lorenz-curve crossing can occur while the scalar entropy still rises, so \(\Vpi\) does not duplicate the local Second-Law test \(\dot S_R\geq0\). Instead, it asks whether the entire accessible distribution has left the stronger class of changes generated by \(\pi\)-preserving stochastic maps.

The exact residual determines how far the true unitary step can depart from that sector. Defining
\begin{align}
\epsilon_{\rm hist}(t,\tau)
&=
\frac12\|r(t,\tau)\|_1
=
\frac12\|p(t+\tau)-K^{(U)}(\tau)p(t)\|_1,
\nonumber\\
\Vpi[p(t)&\rightarrow p(t+\tau)]
\leq
\epsilon_{\rm hist}(t,\tau),
\label{eq:history-bound}
\end{align}
gives the total-variation distance between the actual future record and the future predicted from the present macro-uniform representative alone. The bound is a dynamical specialization of approximate-Lorenz geometry~\cite{horodecki2018approximate}: its radius is not an externally imposed uncertainty, but the exact record-level effect of microscopic information absent from the current record.

This separates two notions that would otherwise be easy to conflate. A nonzero \(\epsilon_{\rm hist}\) means that the present thermodynamic record is not dynamically closed: unresolved microscopic structure changes what will be observed next. A nonzero \(\Vpi\), by contrast, means that this influence has become strong and directed enough to carry the full macrodistribution outside the \(\pi\)-preserving contraction order. Hidden information can therefore remain dynamically active while \(\Vpi=0\); conversely, even \(\Vpi>0\) need not make the entropy rate negative. The local entropy-generation question and the stronger finite-step distributional question are related, but they are not the same. The derivations and numerical checks are given in Notes~\ref{supp:note-relative-majorization}--\ref{supp:note-bound-verification} of the SM~\cite{SuppMat}.

\section{Mixing and engineered thermodynamic return}

We now ask whether this separation occurs in an actual closed many-body dynamics: can microscopic information remain continuously relevant to the thermodynamic record while failing to organize into macroscopic return? Consider an open chain of \(N\) qubits, or equivalently \(Q\) hard-core excitations, governed in the fixed-\(Q\) sector by
\begin{align}
H_{\rm mix}
&=
-J\sum_{j=1}^{N-1}
\left(\sigma_j^+\sigma_{j+1}^-+\sigma_j^-\sigma_{j+1}^+\right)
+V_1\sum_{j=1}^{N-1}n_jn_{j+1}
\nonumber\\
&\quad+V_2\sum_{j=1}^{N-2}n_jn_{j+2},
\label{eq:mixing-H}
\end{align}
with \(V_1=J\) and \(V_2=0.7J\). The chain is divided into four equal cells, and the record \(R=(N_1,N_2,N_3,N_4)\) retains only their excitation numbers, with multiplicity \(\Omega_R=\prod_a\binom{\ell}{N_a}\). The preparation specifies that all excitations lie somewhere in the left half but does not resolve their microscopic positions. It therefore defines one thermodynamic class of fine initial configurations, each of which subsequently follows its own definite unitary history under the same Hamiltonian. The models and preparation projectors are specified in Note~\ref{supp:note-models} of the SM~\cite{SuppMat}.

For \(N=16\) and \(Q=2\), the normalized accessible entropy rises from \(10\%\) to \(90\%\) over \(Jt\simeq0.4\)--\(4.5\). The important point is not merely that the endpoints are ordered: the exactly evaluated instantaneous rate remains positive throughout this first-passage window, with minimum sampled value \(0.143\,k_{\rm B}J\). Thus entropy is being generated throughout the observed relaxation, rather than only ending at a larger value. The complete distribution follows an even stronger ordering. Every consecutive step of width \(\Delta(Jt)=0.1\), and all \(205\) steps obtained after refinement to \(\Delta(Jt)=0.02\), satisfy
\[
p(t)\succ_\pi p(t+\Delta t).
\]
Yet the microscopic information omitted from the current record is far from dynamically inert. For \(\tau=0.1/J\), \(\epsilon_{\rm hist}\) reaches \(1.98\times10^{-2}\), while \(\Vpi\) remains at the numerical floor, \(2.2\times10^{-16}\). The primary relaxation therefore realizes the nontrivial regime
\[
\epsilon_{\rm hist}>0,
\quad
\Vpi=0,
\quad
\dot S_R>0.
\]
Microscopic information changes the future thermodynamic record throughout the relaxation, but its influence does not organize into a return that overcomes the entropy-spreading currents.

The finite-step entropy decomposition gives a complementary view of the same dynamics. For the finite interval \(\tau=0.1/J\), both \(\Delta_\tau S_{\rm mix}/\tau\) and \(\Delta_\tau S_{\rm corr}/\tau\) are generally nonzero, and together reproduce the actual finite-step slope \(\Delta_\tau S_R/\tau\). These quantities should not be confused with instantaneous derivatives: the record-only reference accumulates a finite \(O(\tau^2)\) entropy change even though its true first-order rate vanishes, \(\dot S_{\rm mix}=0\), whereas
\begin{equation}\label{eq:short-time}
\frac{\Delta_\tau S_{\rm corr}}{\tau}
\longrightarrow
\dot S_R
\quad
(\tau\rightarrow0^+).
\end{equation}
The finite-step decomposition is therefore useful for resolved time intervals, but the irreversible entropy accumulation itself is determined locally by \(\dot S_R\).

The fact that return is absent during this mixing window is not imposed by the thermodynamic record. To demonstrate this directly, we keep the chain length, conserved excitation sector, preparation, four-cell record, and local hopping scale fixed, but deliberately reorganize the microscopic dynamics using
\begin{equation}
\begin{aligned}
H_{\rm ret}
&=
-\sum_{j=1}^{N-1}J_j
\left(\sigma_j^+\sigma_{j+1}^-+\sigma_j^-\sigma_{j+1}^+\right),
\end{aligned}
\label{eq:return-H}
\end{equation}
with
\(
J_j=\frac{2J}{N}\sqrt{j(N-j)}.
\)
This changes the interaction structure and, crucially, produces a commensurate spectrum. At
\(
t_{\rm mir}=(\pi N)/(4J),
\)
a fine configuration initially confined to the left half is transferred, up to phase, to its spatial mirror in the right half. The microscopic final state need not coincide with the initial state, but the initial and mirrored thermodynamic sectors have equal multiplicity. The hidden structure can therefore reassemble into a macroscopically low-entropy configuration even though no microscopic information has been destroyed or recreated.

\begin{figure}[t]
\centering
\includegraphics[width=1\linewidth]{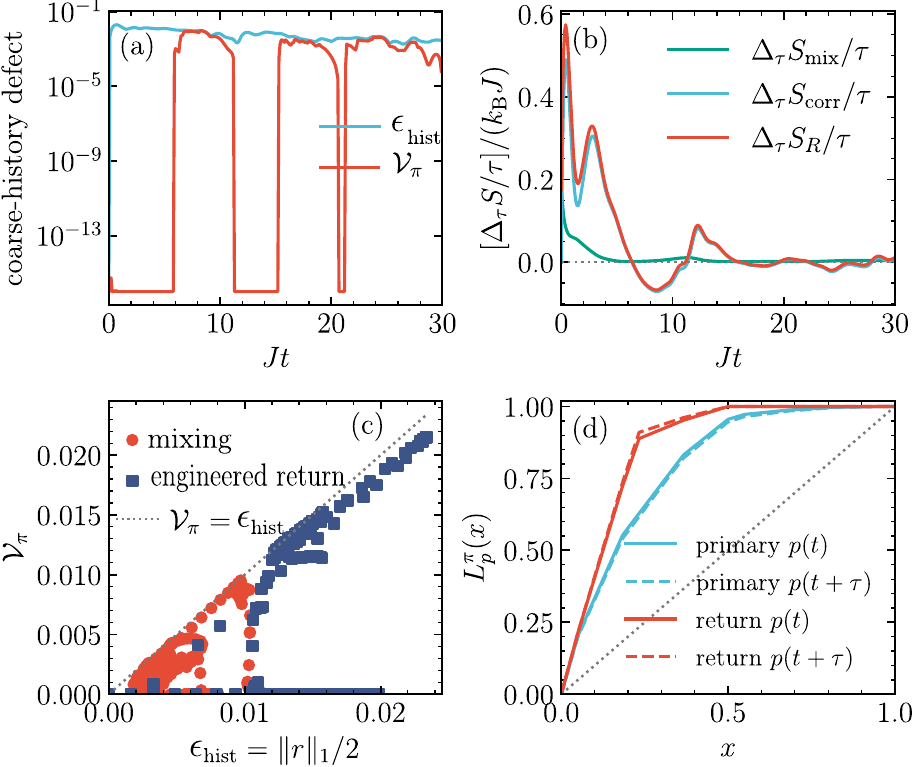}
\caption{
Hidden microscopic structure can remain dynamically active without producing thermodynamically relevant return.
(a) For the exact \(N=16\), \(Q=2\) mixing dynamics, \(\epsilon_{\rm hist}=\|r\|_1/2\) remains nonzero while \(\Vpi\) vanishes throughout primary relaxation.
(b) Finite-step slopes \(\Delta_\tau S_{\rm mix}/\tau\), \(\Delta_\tau S_{\rm corr}/\tau\), and \(\Delta_\tau S_R/\tau\) for \(\tau=0.1/J\). The record-only finite-step contribution is nonzero even though its instantaneous rate satisfies \(\dot S_{\rm mix}=0\).
(c) Direct numerical test of the exact bound \(\Vpi\leq\epsilon_{\rm hist}\) for mixing and engineered-return dynamics. Hidden activity with \(\Vpi=0\) and organized return with \(\Vpi>0\) are both realized.
(d) Representative generalized Lorenz curves. Primary mixing remains inside the \(\pi\)-preserving contraction order, whereas organized return produces a Lorenz-curve crossing.\justifying}
\label{fig:history-defect}
\end{figure}

Figure~\ref{fig:history-defect} brings together the different levels of this distinction. Panel~\figpanel{fig:history-defect}{a} shows the central mixing regime directly: \(\epsilon_{\rm hist}\) remains finite while \(\Vpi\) stays at numerical zero through the primary relaxation, demonstrating that microscopic feedback need not constitute thermodynamic return. Panel~\figpanel{fig:history-defect}{b} resolves the corresponding finite-step entropy balance. The nonzero \(\Delta_\tau S_{\rm mix}/\tau\) is a finite-interval contribution from the record-only reference, not an instantaneous entropy-generation rate; the hidden-state correction supplies the contribution that survives as \(\tau\to0\), while their sum gives the actual finite-step entropy slope.

Panel~\figpanel{fig:history-defect}{c} tests the connection between hidden activity and distributional return without reducing either to a scalar entropy. Every sampled point from both the mixing and engineered dynamics satisfies the exact bound \(\Vpi\leq\epsilon_{\rm hist}\). The mixing trajectory contains many steps with appreciable \(\epsilon_{\rm hist}\) but \(\Vpi=0\), whereas the engineered dynamics also enters the region \(\Vpi>0\): unresolved microscopic information that merely modifies the relaxing distribution in one dynamics becomes sufficiently organized to leave the contraction order in the other. Panel~\figpanel{fig:history-defect}{d} shows this geometrically. During primary mixing the later generalized Lorenz curve remains below the earlier one, whereas organized return produces a crossing. The figure therefore separates, in one view, microscopic activity, finite-step entropy response, and complete-distribution return.

\begin{figure}[t]
\centering
\includegraphics[width=1\linewidth]{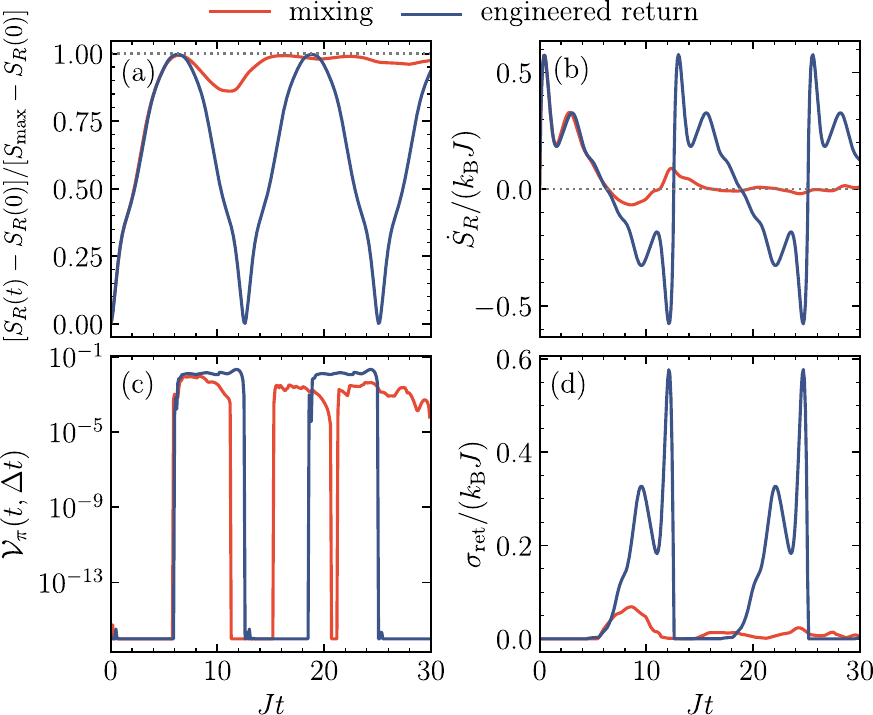}
\caption{
Controlled comparison between thermodynamic mixing and engineered return for the same \(N=16\), \(Q=2\) chain, preparation, and four-cell record.
(a) Normalized accessible entropy.
(b) Exact instantaneous accessible-entropy rate.
(c) Relative-majorization return defect for \(\Delta(Jt)=0.1\).
(d) Entropy-lowering macrocurrent contribution \(\sigma_{\rm ret}\). The mixing dynamics can sustain return-oriented microscopic currents without reversing the entropy accumulation, whereas the engineered spectrum synchronizes them strongly enough that \(\sigma_{\rm ret}\) eventually overtakes \(\sigma_{\rm spread}\) and drives macroscopic return.\justifying}
\label{fig:return-control}
\end{figure}

The purpose of the engineered Hamiltonian is constructive: it shows that the chosen record does not manufacture the arrow. Both Hamiltonians initially spread the thermodynamic record, but only the commensurate control later synchronizes the hidden microscopic structure strongly enough to reconstruct a low-entropy macrostate [Fig.~\ref{fig:return-control}]. For \(N=16\), \(Q=2\), and \(\tau=0.1/J\), the post-relaxation return defect reaches \(2.15\times10^{-2}\), while the finite-step entropy change reaches \(-5.75\times10^{-2}k_{\rm B}\). In the instantaneous current language this reversal has a direct meaning: the mixing window satisfies \(\sigma_{\rm spread}>\sigma_{\rm ret}\), whereas the engineered evolution later develops intervals with \(\sigma_{\rm ret}>\sigma_{\rm spread}\). Return is therefore not the reappearance of information that had vanished; it is a reversal in the thermodynamic orientation of currents carried by information that remained present throughout.

The exact current spectrum reveals what microscopic organization makes this reversal possible. At the stated numerical spectral resolution, the interacting mixing Hamiltonian has no repeated nonzero gap among distinct energies, so \(D_G=1\) and \(\Gamma_{\rm spec}=1\). A free nearest-neighbor chain provides an intermediate control, with \(D_G=26\) and \(\Gamma_{\rm spec}=12.55\), whereas the engineered Hamiltonian has \(D_G=28\) and \(\Gamma_{\rm spec}=20.24\). The distinction is more transparent in the distribution of current power: it occupies an effective \(152.4\) positive-gap channels for mixing, \(11.37\) for the free chain, and only \(2.30\) for engineered return, while the strongest single channel carries \(3.31\%\), \(19.33\%\), and \(63.02\%\), respectively. Gap degeneracy alone is therefore not the mechanism. What matters is whether the prepared state and the thermodynamic current operators use those degenerate gap classes coherently enough to concentrate currents that can contribute to return into a small set of synchronized frequencies.

This hierarchy persists across the exact-diagonalization fixed-\(Q=2\) sequence \(N=8,12,16,20\): \(\Gamma_{\rm spec}=1\) for the tested mixing spectra, while it grows from \(4.30\) to \(17.28\) for the free chain and from \(8.45\) to \(25.99\) for engineered return. We do not infer an asymptotic law from these four finite sizes. More importantly, the spectral description is not just a classification: reconstruction from the gap amplitudes reproduces \(\sigma_{\rm ret}\), \(\sigma_{\rm spread}\), and \(\dot S_R\) to better than \(10^{-15}\) in the central calculation. On the common post-relaxation window \(8\leq Jt\leq11\), the root-mean-square return-oriented entropy-rate contribution is \(0.05229\), \(0.07220\), and \(0.25312\,k_{\rm B}J\) for mixing, free, and engineered dynamics. These values are not determined by \(\Gamma_{\rm spec}\) alone---the affinities and total pair powers also change---but the exact reconstruction establishes that the gap-resolved amplitudes are the microscopic currents whose organization produces thermodynamic return. The full spectral comparison appears in Supplementary Figs.~\ref{fig:S20-spectral-coherence}--\ref{fig:S23-spectral-reconstruction}.

\section{Finite-model tests and the classical free-expansion endpoint}

A single one-dimensional example could reflect a special transport geometry. We therefore repeat the same diagnostics in an \(L=8\), \(Q=3\) interacting Bose--Hubbard chain, an interacting \(4\times4\) hard-core lattice with \(Q=3\), and a number-conserving Floquet unitary. Each model uses a four-region spatial record and an initial thermodynamic class confined to the left half.

\begin{figure}[t]
\centering
\includegraphics[width=1\linewidth]{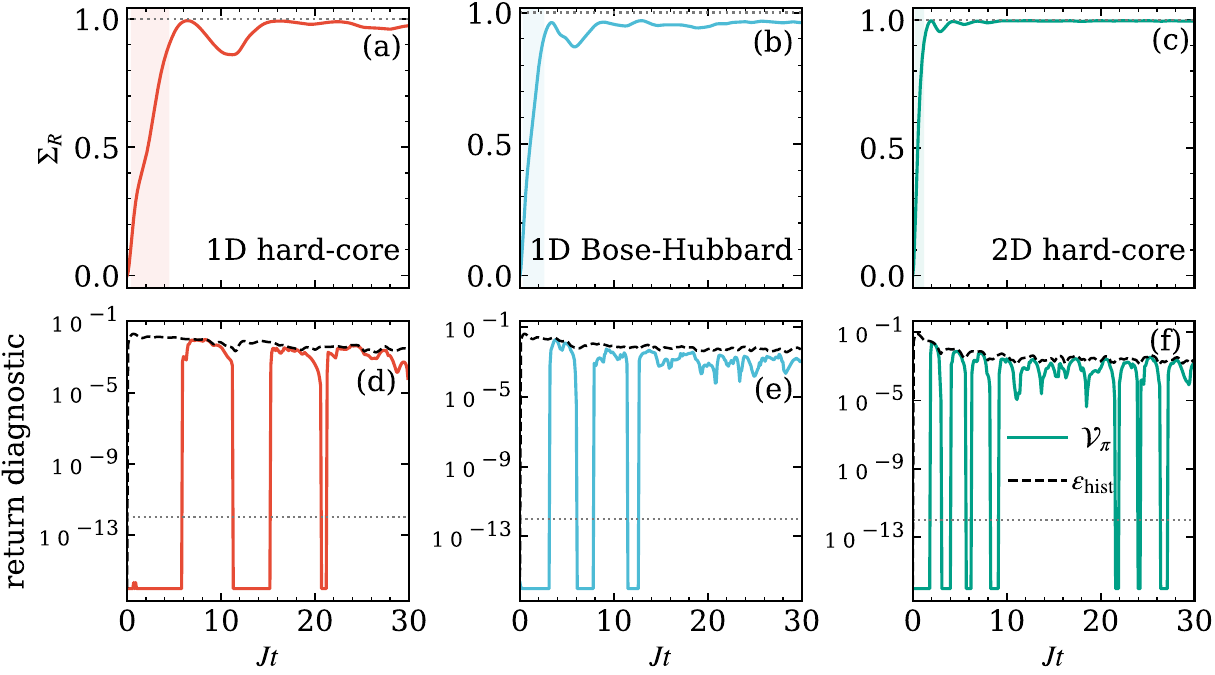}
\caption{Cross-model finite-system test.
(a)--(c) Normalized accessible entropy for the interacting one-dimensional hard-core chain, interacting Bose--Hubbard chain, and interacting two-dimensional hard-core lattice. Shading marks each first \(10\%\)--\(90\%\) relaxation window.
(d)--(f) Relative-majorization defect \(\Vpi\) and history-sensitive distance \(\epsilon_{\rm hist}\). In every tested primary window, hidden structure remains active while the return defect is zero to numerical precision.\justifying}
\label{fig:generality-models}
\end{figure}

The three static systems display the same qualitative separation. Their minimum sampled accessible-entropy rates over the primary windows are approximately \(0.143\), \(0.231\), and \(0.592\,k_{\rm B}J\), while their largest tested return defects are \(2.2\times10^{-16}\), \(0\), and \(0\). In contrast, the corresponding history-sensitive distances reach \(1.98\times10^{-2}\), \(4.10\times10^{-2}\), and \(7.78\times10^{-2}\) [Fig.~\ref{fig:generality-models}]. The Floquet realization gives the same finite-resolution pattern, with \(\epsilon_{\rm hist}\simeq1.17\times10^{-1}\) and numerical-zero \(\Vpi\). These calculations do not establish universality; they show that active hidden structure without detectable return is not tied to one particle statistic, dimensionality, or static Hamiltonian.

The stronger Lorenz order is not necessary for a positive entropy rate. Scanning \(0\leq V_1/J\leq2\) and \(0\leq V_2/J\leq1.5\), all 56 sampled points with \(V_1/J<2\) pass the primary-window relative-majorization test, while seven boundary points at \(V_1/J=2\) develop small defects, the largest being \(1.81\times10^{-4}\). Nevertheless, the minimum sampled accessible-entropy rate over all 63 points remains positive, \(0.0629\,k_{\rm B}J\). The refinement tests therefore support the intended hierarchy: relative majorization defines a strong finite-step contraction order, whereas the local entropy-generation condition \(\dot S_R>0\) can persist beyond it. Record, parameter, and time-step robustness are documented in the SM~\cite{SuppMat}.

\begin{figure}[t]
\centering
\includegraphics[width=1\linewidth]{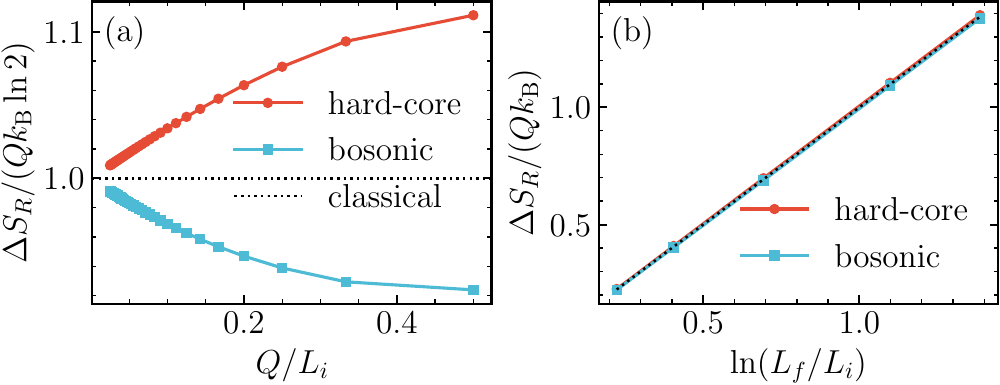}
\caption{Quantum-statistical approach to the dilute classical free-expansion endpoint.
(a) Exact entropy for volume doubling. Hard-core counting approaches \(Qk_{\rm B}\ln2\) from above and bosonic counting from below as \(Q/L_i\) decreases.
(b) Entropy per particle versus \(\ln(L_f/L_i)\) at low filling. Both statistics converge to \(\Delta S_R/(Qk_{\rm B})=\ln(L_f/L_i)\).\justifying}
\label{fig:statistics-limit}
\end{figure}

\begin{figure}[t]
\centering
\includegraphics[width=1\linewidth]{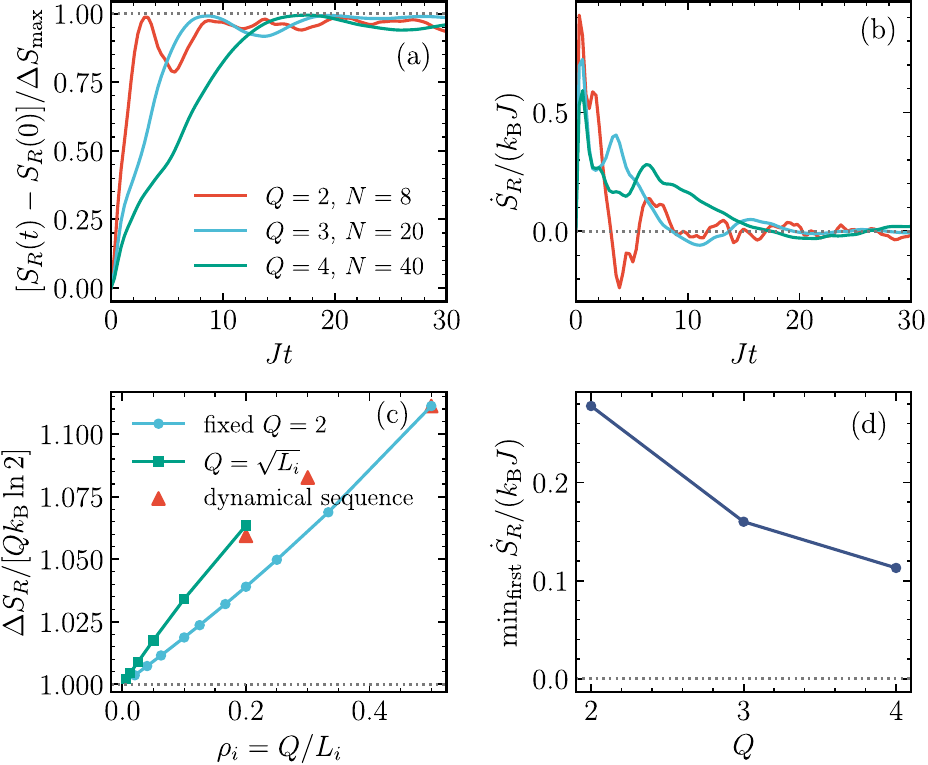}
\caption{
Closed-system free expansion along the finite sequence \((N,Q)=(8,2),(20,3),(40,4)\).
(a) Normalized four-cell entropy.
(b) Instantaneous accessible-entropy rates.
(c) Exact endpoint ratio versus initial filling.
(d) Minimum sampled rate in each primary relaxation window. The endpoint trend is combinatorial; the rate calculations are finite-size dynamical evidence.\justifying}
\label{fig:free-expansion}
\end{figure}

The free-expansion endpoint supplies an independent thermodynamic check. If \(Q\) hard-core excitations expand from \(L_i\) to \(L_f\) sites, the record entropy changes by
\begin{equation}
\Delta S_R^{\rm HC}
=
k_{\rm B}\ln\!\left[
\frac{\binom{L_f}{Q}}{\binom{L_i}{Q}}
\right].
\label{eq:HC-free-expansion}
\end{equation}
For unrestricted bosons, \(\Delta S_R^{\rm B}=k_{\rm B}\ln[\binom{L_f+Q-1}{Q}/\binom{L_i+Q-1}{Q}]\). In the dilute limit \(Q/L_i\to0\) at fixed \(L_f/L_i\), both statistics approach
\begin{equation}
\frac{\Delta S_R^{\rm HC}}{Qk_{\rm B}},
\frac{\Delta S_R^{\rm B}}{Qk_{\rm B}}
\longrightarrow
\ln\!\left(\frac{V_f}{V_i}\right),
\label{eq:classical-free-expansion}
\end{equation}
from opposite finite-density sides [Fig.~\ref{fig:statistics-limit}]. This endpoint is combinatorial and does not depend on the dynamical residual. The simultaneous sequence \((N,Q)=(8,2),(20,3),(40,4)\) separately shows that positive sampled entropy rates persist while the exact endpoint moves toward the dilute classical value [Fig.~\ref{fig:free-expansion}].

\section{Discussion}

The central result is not a new entropy functional. The macro-uniform representative, observational entropy, Bayesian/Petz interpretation, and entropy nondecrease when that representative is the actual initial state are established~\cite{strasberg2021firstsecond,buscemi2023coarse,nagasawa2024generic,bai2024priors,nagasawa2025macroscopicity}. Their role here is to define the state specified by the current record. The new dynamical content begins when the exact microscopic state is allowed to differ from that representative and is propagated without erasing the difference. The resulting residual keeps the complete effect of the unresolved structure on the future macrodistribution. Because the record-only contribution has zero first-order rate, that residual carries the entire instantaneous motion of the record.

This also makes clear where physics enters the thermodynamic description. The record and its entropy define what is observed, but they do not determine the direction of motion. The Hamiltonian does. Through the exact blocks \(P_RHP_{R'}\), together with the hidden structure in \(\chi_t\), it creates the currents that either spread probability or drive it back toward a more ordered record. The local entropy balance therefore turns irreversibility into a physical competition: hidden currents may be active in both directions, but entropy is generated while the spreading contribution remains larger than the return-oriented one. The amount of information hidden from a record is therefore not enough to determine irreversibility; what matters is how the actual microscopic dynamics organizes that information.

Relative majorization asks a different, stronger finite-step question about the complete distribution. The regime \(\epsilon_{\rm hist}>0\) with \(\Vpi=0\) shows that microscopic structure can influence future observables without becoming organized strongly enough to leave the multiplicity-preserving contraction order. Conversely, leaving that order need not immediately make the instantaneous entropy rate negative. The local Second-Law condition and the finite-step Lorenz ordering therefore probe different levels of thermodynamic behavior.

The spectral decomposition identifies one concrete way in which the Hamiltonian can organize the hidden currents. It separates the pairwise transition content \(P_{\rm pair}\) from coherent addition among transitions sharing the same gap, measured by \(\Gamma_{\rm spec}\). Gap degeneracy is neither a sufficient nor a complete criterion for entropy decrease: nondegenerate gaps forbid equal-gap enhancement but do not eliminate current power, while a large \(\Gamma_{\rm spec}\) does not determine the sign of the current--affinity product. The controlled comparison nevertheless exposes a clear hierarchy. Mixing distributes the thermodynamic current over many frequencies; the free chain retains moderate equal-gap organization; the commensurate control concentrates most current power into a few synchronized channels and produces macroscopic return. The controls differ in more than one Hamiltonian parameter, so we do not claim a one-parameter causal isolation. Their purpose is instead to show directly that the same thermodynamic record can display relaxation or return depending on how the microscopic dynamics organizes the actual current-carrying amplitudes.

These finite-model tests show the range of the result rather than establishing universality. The hard-core, Bose--Hubbard, two-dimensional, and Floquet examples all show positive entropy-generation rates over the reported primary windows, and the engineered control shows that recurrence and return remain possible. No finite-dimensional system is claimed to have monotonically increasing record entropy for all times, and no thermodynamic-limit theorem is inferred from the tested size sequences. A broader result would have to derive both suppression of return-oriented current power and a positive spreading margin from locality, operator spreading, transport, and record resolution. Generic effective-dimension and gap-concentration bounds remain valuable sufficient conditions, but their finite-time norm bounds are quantitatively conservative for the systems studied here and are not used to establish the observed windows.

The classical counterpart clarifies which part of the structure is specifically quantum. The stochastic cell-transition matrix is the established Ulam object, and the relevant--irrelevant split belongs to the Zwanzig--Mori lineage. What is added is the exact residual completion and its thermodynamic use: a multiplicity-preserving record-only evolution, a retained vector that localizes every departure from it, and a distributional return bound. Quantum mechanics adds the quadratic short-time onset between orthogonal macrospaces and genuinely quantum contributions to \(\chi_t\), but the finite-step contraction--residual structure follows more generally from reversible microscopic dynamics and a representative uniform with respect to the invariant microscopic measure.

The present paper deliberately stops before assigning heat or work to the autonomous relaxation. Heat and work are boundary-transfer modes and become meaningful only after a thermodynamic boundary, control class, and comparator have been specified. A companion study applies the same restricted-information framework to matched finite-particle quantum and classical Hamiltonian dynamics and shows explicitly how microscopic physics changes thermodynamic irreversibility: the quantum system can relax more strongly in the chosen spatial record while a larger fraction of the hidden information remains resolvable once energy is retained. This produces a smaller residual irreversibility and, in a specified cycle, a corresponding work and efficiency advantage~\cite{ahmadi2026finiteparticle}. Keeping that operational question separate sharpens the present claim. Within the chosen record and observation window, thermodynamic irreversibility is not the destruction of microscopic information, but the failure of the information that survives to organize currents that restore macroscopic order.

\section{Acknowledgments}
BA acknowledges support from IRA Program (project no. FENG.02.01-IP.05-0006/23) financed by the FENG program 2021-2027, Priority FENG.02, Measure FENG.02.01., with the support of the FNP.

\bibliography{References}

\clearpage
\onecolumngrid


\setcounter{section}{0}
\setcounter{subsection}{0}
\setcounter{equation}{0}
\setcounter{figure}{0}
\setcounter{table}{0}

\renewcommand{\thesection}{\arabic{section}}
\renewcommand{\thesubsection}{\Alph{subsection}}

\makeatletter
\renewcommand{\p@subsection}{\thesection}
\makeatother

\renewcommand{\theequation}{S\arabic{equation}}
\renewcommand{\thefigure}{S\arabic{figure}}
\renewcommand{\thetable}{S\arabic{table}}

\newcommand{\suppnote}[1]{%
    \refstepcounter{section}%
    \setcounter{subsection}{0}%
    \section*{Supplementary Note \arabic{section}. #1}%
    \addcontentsline{toc}{section}{Supplementary Note \arabic{section}. #1}%
}

\newcommand{\suppsubsection}[1]{%
    \refstepcounter{subsection}%
    \subsection*{\Alph{subsection}. #1}%
    \addcontentsline{toc}{subsection}{\Alph{subsection}. #1}%
}

\section*{Supplementary Materials for ``Thermodynamic Irreversibility from Inaccessible Endogenous Quantum Histories''}

This Supplementary Material develops the derivations, numerical definitions, robustness tests, and convergence checks supporting the main text. The foundational setting is one isolated system whose microscopic state evolves unitarily while thermodynamic quantities are defined relative to a restricted record. No bath, ancillary stream, partial trace, or information-destroying microscopic process is invoked as the origin of irreversibility. Note~\ref{supp:note-coarse-entropy} reviews the representative and entropy in their observational, Bayesian, and maximum-entropy lineage. Note~\ref{supp:note-projection-classical} gives the precise projection-operator interpretation and proves the exact classical measure-preserving counterpart. Notes~\ref{supp:note-macrocurrents} and~\ref{supp:note-spectral-return} develop the instantaneous current balance and its state-resolved spectral organization.

\suppnote{Thermodynamic coarse graining and the accessible entropy}
\label{supp:note-coarse-entropy}

The construction below combines Boltzmannian macrospace multiplicity and Gibbsian fine-state preservation with the Jaynesian maximum-entropy representative. For one projective partition its entropy is observational entropy~\cite{boltzmann1877beziehung,gibbs1902elementary,jaynes1957information,safranek2019thermodynamics,safranek2019thermalization}. The same representative is established in modern information-theoretic treatments as the coarse-grained state inferred from the record by Bayesian retrodiction or Petz recovery, and conventional observational entropy corresponds to a uniform microscopic prior~\cite{buscemi2023coarse,bai2024priors,nagasawa2025macroscopicity}. These antecedents are part of the definition and are not claimed as new here.

\suppsubsection{Fine histories, fine initial states, and the accessible record}
\label{supp:fine-histories}

For a fixed autonomous Hamiltonian \(H\), one exact fine initial state determines one exact microscopic history. If the initial state associated with history \(\gamma\) is \(\rho_{\gamma,0}\), then
\begin{equation}
\gamma
=
\left\{
\rho_\gamma(s)
=
e^{-iHs}
\rho_{\gamma,0}
e^{iHs}
:
0\leq s\leq T
\right\}.
\end{equation}
Thus \(\gamma\) is the complete microscopic trajectory of the same isolated system over the interval of interest. Once both \(H\) and \(\rho_{\gamma,0}\) are fixed, there is no microscopic branching. In particular, if \(\rho_{\gamma_1,0}\neq\rho_{\gamma_2,0}\), then the corresponding histories are distinct because unitary evolution is invertible.

A thermodynamic preparation, however, generally does not determine the exact fine initial state completely. Let \(R_0\) denote the accessible preparation record and \(\mathcal X_{R_0}\) the set of fine initial states compatible with it. Each \(x\in\mathcal X_{R_0}\) determines one definite microscopic history,
\begin{equation}
\gamma_x
=
\left\{
\rho_x(s)
=
e^{-iHs}
\rho_{x,0}
e^{iHs}
:
0\leq s\leq T
\right\}.
\end{equation}
The multiplicity of histories in the autonomous examples therefore arises solely from the multiplicity of microscopic initial states compatible with the same thermodynamic preparation. It does not arise from different Hamiltonians, stochastic branching, or several possible evolutions of one exact initial state.

More generally, let \(R_t\) denote the thermodynamic information retained at time \(t\), and let \(\mathfrak R\) denote the rule that extracts this accessible information from a complete microscopic history. Histories that are indistinguishable under that record form the compatibility class
\begin{equation}
\Cclass_{R_t}
=
\left\{
\gamma:
\mathfrak R[\gamma]
=
R_t
\right\}.
\end{equation}
The exact microscopic history remains definite. The class \(\Cclass_{R_t}\) specifies only which microscopic distinctions are unresolved by the chosen thermodynamic description.

In the many-qubit calculations this distinction is especially explicit. A fine product-state preparation is specified by a microscopic occupation pattern \(x\), with \(\rho_{x,0}=|x\rangle\langle x|\). Two different patterns can possess the same coarse record, for example the same numbers of excitations in four spatial cells, while generating different unitary trajectories under the same Hamiltonian \(H\). The thermodynamic description retains the coarse record and leaves the exact occupation pattern unresolved.

We finally distinguish \(R_t\) from the symbol \(R\) used to label the macrospaces below. The former denotes the thermodynamic information accessible at time \(t\), whereas \(R\) labels one possible value of the chosen macrovariable. When the exact state is not confined to a single macrospace, the thermodynamic description is therefore specified by the distribution \(p_R(t)\) over the possible values \(R\).

\suppsubsection{Macroprojectors and the thermodynamic representative}
\label{supp:macroprojectors}

Let \(\mathcal H\) denote the Hilbert space of the isolated system or, when exact conserved quantities are fixed, the corresponding invariant sector. The thermodynamic record partitions this space into mutually orthogonal macrospaces labeled by \(R\), with projectors \(P_R\) satisfying
\begin{equation}
P_RP_{R'}
=
\delta_{RR'}P_R,
\quad
\sum_RP_R
=
I_{\mathcal H}.
\end{equation}
The first relation expresses the mutual exclusivity of distinct thermodynamic alternatives, while the second ensures that the macrospaces exhaust the entire Hilbert sector relevant to the thermodynamic description.

The dimension of macrospace \(R\),
\begin{equation}
\Omega_R
=
\Tr P_R,
\end{equation}
is its microscopic multiplicity: the number of mutually orthogonal microscopic directions compatible with the same thermodynamic record. The total dimension of the partitioned sector is
\begin{equation}
\Omega_{\rm tot}
=
\sum_R\Omega_R
=
\dim\mathcal H.
\end{equation}
For an exact microscopic state \(\rho\), the probability assigned to the accessible outcome \(R\) is
\begin{equation}
p_R
=
\Tr(P_R\rho).
\end{equation}
The quantities \(\Omega_R\) and \(p_R\) therefore have distinct roles. The multiplicity \(\Omega_R\) is fixed by the chosen thermodynamic partition and quantifies how much microscopic structure lies beneath record \(R\), whereas \(p_R\) depends on the exact microscopic state and gives its current probability weight in that macrospace.

The state containing exactly the accessible probabilities \(\{p_R\}\) and no additional resolved information within the macrospaces is
\begin{equation}
\cG[\rho]
=
\sum_R
\frac{p_R}{\Omega_R}
P_R.
\end{equation}
Within each macrospace, \(\cG[\rho]\) assigns the total probability \(p_R\) uniformly over the \(\Omega_R\) microscopic directions that the thermodynamic record does not distinguish. This uniformity is a property of the thermodynamic representative, not an assertion that the exact microscopic state \(\rho\) is uniform within that macrospace.

Crucially, \(\cG[\rho]\) is defined on the same physical Hilbert space as \(\rho\). It is not obtained by tracing over a bath, environment, ancillary system, or any other physical subsystem. The exact state retains all microscopic structure; \(\cG[\rho]\) represents only the information retained by the specified thermodynamic record.

The map \(\cG\) preserves normalization,
\begin{equation}
\Tr\cG[\rho]
=
\sum_Rp_R
=
1,
\end{equation}
and is idempotent,
\begin{equation}
\cG[\cG[\rho]]
=
\cG[\rho].
\end{equation}
It also preserves every accessible macroprobability,
\begin{equation}
\Tr
\left[
P_R\cG[\rho]
\right]
=
p_R.
\end{equation}
Hence applying \(\cG\) removes only distinctions that are absent from the chosen record; applying it again removes nothing further. Two microscopic states \(\rho\) and \(\rho'\) that produce the same thermodynamic representative,
\begin{equation}
\cG[\rho]
=
\cG[\rho'],
\end{equation}
are therefore indistinguishable with respect to that thermodynamic record, even though they may remain microscopically distinct.

\suppsubsection{Accessible entropy}
\label{supp:accessible-entropy}

The entropy associated with the restricted thermodynamic description is defined as the von Neumann entropy of its representative,
\begin{equation}
S_R(\rho)
=
-k_{\rm B}
\Tr
\left[
\cG[\rho]
\ln\cG[\rho]
\right].
\end{equation}
Since \(\cG[\rho]\) has eigenvalue \(p_R/\Omega_R\) with multiplicity \(\Omega_R\) inside macrospace \(R\), this becomes
\begin{equation}
S_R
=
-k_{\rm B}
\sum_R
p_R
\ln
\left(
\frac{p_R}{\Omega_R}
\right)
=
-k_{\rm B}\sum_Rp_R\ln p_R
+
k_{\rm B}\sum_Rp_R\ln\Omega_R.
\end{equation}
The first contribution is the Shannon entropy of the accessible macrorecord, while the second accounts for the unresolved microscopic multiplicity within the occupied macrospaces. Thus \(S_R\) contains both uncertainty over the accessible alternatives and the microscopic multiplicity hidden beneath each alternative.

The macrospace dimensions define the multiplicity-weighted reference distribution
\begin{equation}
\pi_R
=
\frac{\Omega_R}{\Omega_{\rm tot}}.
\end{equation}
Equivalently, \(\pi\) is the macrodistribution generated by uniform microscopic weight over the chosen Hilbert sector. It is defined solely by the thermodynamic partition and the sector dimension; it is not, by definition, either the macrodistribution of an infinite-time averaged orbit or an energy-matched finite-temperature Gibbs state. Those reference distributions depend on additional dynamical or energetic information and need not coincide with \(\pi\).

Using \(\Omega_R=\Omega_{\rm tot}\pi_R\), the accessible entropy can be written exactly as
\begin{equation}
S_R
=
k_{\rm B}
\left[
\ln\Omega_{\rm tot}
-
D(p\Vert\pi)
\right],
\quad
D(p\Vert\pi)
=
\sum_R
p_R
\ln
\left(
\frac{p_R}{\pi_R}
\right).
\end{equation}
This representation makes the thermodynamic meaning of the construction particularly transparent. At fixed Hilbert sector, \(\Omega_{\rm tot}\) is constant, so increasing accessible entropy is exactly equivalent to decreasing the distinguishability of the macrodistribution \(p\) from the multiplicity-defined reference \(\pi\),
\begin{equation}
S_R(t_2)
\geq
S_R(t_1)
\quad\Longleftrightarrow\quad
D[p(t_2)\Vert\pi]
\leq
D[p(t_1)\Vert\pi].
\end{equation}
When \(\pi\) is also the appropriate thermodynamic equilibrium macrodistribution for the imposed macroscopic constraints, this contraction has the usual interpretation as approach toward equilibrium; that additional identification is not contained in the algebraic identity itself.

The maximum accessible entropy is therefore
\begin{equation}
S_{\max}
=
k_{\rm B}\ln\Omega_{\rm tot},
\end{equation}
and is attained when \(p_R=\pi_R\) for every macrospace. At this point the probability per microscopic direction, \(p_R/\Omega_R\), is uniform throughout the chosen Hilbert sector.

It is essential to distinguish \(S_R\) from the microscopic von Neumann entropy of the exact isolated state. An exact pure state may satisfy \(S_{\rm vN}(\rho)=-k_{\rm B}\Tr(\rho\ln\rho)=0\) while \(S_R>0\). The latter does not signify destruction of microscopic information: it quantifies the information and multiplicity left unresolved by the specified thermodynamic record. Consequently, neither positive \(S_R\) nor an increase of \(S_R\) requires the exact microscopic state to become mixed or to lose information.

\suppsubsection{Information-theoretic status and observational deficit}
\label{supp:coarse-state-deficit}

For the projective record considered here, the representative in Eq.~(S8) is the coarse-grained state obtained by applying the Petz recovery map associated with the maximally mixed prior to the quantum-to-classical record channel~\cite{buscemi2023coarse}. Equivalently, it is the Bayesian-retrodicted state inferred from the outcome probabilities and the projectors. The prior is uniform on the chosen invariant Hilbert sector; its induced macrodistribution is therefore \(\pi_R=\Omega_R/\Omega_{\rm tot}\). General-prior observational entropies replace this uniform microscopic prior by another reference state~\cite{bai2024priors}, while recent work unifies the measurement-based and Jaynesian constructions through informational priors~\cite{schindler2025unification}.

The entropy excess over the exact von Neumann entropy has a simple relative-entropy form. Since \(\ln\cG[\rho]\) is constant inside each macrospace and \(\rho\) and \(\cG[\rho]\) have the same macroprobabilities,
\begin{equation}
\Tr\!\left[\rho\ln\cG[\rho]\right]
=
\Tr\!\left[\cG[\rho]\ln\cG[\rho]\right].
\end{equation}
Consequently,
\begin{equation}
\begin{aligned}
D\!\left(\rho\Vert\cG[\rho]\right)
&=
\Tr(\rho\ln\rho)
-
\Tr\!\left[\rho\ln\cG[\rho]\right]
\\
&=
\frac{S_R(\rho)-S_{\rm vN}(\rho)}{k_{\rm B}}.
\end{aligned}
\label{eq:supp-observational-deficit}
\end{equation}
This quantity is the observational deficit or relative-entropy microscopicity associated with the chosen inferential frame~\cite{buscemi2023coarse,nagasawa2025macroscopicity}. It is a scalar measure of how much of the present microscopic state cannot be recovered from the current record. It must be distinguished from the vector \(r(t,\tau)\) introduced below: \(r\) records how the unresolved component is dynamically converted into changes of the next macrodistribution. States having the same scalar deficit can therefore generate different residual vectors, and a nonzero deficit need not produce a relative-majorization violation over a particular step.

\suppnote{Exact finite-time coarse propagation of a closed unitary system}
\label{supp:note-coarse-propagation}

The separation into thermodynamically retained and unresolved structure is a particular relevant--irrelevant projection in the sense of Zwanzig and Mori~\cite{zwanzig1960ensemble,zwanzig1961memory,mori1965transport}. When the exact state already equals its macro-uniform representative, the observational entropy is known not to decrease under unitary evolution, and generic unitaries increase it strictly~\cite{strasberg2021firstsecond,nagasawa2024generic}. The exact finite-step identity derived here uses that established macrostate evolution as a baseline for an arbitrary current microscopic state and retains its missing contribution explicitly. Its precise block-propagator relation to memory and source terms is developed in Note~\ref{supp:note-projection-classical}.

\suppsubsection{Decomposition of the microscopic state}
\label{supp:state-decomposition}

Let the exact state of the closed system at time \(t\) be \(\rho_t\). We decompose it exactly into the thermodynamic representative determined by the current accessible record and a complementary microscopic contribution,
\begin{equation}
\rho_t
=
\cG[\rho_t]
+
\chi_t,
\quad
\chi_t
=
\rho_t-\cG[\rho_t].
\end{equation}
Because \(\cG[\rho_t]\) reproduces all current macroprobabilities, the remainder satisfies
\begin{equation}
\Tr(P_R\chi_t)
=
0
\quad
\forall R,
\quad
\cG[\chi_t]
=
0.
\end{equation}
Thus \(\chi_t\) contains precisely microscopic structure that is invisible to the present thermodynamic record. It may include nonuniform structure within a macrospace, coherences between microscopic basis states, and phase relations that do not affect the current values of \(p_R(t)\). Importantly, invisibility to the record at time \(t\) does not imply dynamical irrelevance: subsequent unitary evolution can convert information contained in \(\chi_t\) into changes of future macroprobabilities.

Let
\begin{equation}
U_\tau
=
e^{-iH\tau}
\end{equation}
be the exact propagator over a finite interval \(\tau\), with \(\hbar=1\). The exact future macroprobability is
\begin{equation}
p_R(t+\tau)
=
\Tr
\left[
P_R
U_\tau
\rho_t
U_\tau^\dagger
\right].
\end{equation}
Substituting \(\rho_t=\cG[\rho_t]+\chi_t\) gives
\begin{equation}\label{pRS}
\begin{aligned}
p_R(t+\tau)
={}&
\Tr
\left[
P_R
U_\tau
\cG[\rho_t]
U_\tau^\dagger
\right]
+
\Tr
\left[
P_R
U_\tau
\chi_t
U_\tau^\dagger
\right].
\end{aligned}
\end{equation}
Using
\begin{equation}
\cG[\rho_t]
=
\sum_{R'}
\frac{p_{R'}(t)}{\Omega_{R'}}
P_{R'},
\end{equation}
we obtain the exact finite-step decomposition
\begin{equation}
p(t+\tau)
=
K^{(U)}(\tau)p(t)
+
r(t,\tau),
\end{equation}
with
\(
K^{(U)}_{RR'}(\tau)
=
\frac{1}{\Omega_{R'}}
\Tr
\left[
P_R
U_\tau
P_{R'}
U_\tau^\dagger
\right],
\)
and
\(
r_R(t,\tau)
=
\Tr
\left[
P_R
U_\tau
\chi_t
U_\tau^\dagger
\right].
\)
These two terms have distinct meanings. For a source macrospace \(R'\), the factor \(1/\Omega_{R'}\) averages uniformly over the microscopic directions that the thermodynamic description does not distinguish within that macrospace. Consequently, \(K^{(U)}_{RR'}(\tau)\) is the probability of reaching macrospace \(R\) after the unitary step when the source macrospace \(R'\) is represented only by its thermodynamic state \(P_{R'}/\Omega_{R'}\). The vector \(K^{(U)}(\tau)p(t)\) is therefore the future macrodistribution predicted from the current thermodynamic representative \(\cG[\rho_t]\) alone.

The correction \(r(t,\tau)\), by contrast, is generated entirely by \(\chi_t\). It gives the difference between this macro-uniform prediction and the exact future macrodistribution and therefore isolates the influence of microscopic information that is absent from the present accessible probabilities. In particular, \(r(t,\tau)\neq0\) means that information unresolved by the thermodynamic record at time \(t\) remains capable of influencing later thermodynamic observables.

The decomposition is an exact identity. No stochastic dynamics, Markov approximation, equilibration hypothesis, or loss of microscopic information has been introduced. The matrix \(K^{(U)}\) is not postulated as a fundamental microscopic evolution law; it is the coarse propagation induced by the same exact unitary \(U_\tau\) when the state at time \(t\) is replaced by the thermodynamic representative containing no information beyond the accessible record.

If \(\chi_t=0\), then \(\rho_t=\cG[\rho_t]\) is a macroscopic state for the chosen record and \(r(t,\tau)=0\). In that case Eq.~\eqref{pRS} reduces to the established observational-entropy evolution of a macro-uniform state. The role of Eq.~\eqref{pRS} is therefore not to rediscover that H theorem, but to embed it into an exact identity for an arbitrary present state. The residual is the complete correction that is absent when the initial state of the step is already determined by the record.

\suppsubsection{Positivity and stochasticity of \texorpdfstring{\(K^{(U)}\)}{K(U)}}
\label{supp:K-stochastic}

The matrix \(K^{(U)}(\tau)\) is nonnegative and column stochastic for every unitary \(U_\tau\), every finite \(\tau\), and every projective thermodynamic partition.

To establish positivity, define
\begin{equation}
A_{RR'}
=
P_RU_\tau P_{R'}.
\end{equation}
Using \(P_R^2=P_R\) and cyclicity of the trace,
\begin{equation}
\begin{aligned}
\Tr
\left[
P_R
U_\tau
P_{R'}
U_\tau^\dagger
\right]
&=
\Tr
\left[
P_R
U_\tau
P_{R'}
U_\tau^\dagger
P_R
\right]
=
\Tr
\left[
A_{RR'}A_{RR'}^\dagger
\right]
\geq
0.
\end{aligned}
\end{equation}
Since \(\Omega_{R'}>0\) for every macrospace included in the partition,
\begin{equation}
K^{(U)}_{RR'}(\tau)
\geq
0.
\end{equation}
The column sums follow from completeness of the macroprojectors:
\begin{equation}
\begin{aligned}
\sum_R
K^{(U)}_{RR'}(\tau)
&=
\frac{1}{\Omega_{R'}}
\sum_R
\Tr
\left[
P_R
U_\tau
P_{R'}
U_\tau^\dagger
\right]
\\
&=
\frac{1}{\Omega_{R'}}
\Tr
\left[
U_\tau
P_{R'}
U_\tau^\dagger
\right]
\\
&=
\frac{1}{\Omega_{R'}}
\Tr P_{R'}
=
1.
\end{aligned}
\end{equation}
Hence \(K^{(U)}(\tau)\) is column stochastic when probability distributions are represented as column vectors. This stochasticity is not an additional dynamical assumption: it follows directly from the unitary evolution and the complete projective partition of the chosen Hilbert sector.

\suppsubsection{Exact preservation of the multiplicity-defined reference}
\label{supp:K-pi}

The multiplicity-defined reference distribution
\begin{equation}
\pi_R
=
\frac{\Omega_R}{\Omega_{\rm tot}}
\end{equation}
is an exact fixed point of the coarse matrix \(K^{(U)}(\tau)\). Indeed,
\begin{equation}
\begin{aligned}
\sum_{R'}
K^{(U)}_{RR'}(\tau)\pi_{R'}
&=
\frac{1}{\Omega_{\rm tot}}
\sum_{R'}
\Tr
\left[
P_R
U_\tau
P_{R'}
U_\tau^\dagger
\right]
\\
&=
\frac{1}{\Omega_{\rm tot}}
\Tr
\left[
P_R
U_\tau
\left(
\sum_{R'}P_{R'}
\right)
U_\tau^\dagger
\right]
\\
&=
\frac{1}{\Omega_{\rm tot}}
\Tr
\left[
P_R
U_\tau
U_\tau^\dagger
\right]
=
\frac{\Tr P_R}{\Omega_{\rm tot}}
=
\frac{\Omega_R}{\Omega_{\rm tot}}
=
\pi_R.
\end{aligned}
\end{equation}
Therefore
\begin{equation}
K^{(U)}(\tau)\pi
=
\pi
\end{equation}
for every unitary \(U_\tau\), every finite propagation interval \(\tau\), and every projective thermodynamic partition of the chosen Hilbert sector.

This preservation law is purely kinematic. It follows from unitarity, completeness of the macroprojectors, and the definition \(\pi_R=\Omega_R/\Omega_{\rm tot}\); it requires neither detailed balance, Markovianity, chaos, thermalization, nor an external environment. In particular, it should not be interpreted as a statement that the actual distribution \(p(t)\) must dynamically approach \(\pi\). Whether the complete trajectory contracts toward \(\pi\) depends additionally on the history-sensitive contribution \(r(t,\tau)\).

\suppsubsection{Normalization of the history-sensitive correction}
\label{supp:history-correction}

The history-sensitive correction carries no net probability. Summing its definition over all macrospaces and using \(\sum_RP_R=I_{\mathcal H}\) gives
\begin{equation}
\begin{aligned}
\sum_R
r_R(t,\tau)
&=
\Tr
\left[
U_\tau
\chi_t
U_\tau^\dagger
\right]
\\
&=
\Tr\chi_t
\\
&=
0,
\end{aligned}
\end{equation}
where \(\Tr\chi_t=\Tr\rho_t-\Tr\cG[\rho_t]=0\). Equivalently, the same result follows because both \(p(t+\tau)\) and \(K^{(U)}(\tau)p(t)\) are normalized probability distributions. Thus \(r(t,\tau)\) redistributes probability among the accessible macrostates without changing the total probability.

A natural measure of its magnitude is
\begin{equation}
\epsilon_{\rm hist}(t,\tau)
=
\frac{1}{2}
\left\|
r(t,\tau)
\right\|_1.
\end{equation}
Using the exact finite-step decomposition,
\begin{equation}
\epsilon_{\rm hist}(t,\tau)
=
\frac{1}{2}
\left\|
p(t+\tau)
-
K^{(U)}(\tau)p(t)
\right\|_1.
\end{equation}
The factor \(1/2\) makes \(\epsilon_{\rm hist}\) exactly the total-variation distance between the true future macrodistribution and the distribution predicted from the present thermodynamic representative alone.

Accordingly, \(\epsilon_{\rm hist}(t,\tau)>0\) has a precise operational meaning: the current accessible probabilities \(p_R(t)\) are insufficient by themselves to determine the future accessible probabilities at \(t+\tau\). Microscopic information contained in \(\chi_t\), although absent from the present thermodynamic record, remains dynamically active and affects the later record.

This observation must be kept distinct from thermodynamic return. A nonzero history-sensitive correction need not oppose thermodynamic relaxation; it may alter the future macrodistribution while remaining compatible with further contraction toward the multiplicity-defined reference. Thus
\begin{equation}
\epsilon_{\rm hist}(t,\tau)
>
0
\end{equation}
signals dynamical dependence on inaccessible microscopic information, not by itself an entropy decrease or a violation of relative majorization. The stronger question of whether this microscopic influence becomes organized into thermodynamically relevant return is addressed in the following Supplementary Notes.

\suppsubsection{Short-time limit and vanishing baseline rate}
\label{supp:short-time-baseline}

The finite-step baseline is nondecreasing, but it has no nonzero first-order rate. Since \(U_0=I\),
\begin{equation}
K^{(U)}_{RR'}(0)
=
\delta_{RR'}.
\end{equation}
Differentiating the exact matrix element gives
\begin{equation}
\begin{aligned}
\left.
\partial_\tau K^{(U)}_{RR'}(\tau)
\right|_{\tau=0}
&=
-\frac{i}{\Omega_{R'}}
\Tr\!\left[
P_R[H,P_{R'}]
\right]
\\
&=
0.
\end{aligned}
\label{eq:supp-K-first-derivative}
\end{equation}
For \(R\neq R'\), both traces vanish by projector orthogonality and cyclicity; for \(R=R'\), they cancel. Hence
\begin{equation}
K^{(U)}(\tau)
=
I+O(\tau^2).
\label{eq:supp-K-quadratic}
\end{equation}
More explicitly, for \(R\neq R'\),
\begin{equation}
K^{(U)}_{RR'}(\tau)
=
\frac{\tau^2}{\Omega_{R'}}
\Tr\!\left(
P_RHP_{R'}H
\right)
+O(\tau^3),
\label{eq:supp-K-offdiagonal-second}
\end{equation}
while the diagonal second-order term follows from column normalization.

Let \(m>1\) be the number of macrostates (the case \(m=1\) is trivial) and set
\begin{equation}
\delta_\tau
=
\frac12
\left\|
K^{(U)}(\tau)p-p
\right\|_1
=
O(\tau^2).
\end{equation}
The record entropy is the sum of the Shannon entropy and the linear multiplicity term. For interior distributions, differentiability immediately gives \(\Delta_\tau S_{\rm mix}=O(\tau^2)\). The conclusion also holds on the boundary of the probability simplex. The Fannes--Audenaert continuity bound gives~\cite{audenaert2007continuity}
\begin{equation}
\left|
H[K^{(U)}(\tau)p]-H(p)
\right|
\leq
\delta_\tau\ln(m-1)
+
h_2(\delta_\tau),
\end{equation}
where \(h_2\) is the binary entropy, while the multiplicity term changes by \(O(\delta_\tau)\). Therefore
\begin{equation}
\Delta_\tau S_{\rm mix}
=
O\!\left(
\tau^2|\ln\tau|
\right),
\quad
\dot S_{\rm mix}
\equiv
\lim_{\tau\rightarrow0}
\frac{\Delta_\tau S_{\rm mix}}{\tau}
=
0.
\label{eq:supp-Smix-zero}
\end{equation}
The logarithm is needed only when zero-probability macrostates are present.

The residual has a nontrivial linear term. Since \(r_R(t,0)=\Tr(P_R\chi_t)=0\),
\begin{equation}
\left.
\partial_\tau r_R(t,\tau)
\right|_{\tau=0}
=
-i\Tr\!\left[
P_R[H,\chi_t]
\right].
\label{eq:supp-r-first-derivative}
\end{equation}
The representative makes no contribution to the instantaneous record velocity:
\begin{equation}
-i\Tr\!\left[
P_R[H,\cG[\rho_t]]
\right]
=
0,
\end{equation}
because \(\cG[\rho_t]\) is a linear combination of the mutually orthogonal macroprojectors. Consequently,
\begin{equation}
\dot p_R(t)
=
-i\Tr\!\left[
P_R[H,\chi_t]
\right],
\quad
\lim_{\tau\rightarrow0}
\frac{r(t,\tau)}{\tau}
=
\dot p(t).
\label{eq:supp-r-over-tau}
\end{equation}
At times where \(S_R(t)\) is differentiable, the exact entropy balance and Eq.~\eqref{eq:supp-Smix-zero} also give
\begin{equation}
\lim_{\tau\rightarrow0}
\frac{\Delta_\tau S_{\rm corr}}{\tau}
=
\dot S_R(t).
\label{eq:supp-Scorr-rate}
\end{equation}
Thus the macro-uniform baseline is a genuinely finite-step contraction, whereas the complete instantaneous thermodynamic motion is history sensitive.

\suppnote{Projection-operator structure and the exact classical measure-preserving analogue}
\label{supp:note-projection-classical}

This note places the finite-step decomposition in its two closest mathematical lineages. First, it identifies precisely how the thermodynamic projection is related to Mori--Zwanzig projection theory, including modern exact discrete-memory and memory-embedding constructions~\cite{aristoff2023coarse,espanol2026memory}. Second, it derives the exact classical counterpart for invertible measure-preserving dynamics and separates that identity from the standard Ulam finite-state approximation. Maximum-entropy assignment maps can also induce nonlinear and non-Markovian effective dynamics~\cite{castillo2025coarse}. The purpose is therefore not to claim that projection operators, exact memory representations, assignment maps, or cell-transition matrices are new. It is to identify the additional thermodynamic structure obtained when the unresolved component is retained as the present-state finite-step residual and the retained coordinates form a complete probability distribution over thermodynamic cells.

\suppsubsection{Readout, embedding, and the thermodynamic projection}
\label{supp:projection-readout-embedding}

Define the linear readout map \(\mathsf C\) from operators on \(\mathcal H\) to macroprobability vectors by
\begin{equation}
(\mathsf C X)_R
=
\Tr(P_RX),
\end{equation}
and the embedding map \(\mathsf E\) from vectors to operators by
\begin{equation}
\mathsf E x
=
\sum_R
\frac{x_R}{\Omega_R}
P_R.
\end{equation}
On the macroprobability space,
\begin{equation}
\mathsf C\mathsf E
=
I,
\end{equation}
whereas on operator space
\begin{equation}
\mathbb P
\equiv
\mathsf E\mathsf C
=
\cG,
\quad
\mathbb Q
\equiv
\mathcal I-\mathbb P.
\end{equation}
Here \(\mathcal I\) is the identity superoperator. Since \(\mathsf C\mathsf E=I\),
\begin{equation}
\mathbb P^2
=
\mathbb P,
\quad
\mathbb Q^2
=
\mathbb Q,
\quad
\mathbb P\mathbb Q
=
\mathbb Q\mathbb P
=
0.
\end{equation}
Thus \(\cG\) is a genuine linear projection superoperator onto the subspace spanned by the normalized macroprojectors \(P_R/\Omega_R\). For the exact state,
\begin{equation}
\mathbb P\rho_t
=
\cG[\rho_t]
=
\mathsf E p(t),
\quad
\mathbb Q\rho_t
=
\chi_t.
\end{equation}
Let the unitary channel over the interval \(\tau\) be
\begin{equation}
\mathscr U_\tau[X]
=
U_\tau XU_\tau^\dagger.
\end{equation}
Then the exact finite-step quantities introduced in the main text have the operator identities
\begin{equation}
K^{(U)}(\tau)
=
\mathsf C\mathscr U_\tau\mathsf E,
\quad
r(t,\tau)
=
\mathsf C\mathscr U_\tau\mathbb Q\rho_t.
\label{eq:supp-block-coordinate}
\end{equation}
Equivalently, after re-embedding the two vectors,
\begin{equation}
\begin{aligned}
\mathsf E K^{(U)}(\tau)p(t)
&=
\mathbb P\mathscr U_\tau\mathbb P\rho_t,
\\
\mathsf E r(t,\tau)
&=
\mathbb P\mathscr U_\tau\mathbb Q\rho_t.
\end{aligned}
\label{eq:supp-block-propagator}
\end{equation}
The matrix \(K^{(U)}\) is therefore the coordinate representation of the exact finite-time \(\mathbb P\to\mathbb P\) block of the microscopic propagator, whereas \(r\) is the record-level response generated by the component lying in the \(\mathbb Q\) sector at the beginning of the step.

Although \(K^{(U)}(\tau)\) is stochastic for every finite \(\tau\), the family \(\{K^{(U)}(\tau)\}_\tau\) does not generally form a Markov semigroup. For autonomous evolution and two positive intervals \(\tau_1\) and \(\tau_2\),
\begin{equation}
\begin{aligned}
&K^{(U)}(\tau_2+\tau_1)
-
K^{(U)}(\tau_2)K^{(U)}(\tau_1)
\\
&\quad=
\mathsf C
\mathscr U_{\tau_2}
\mathbb Q
\mathscr U_{\tau_1}
\mathsf E.
\end{aligned}
\label{eq:supp-K-composition-defect}
\end{equation}
The right-hand side is the record-level influence, during the second interval, of unresolved structure generated from a macro-uniform representative during the first interval. Acting on an initial macrodistribution \(p\), it is precisely the history-sensitive residual of the second step for the exact intermediate state \(\mathscr U_{\tau_1}\mathsf E p\). By contrast, the product \(K^{(U)}(\tau_2)K^{(U)}(\tau_1)\) re-embeds the intermediate macrodistribution through \(\mathsf E\) and therefore inserts the thermodynamic projection between the two steps. Stochasticity of each finite-time block consequently does not amount to postulating an autonomous Markov process for the exact macrotrajectory.

\suppsubsection{Precise relation to Mori--Zwanzig memory and source terms}
\label{supp:projection-MZ}

The relevant--irrelevant split above belongs to the projection-operator lineage of Zwanzig and Mori~\cite{zwanzig1960ensemble,zwanzig1961memory,mori1965transport,grabert1982projection}. Exact discrete Mori--Zwanzig equations and Markov-renewal representations likewise show how coarse transition probabilities can retain memory beyond a Markov state model~\cite{aristoff2023coarse}, while Hamiltonian coarse graining generically produces memory that may be embedded through auxiliary variables~\cite{espanol2026memory}. Because \(\mathbb P\) acts directly on the state, its closest antecedent is Zwanzig's ensemble-density projection; Mori's construction supplies the related observable-space generalized-Langevin form. The finite-step objects must nevertheless not be identified term by term with the usual memory and fluctuating-force terms.

For autonomous evolution, define the Liouville superoperator
\begin{equation}
\mathcal L X
=
-i[H,X],
\quad
\mathscr U_s
=
e^{s\mathcal L},
\end{equation}
and restart the time coordinate at the present state by writing
\begin{equation}
\varrho(s)
=
\mathscr U_s\rho_t,
\quad
0\leq s\leq\tau.
\end{equation}
The projected components obey
\begin{equation}
\begin{aligned}
\frac{d}{ds}\mathbb P\varrho(s)
&=
\mathbb P\mathcal L\mathbb P\varrho(s)
+
\mathbb P\mathcal L\mathbb Q\varrho(s),
\\
\frac{d}{ds}\mathbb Q\varrho(s)
&=
\mathbb Q\mathcal L\mathbb P\varrho(s)
+
\mathbb Q\mathcal L\mathbb Q\varrho(s).
\end{aligned}
\end{equation}
Solving the second equation and substituting it into the first gives the exact Zwanzig time-convolution form
\begin{equation}
\begin{aligned}
\frac{d}{ds}\mathbb P\varrho(s)
={}&
\mathbb P\mathcal L\mathbb P\varrho(s)
\\
&+
\int_0^sdu\,
\mathbb P\mathcal L\mathbb Q
 e^{(s-u)\mathbb Q\mathcal L\mathbb Q}
\mathbb Q\mathcal L\mathbb P\varrho(u)
\\
&+
\mathbb P\mathcal L\mathbb Q
 e^{s\mathbb Q\mathcal L\mathbb Q}
\mathbb Q\rho_t.
\end{aligned}
\label{eq:supp-NZ}
\end{equation}
The second line is the memory contribution generated by excursions from the retained sector into the unresolved sector and back. The third line is the inhomogeneous source generated by the unresolved component already present at the chosen initial time of the projection equation.

Equation~\eqref{eq:supp-block-propagator} shows why the finite-step correspondence is more subtle than the schematic identification \(r\leftrightarrow\) memory. The exact block
\begin{equation}
\mathbb P e^{\tau\mathcal L}\mathbb P
\end{equation}
contains all \(\mathbb P\to\mathbb Q\to\mathbb P\) excursions that occur during the interval when the initial state is the thermodynamic representative. Those excursions are precisely what appears as a memory kernel when the \(\mathbb Q\) sector is eliminated from a time-local microscopic equation. They are therefore already resummed inside \(K^{(U)}(\tau)\), even when \(r(t,\tau)=0\).

By contrast,
\begin{equation}
\mathbb P e^{\tau\mathcal L}\mathbb Q\rho_t
\end{equation}
is the finite-time response to the unresolved component present at time \(t\). In the projected differential equation, this response is generated by the inhomogeneous source in Eq.~\eqref{eq:supp-NZ}, together with its subsequent propagation through the projected dynamics. The closest Mori--Zwanzig analogue of \(r(t,\tau)\) is therefore this finite-time source response, not the memory kernel as a whole and not, without additional structure, Mori's fluctuating force.

A vanishing residual at one state and one interval,
\begin{equation}
r(t,\tau)
=
0,
\end{equation}
means only that the current \(\mathbb Q\)-sector component produces no endpoint change of the chosen macroprobabilities over that step. It does not by itself imply a Markovian coarse process. A Markov approximation in projection theory concerns the temporal localization or rapid decay of the memory kernel. Exact autonomous macropropagation requires a stronger, state-independent closure over the relevant class of states and time intervals.

Finally, a generic projection operator need not preserve positivity, normalization, or a distinguished multiplicity distribution on its retained coordinates. The thermodynamic projection used here is special: its coordinates form a complete probability distribution, its embedding is uniform inside mutually orthogonal macrospaces, and the exact \(\mathbb P\to\mathbb P\) block is consequently represented by a stochastic matrix satisfying \(K^{(U)}\pi=\pi\). These additional properties, rather than the existence of a relevant--irrelevant split itself, produce the entropy-contraction and relative-majorization results.

\suppsubsection{Exact classical construction on a finite invariant measure space}
\label{supp:classical-exact-construction}

The same kinematic architecture holds for invertible measure-preserving classical dynamics. Let \((\Gamma,\Sigma,\mu)\) be an invariant measurable phase-space region with
\begin{equation}
0
<
\mu(\Gamma)
<
\infty,
\end{equation}
and let \(\Phi^\tau:\Gamma\to\Gamma\) be an invertible measurable map satisfying
\begin{equation}
\mu\!\left[(\Phi^\tau)^{-1}A\right]
=
\mu(A)
\end{equation}
for every measurable \(A\subseteq\Gamma\). Hamiltonian flow on a bounded invariant region or on a finite invariant energy surface equipped with its microcanonical measure is the principal example. Equivalently, one may begin with an invariant probability measure. The finiteness assumption is essential for defining the normalized reference distribution below; the unrestricted Liouville measure of an unbounded phase space must not be normalized in this way.

Let \(\{C_R\}_R\) be a finite measurable partition of \(\Gamma\), with pairwise disjoint cells whose union is \(\Gamma\), and
\begin{equation}
0
<
\mu(C_R)
<
\infty.
\end{equation}
Assigning every boundary point to a definite cell is immaterial for absolutely continuous densities but is required for the measure-level extension below, where the exact state may be singular with respect to \(\mu\).
For a normalized fine density \(f_t\), define
\begin{equation}
p_R(t)
=
\int_{C_R}f_t\,d\mu.
\end{equation}
The classical thermodynamic representative is the piecewise-uniform density
\begin{equation}
\cG_{\rm cl}[f_t]
=
\sum_R
\frac{p_R(t)}{\mu(C_R)}
\mathbf 1_{C_R},
\end{equation}
and the unresolved component is
\begin{equation}
\chi_t^{\rm cl}
=
f_t-\cG_{\rm cl}[f_t].
\end{equation}
For every cell,
\begin{equation}
\int_{C_R}\chi_t^{\rm cl}\,d\mu
=
0.
\end{equation}
Let \(\mathscr P_\tau\) be the Perron--Frobenius operator. For the invertible measure-preserving map considered here,
\begin{equation}
(\mathscr P_\tau f)(x)
=
f\!\left[(\Phi^\tau)^{-1}x\right].
\end{equation}
The exact future cell probability is
\begin{equation}
p_R(t+\tau)
=
\int_{C_R}\mathscr P_\tau f_t\,d\mu.
\end{equation}
Substitution of \(f_t=\cG_{\rm cl}[f_t]+\chi_t^{\rm cl}\) gives
\begin{equation}
p(t+\tau)
=
K^{(\Phi)}(\tau)p(t)
+
r^{\rm cl}(t,\tau),
\label{eq:supp-classical-finite-step}
\end{equation}
where
\begin{equation}
\begin{aligned}
K^{(\Phi)}_{RR'}(\tau)
&=
\frac{
\mu\!\left[
C_{R'}
\cap
(\Phi^\tau)^{-1}C_R
\right]
}{
\mu(C_{R'})
}
\\
&=
\frac{
\mu\!\left[
C_R
\cap
\Phi^\tau C_{R'}
\right]
}{
\mu(C_{R'})
},
\end{aligned}
\label{eq:supp-classical-K}
\end{equation}
and
\begin{equation}
\begin{aligned}
r_R^{\rm cl}(t,\tau)
&=
\int_{C_R}
\mathscr P_\tau\chi_t^{\rm cl}
\,d\mu
\\
&=
\int_{(\Phi^\tau)^{-1}C_R}
\chi_t^{\rm cl}
\,d\mu.
\end{aligned}
\label{eq:supp-classical-r}
\end{equation}
The first form of Eq.~\eqref{eq:supp-classical-K} does not rely on manipulating images of sets and makes the source-to-destination convention explicit. The second follows from invertibility and measure preservation. Equation~\eqref{eq:supp-classical-finite-step} is exact for the fixed partition and finite step; no Markov approximation or cell randomization has been introduced.

The density notation is not essential. For an arbitrary probability measure \(\nu_t\), including a singular point measure, set \(p_R(t)=\nu_t(C_R)\), define its cell-uniform representative by \(\overline{\nu}_t(A)=\sum_Rp_R(t)\mu(A\cap C_R)/\mu(C_R)\), and let \(\xi_t=\nu_t-\overline{\nu}_t\). The same finite-step identity holds with \(r_R^{\rm cl}(t,\tau)=\xi_t[(\Phi^\tau)^{-1}C_R]\). Thus the probability-level construction does not require absolute continuity of the exact classical state; the density form above is used only for notational economy.

\suppsubsection{Classical stochasticity, invariant reference, and thermodynamic contraction}
\label{supp:classical-contraction}

The matrix \(K^{(\Phi)}\) is nonnegative. Its column sums are
\begin{equation}
\sum_RK^{(\Phi)}_{RR'}(\tau)
=
\frac{
\mu\!\left[
C_{R'}
\cap
(\Phi^\tau)^{-1}\Gamma
\right]
}{
\mu(C_{R'})
}
=
1.
\end{equation}
Define the invariant cell-weight distribution
\begin{equation}
\pi_R^{\rm cl}
=
\frac{
\mu(C_R)
}{
\mu(\Gamma)
}.
\end{equation}
Then
\begin{equation}
\begin{aligned}
\sum_{R'}
K^{(\Phi)}_{RR'}(\tau)
\pi_{R'}^{\rm cl}
&=
\frac{1}{\mu(\Gamma)}
\sum_{R'}
\mu\!\left[
C_{R'}
\cap
(\Phi^\tau)^{-1}C_R
\right]
\\
&=
\frac{
\mu\!\left[(\Phi^\tau)^{-1}C_R\right]
}{
\mu(\Gamma)
}
\\
&=
\pi_R^{\rm cl}.
\end{aligned}
\end{equation}
Hence
\begin{equation}
K^{(\Phi)}(\tau)\pi^{\rm cl}
=
\pi^{\rm cl}.
\label{eq:supp-classical-pi}
\end{equation}
Moreover,
\begin{equation}
\sum_Rr_R^{\rm cl}(t,\tau)
=
0.
\end{equation}
The proofs are the measure-theoretic counterparts of the quantum proofs based on completeness and unitarity.

To write an absolute classical coarse entropy with a dimensionless logarithm, choose a fixed reference phase-space volume \(\mu_0>0\) and define
\begin{equation}
\Omega_R^{\rm cl}
=
\frac{\mu(C_R)}{\mu_0},
\quad
\Omega_{\rm tot}^{\rm cl}
=
\frac{\mu(\Gamma)}{\mu_0}.
\end{equation}
Changing \(\mu_0\) adds only a state-independent constant. The record entropy is
\begin{equation}
\begin{aligned}
S_R^{\rm cl}(f)
&=
-k_{\rm B}
\int_\Gamma
\cG_{\rm cl}[f]
\ln
\left(
\mu_0\cG_{\rm cl}[f]
\right)
\,d\mu
\\
&=
-k_{\rm B}
\sum_Rp_R
\ln
\left(
\frac{p_R}{\Omega_R^{\rm cl}}
\right)
\\
&=
k_{\rm B}
\left[
\ln\Omega_{\rm tot}^{\rm cl}
-
D(p\Vert\pi^{\rm cl})
\right].
\end{aligned}
\label{eq:supp-classical-entropy}
\end{equation}
Equation~\eqref{eq:supp-classical-entropy} is the single-partition classical observational-entropy form~\cite{safranek2020classical}; all entropy changes and ordering statements below are independent of the arbitrary reference volume \(\mu_0\). For a singular exact probability measure, the second line of Eq.~\eqref{eq:supp-classical-entropy}, expressed entirely through the cell probabilities, defines the same record entropy without requiring a fine density.

For
\begin{equation}
q_0
=
K^{(\Phi)}p,
\end{equation}
the data-processing inequality and Eq.~\eqref{eq:supp-classical-pi} give
\begin{equation}
D(q_0\Vert\pi^{\rm cl})
\leq
D(p\Vert\pi^{\rm cl}),
\end{equation}
and therefore
\begin{equation}
\Delta_\tau S_{{\rm mix},{\rm cl}}
\geq
0.
\end{equation}
Writing \(q=q_0+r^{\rm cl}\) gives the same exact entropy balance as in the quantum construction. Because the proof of the Lorenz-curve bound is purely distributional, it also gives
\begin{equation}
\mathcal V_{\pi^{\rm cl}}
\left[
p(t)\rightarrow p(t+\tau)
\right]
\leq
\frac12
\left\|
r^{\rm cl}(t,\tau)
\right\|_1.
\label{eq:supp-classical-history-bound}
\end{equation}
Thus the complete finite-step contraction--return theorem has an exact classical measure-preserving counterpart. The quantum and classical constructions differ in their microscopic state spaces and in the structures that can reside in the unresolved component, but not in this kinematic probability-level architecture.

\suppsubsection{Relation to the Ulam finite-state approximation}
\label{supp:classical-Ulam}

The matrix in Eq.~\eqref{eq:supp-classical-K} is the piecewise-constant cell-transition matrix associated with Ulam's finite-rank approximation of the Perron--Frobenius operator~\cite{ulam1960collection,froyland1999ulam}. In the column-stochastic convention used here, its entry is the fraction of the invariant measure in source cell \(C_{R'}\) that reaches destination cell \(C_R\) in one step.

Introduce the classical readout and embedding maps
\begin{equation}
(\mathsf C_{\rm cl}f)_R
=
\int_{C_R}f\,d\mu,
\quad
\mathsf E_{\rm cl}p
=
\sum_R
\frac{p_R}{\mu(C_R)}
\mathbf 1_{C_R}.
\end{equation}
Then
\begin{equation}
\cG_{\rm cl}
=
\mathsf E_{\rm cl}\mathsf C_{\rm cl},
\quad
K^{(\Phi)}
=
\mathsf C_{\rm cl}
\mathscr P_\tau
\mathsf E_{\rm cl}.
\end{equation}
The standard Ulam representation closes the dynamics on piecewise-uniform densities and iterates the finite-state rule \(p_{n+1}=K^{(\Phi)}p_n\). As a numerical or effective representation, this does not physically alter the underlying Hamiltonian trajectory. It does, however, omit from the finite-state state vector the nonuniform within-cell component generated by the exact evolution. Correspondingly, for autonomous evolution,
\begin{equation}
\begin{aligned}
&K^{(\Phi)}(\tau_2+\tau_1)
-
K^{(\Phi)}(\tau_2)K^{(\Phi)}(\tau_1)
\\
&\quad=
\mathsf C_{\rm cl}
\mathscr P_{\tau_2}
\left(
\mathcal I-\cG_{\rm cl}
\right)
\mathscr P_{\tau_1}
\mathsf E_{\rm cl}.
\end{aligned}
\label{eq:supp-Ulam-composition-defect}
\end{equation}
Thus the finite-state iteration inserts a cell-uniform projection at the intermediate time, while the one-shot exact retained block carries the within-cell structure generated during the first interval into the second. Acting on an initial cell distribution, the right-hand side is exactly the classical history-sensitive residual of the second step generated by that intermediate within-cell structure.

The exact construction retains that component and identifies the closure defect explicitly:
\begin{equation}
\begin{aligned}
r^{\rm cl}(t,\tau)
&=
\mathsf C_{\rm cl}
\mathscr P_\tau
\left(
\mathcal I-\cG_{\rm cl}
\right)
f_t
\\
&=
p(t+\tau)
-
K^{(\Phi)}(\tau)p(t).
\end{aligned}
\label{eq:supp-Ulam-closure-defect}
\end{equation}
Accordingly, \(K^{(\Phi)}\) itself is not the new element. The additional result is the exact completion of the cell-uniform transition description by a retained residual, followed by the thermodynamic consequences of invariant-reference preservation: nonnegative representative-driven entropy change, localization in the residual of every departure from that fixed-partition contraction baseline, and the relative-majorization bound in Eq.~\eqref{eq:supp-classical-history-bound}.

\suppsubsection{Common kinematic structure and quantum-specific content}
\label{supp:classical-quantum-scope}

The finite-step contraction--return architecture is therefore a consequence of reversible microscopic evolution, a complete thermodynamic partition, and a representative uniform with respect to the invariant microscopic measure inside each macrocell. In the quantum case, the invariant measure is replaced by Hilbert-space dimension, and \(\chi_t\) can contain genuinely quantum coherences in addition to unresolved nonuniform populations. The short-time identity \(\partial_\tau K^{(U)}(0)=0\) is not part of this shared backbone: it reflects the quadratic onset of transitions between orthogonal quantum macrospaces, whereas a classical flow can carry probability across a cell boundary at first order in time. The interacting quantum models, particle-statistical comparisons, Floquet realization, and engineered rephasing control are quantum applications of the common finite-step architecture.

The classical theorem above is finite-step and finite-measure. Related nondestructive classical entropy constructions likewise retain the exact Hamiltonian evolution and recover echo reversal, although they use a designated subsystem--heat-bath organization rather than the present cell record on one phase space~\cite{ding2025hamiltonian}. Rigorous derivations of dissipative GENERIC dynamics from Hamiltonian models employ still different bath and compression architectures~\cite{mielke2025generic}. The present theorem does not establish that a generic classical Hamiltonian suppresses \(r^{\rm cl}\), nor does it prove irreversible relaxation in a thermodynamic limit. Just as in the quantum case, the remaining dynamical question is when the exact unresolved component fails to organize a thermodynamically relevant return within the specified observation and control window.

\suppnote{Relative majorization and the finite-step Second Law}
\label{supp:note-relative-majorization}

We use relative majorization and its generalized Lorenz-curve characterization for stochastic transformations preserving a reference distribution~\cite{renes2016relative,buscemi2017lorenz}.

\suppsubsection{\texorpdfstring{\(\pi\)-preserving}{pi-preserving} contraction}
\label{supp:equilibrium-contraction}

The exact coarse propagation separates the future macrodistribution into the contribution determined by the current thermodynamic representative and the history-sensitive correction. Define
\begin{equation}
q_0
=
K^{(U)}(\tau)p,
\end{equation}
where \(p\) denotes the accessible distribution at the beginning of the finite step. Since \(K^{(U)}(\tau)\) is stochastic and preserves the multiplicity-defined reference, \(K^{(U)}(\tau)\pi=\pi\), the classical data-processing inequality gives
\begin{equation}
D(q_0\Vert\pi)
=
D
\left[
K^{(U)}(\tau)p
\Vert
K^{(U)}(\tau)\pi
\right]
\leq
D(p\Vert\pi).
\end{equation}
Using \(S_R(p)=k_{\rm B}[\ln\Omega_{\rm tot}-D(p\Vert\pi)]\), this is equivalent to
\begin{equation}
\Delta_\tau S_{\rm mix}
\equiv
S_R(q_0)-S_R(p)
\geq
0.
\end{equation}
Thus the part of every finite unitary step predicted from the present thermodynamic representative alone can never decrease the accessible entropy. This result is exact and requires no detailed balance, Markov approximation, thermalization hypothesis, chaoticity assumption, or external environment.

The actual future distribution contains in addition the history-sensitive contribution,
\begin{equation}
q
=
p(t+\tau)
=
q_0+r.
\end{equation}
We therefore define the entropy change associated with this correction by
\begin{equation}
\Delta_\tau S_{\rm corr}
\equiv
S_R(q)-S_R(q_0).
\end{equation}
The complete finite-step entropy change then decomposes exactly as
\begin{equation}
\Delta_\tau S_R
=
\Delta_\tau S_{\rm mix}
+
\Delta_\tau S_{\rm corr},
\quad
\Delta_\tau S_{\rm mix}
\geq
0.
\end{equation}
The history-sensitive term may have either sign. Consequently, any finite-step decrease of the accessible entropy must arise through \(r\), and can occur only when its entropy-lowering contribution is sufficiently large to overcome the nonnegative \(\pi\)-preserving contribution,
\begin{equation}
\Delta_\tau S_R
<
0
\quad\Longrightarrow\quad
\Delta_\tau S_{\rm corr}
<
-\Delta_\tau S_{\rm mix}.
\end{equation}
This statement does not assert that a nonzero history-sensitive correction is itself thermodynamic return. It isolates only the unique part of the exact finite-step evolution through which such a return can occur.

\suppsubsection{Relative majorization}
\label{supp:relative-majorization}

The entropy inequality above compares one scalar functional of the accessible distribution. A stronger ordering can be imposed directly on the complete macrodistributions. For probability distributions \(p\) and \(q\) relative to the same full-support reference \(\pi\), we write
\begin{equation}
p
\succ_\pi
q
\end{equation}
when there exists a stochastic matrix \(M\) such that
\begin{equation}
q
=
Mp,
\quad
M\pi
=
\pi.
\end{equation}
Thus \(p\succ_\pi q\) means that \(q\) can be obtained from \(p\) by a probabilistic redistribution that leaves the multiplicity-defined reference unchanged. Relative majorization therefore constrains the complete accessible distribution rather than only its entropy.

Because \(M\) is stochastic and \(\pi\)-preserving, the data-processing inequality gives
\begin{equation}
D(q\Vert\pi)
=
D(Mp\Vert M\pi)
\leq
D(p\Vert\pi).
\end{equation}
Hence
\begin{equation}
p
\succ_\pi
q
\quad\Longrightarrow\quad
S_R(q)
\geq
S_R(p).
\end{equation}
Relative majorization therefore supplies a sufficient finite-step Second-Law condition. The implication is stronger than monotonicity of \(S_R\): an entropy increase does not, in general, require that the full relative-majorization ordering hold.

\suppsubsection{Generalized Lorenz curves}
\label{supp:lorenz-curves}

Relative majorization can be tested without explicitly constructing the stochastic matrix \(M\). For a distribution \(p\), order the macrostates by decreasing likelihood ratio,
\begin{equation}
\frac{p_{R_1}}{\pi_{R_1}}
\geq
\frac{p_{R_2}}{\pi_{R_2}}
\geq
\cdots.
\end{equation}
Since \(\pi_R=\Omega_R/\Omega_{\rm tot}\), this is equivalently an ordering by \(p_R/\Omega_R\), namely by probability per available microscopic direction up to the common factor \(\Omega_{\rm tot}\).

For this ordering, define the cumulative reference weight and cumulative probability by
\begin{equation}
X_k
=
\sum_{j=1}^{k}
\pi_{R_j},
\quad
Y_k
=
\sum_{j=1}^{k}
p_{R_j}.
\end{equation}
The generalized Lorenz curve \(L_p^\pi(x)\) is the piecewise-linear curve passing through
\begin{equation}
(0,0),
\quad
(X_1,Y_1),
\quad
\ldots,
\quad
(1,1).
\end{equation}
Constructing the corresponding curve for \(q\) using its own decreasing likelihood-ratio ordering, relative majorization is equivalent to
\begin{equation}
p
\succ_\pi
q
\quad\Longleftrightarrow\quad
L_p^\pi(x)
\geq
L_q^\pi(x)
\quad
\forall x\in[0,1].
\end{equation}
Thus \(p\succ_\pi q\) precisely when the Lorenz curve of \(p\) nowhere lies below that of \(q\).

To quantify departures from this ordering, we define the relative-majorization return defect
\begin{equation}
\Vpi(p\rightarrow q)
=
\sup_{0\leq x\leq1}
\left[
L_q^\pi(x)
-
L_p^\pi(x)
\right]_+.
\end{equation}
It follows immediately that
\begin{equation}
\Vpi(p\rightarrow q)
=
0
\quad\Longleftrightarrow\quad
p
\succ_\pi
q.
\end{equation}
Applied to two consecutive points of the thermodynamic trajectory, this gives the finite-step implication
\begin{equation}
\Vpi
\left[
p(t)\rightarrow p(t+\tau)
\right]
=
0
\quad\Longrightarrow\quad
S_R(t+\tau)
\geq
S_R(t).
\end{equation}
The converse does not hold in general. A Lorenz-curve crossing may occur while the scalar entropy still increases, so \(\Vpi>0\) identifies failure of the stronger \(\pi\)-preserving ordering rather than, by itself, an entropy decrease. By contraposition,
\begin{equation}
S_R(t+\tau)
<
S_R(t)
\quad\Longrightarrow\quad
\Vpi
\left[
p(t)\rightarrow p(t+\tau)
\right]
>
0.
\end{equation}
Thus every finite-step entropy decrease necessarily lies outside the relative-majorization contraction sector, although leaving that sector need not immediately reverse the entropy.

\suppsubsection{Instantaneous limit}
\label{supp:instantaneous-limit}

The finite-step ordering also yields a local Second-Law statement. Suppose that, at a given time \(t\),
\begin{equation}
p(t)
\succ_\pi
p(t+\tau)
\end{equation}
for every sufficiently small \(\tau>0\). Relative majorization then implies
\begin{equation}
S_R(t+\tau)
-
S_R(t)
\geq
0
\end{equation}
for all such positive steps. Dividing by \(\tau>0\),
\begin{equation}
\frac{
S_R(t+\tau)-S_R(t)
}{
\tau
}
\geq
0.
\end{equation}
If \(S_R(t)\) is differentiable, taking the right-hand limit \(\tau\rightarrow0^+\) gives
\begin{equation}
\dot S_R(t)
\geq
0.
\end{equation}
Exact local relative majorization is therefore sufficient for a nonnegative instantaneous accessible-entropy rate. The converse is not required: \(\dot S_R(t)\geq0\) may hold even when some of the stronger relative-majorization constraints are violated. This distinction is important in the strong-coupling calculations discussed below, where small finite-step Lorenz-order violations coexist with a positive exact instantaneous accessible-entropy rate.

\suppnote{Bound on thermodynamically relevant history return}
\label{supp:note-history-bound}

We now relate the magnitude of the history-sensitive correction \(r(t,\tau)\) to the strongest possible violation of the relative-majorization ordering. The generalized Lorenz curve admits the variational representation
\begin{equation}
L_p^\pi(x)
=
\max_{\substack{
0\leq f_R\leq1
\\
\sum_R\pi_Rf_R=x
}}
\sum_R
f_Rp_R,
\end{equation}
where \(x\in[0,1]\). The auxiliary variables \(f_R\) select, possibly fractionally, a total reference weight \(x\). This representation is equivalent to the piecewise-linear construction obtained by ordering the macrostates according to decreasing \(p_R/\pi_R\).

For a finite unitary step, write
\begin{equation}
q
=
p(t+\tau)
=
q_0+r,
\quad
q_0
=
K^{(U)}(\tau)p(t),
\end{equation}
and suppress the explicit time arguments temporarily. Applying the variational representation to \(q\) gives
\begin{equation}
\begin{aligned}
L_q^\pi(x)
&=
\max_{\substack{
0\leq f_R\leq1
\\
\sum_R\pi_Rf_R=x
}}
\sum_R
f_R
\left(
q_{0,R}+r_R
\right)
\\
&\leq
L_{q_0}^\pi(x)
+
\max_{0\leq f_R\leq1}
\sum_R
f_Rr_R.
\end{aligned}
\end{equation}
In the second line, the constraint \(\sum_R\pi_Rf_R=x\) has been removed only from the term involving \(r\), thereby enlarging the optimization domain and providing an upper bound.

The history-sensitive correction has zero total weight,
\begin{equation}
\sum_Rr_R
=
0.
\end{equation}
For the unconstrained maximization over \(0\leq f_R\leq1\), the optimal choice is therefore \(f_R=1\) whenever \(r_R>0\) and \(f_R=0\) whenever \(r_R<0\), with arbitrary values when \(r_R=0\). Hence
\begin{equation}
\max_{0\leq f_R\leq1}
\sum_Rf_Rr_R
=
\sum_{r_R>0}r_R.
\end{equation}
Because the positive and negative parts of \(r\) have equal total weight,
\begin{equation}
\sum_{r_R>0}r_R
=
-\sum_{r_R<0}r_R
=
\frac{1}{2}
\sum_R|r_R|
=
\frac{1}{2}\|r\|_1.
\end{equation}
It follows that, for every \(x\in[0,1]\),
\begin{equation}
L_q^\pi(x)
\leq
L_{q_0}^\pi(x)
+
\frac{1}{2}\|r\|_1.
\end{equation}
The distribution \(q_0\) is generated from \(p\) by the stochastic \(\pi\)-preserving map \(K^{(U)}\). Therefore
\begin{equation}
p
\succ_\pi
q_0,
\end{equation}
and the corresponding Lorenz curves satisfy
\begin{equation}
L_{q_0}^\pi(x)
\leq
L_p^\pi(x)
\quad
\forall x\in[0,1].
\end{equation}
Combining the two inequalities yields the pointwise bound
\begin{equation}
L_q^\pi(x)
-
L_p^\pi(x)
\leq
\frac{1}{2}
\|r\|_1
\quad
\forall x\in[0,1].
\end{equation}
Taking the positive supremum over \(x\) gives the exact finite-step relation
\begin{equation}
\Vpi
\left[
p(t)\rightarrow p(t+\tau)
\right]
\leq
\frac{1}{2}
\left\|
r(t,\tau)
\right\|_1
=
\epsilon_{\rm hist}(t,\tau).
\end{equation}
Thus the history-sensitive correction bounds from above the largest possible violation of the \(\pi\)-preserving relative-majorization ordering. Since the macro-uniform distribution \(q_0=K^{(U)}p\) is already relatively majorized by \(p\), every finite-step violation of that ordering must be generated through the correction \(r\).

Approximate-majorization theory already relates total-variation neighbourhoods to extremal Lorenz curves and majorization distances~\cite{horodecki2018approximate}. The result above should therefore be understood as a dynamical specialization: the radius is not an arbitrary approximation tolerance but the exact residual generated by microscopic information absent from the present record, and the reference is the multiplicity distribution preserved by the retained finite-time block.

The logical implication is one-sided. If the present macroprobabilities completely determine the future macrodistribution, then \(r=0\), so
\begin{equation}
\epsilon_{\rm hist}
=
0
\quad\Longrightarrow\quad
\Vpi
=
0.
\end{equation}
The converse does not hold:
\begin{equation}
\epsilon_{\rm hist}
>
0
\quad
\centernot\Longrightarrow
\quad
\Vpi
>
0.
\end{equation}
Microscopic information absent from the current thermodynamic record can therefore remain dynamically active while producing no violation of the stronger relative-majorization order. In that regime, \(r\neq0\) changes the future accessible distribution, but its effect remains compatible with thermodynamic contraction. The primary mixing dynamics studied in the main text realizes precisely this separation, with \(\epsilon_{\rm hist}>0\) while \(\Vpi\) remains zero to numerical precision. Thermodynamically relevant return therefore requires not merely the survival or dynamical activity of inaccessible microscopic information, but its organization into a correction capable of overcoming the \(\pi\)-preserving ordering.

\suppnote{Exact macroprobability currents and instantaneous entropy balance}
\label{supp:note-macrocurrents}

\suppsubsection{Quantum probability currents between macrospaces}
\label{supp:macro-currents}

The finite-step decomposition identifies the history-sensitive contribution to the evolution of the complete accessible distribution. A complementary instantaneous description is obtained by resolving the exact unitary dynamics into probability currents between thermodynamic macrospaces.

The probability associated with macrospace \(R\) is
\begin{equation}
p_R(t)
=
\Tr(P_R\rho_t).
\end{equation}
For autonomous unitary evolution, with \(\hbar=1\),
\begin{equation}
\dot\rho_t
=
-i[H,\rho_t],
\end{equation}
and therefore
\begin{equation}
\dot p_R
=
-i
\Tr
\left[
P_R[H,\rho_t]
\right].
\end{equation}
Inserting the resolution of identity \(\sum_{R'}P_{R'}=I_{\mathcal H}\) gives
\begin{equation}
\begin{aligned}
\dot p_R
={}&
-i
\sum_{R'}
\Bigg\{
\Tr
\left[
P_RHP_{R'}\rho_t
\right]
\\
&
-
\Tr
\left[
P_R\rho_tP_{R'}H
\right]
\Bigg\}.
\end{aligned}
\end{equation}
The two terms inside the braces are complex conjugates. Defining the current from macrospace \(R'\) into macrospace \(R\) by
\begin{equation}
\begin{aligned}
J_{RR'}(t)
&=
2
\operatorname{Im}
\Tr
\left[
P_RHP_{R'}\rho_t
\right]
\\
&=
2
\operatorname{Im}
\Tr
\left[
P_RHP_{R'}\chi_t
\right],
\end{aligned}
\label{eq:supp-current-chi}
\end{equation}
one obtains the exact continuity equation
\begin{equation}
\dot p_R(t)
=
\sum_{R'}
J_{RR'}(t).
\end{equation}
With this convention, \(J_{RR'}>0\) denotes net probability flow into \(R\) from \(R'\). Hermiticity of \(H\) and \(\rho_t\), together with cyclicity of the trace, gives
\begin{equation}
J_{RR'}(t)
=
-J_{R'R}(t),
\end{equation}
and hence \(J_{RR}=0\). The unitary evolution therefore induces an exact antisymmetric probability-current network on the thermodynamic macrospaces. Probability conservation follows immediately:
\begin{equation}
\sum_R\dot p_R
=
\sum_{R,R'}J_{RR'}
=
0.
\end{equation}
No classical stochastic process is introduced in this construction. The second line of Eq.~\eqref{eq:supp-current-chi} follows because the macro-uniform representative carries no inter-macrospace current. Thus every instantaneous record current is generated by the off-diagonal macrospace blocks of \(\chi_t\). Nonuniform structure confined inside one macrospace can still affect a later finite step by first evolving into such blocks, but it does not itself carry an instantaneous macrocurrent.

\suppsubsection{Accessible-entropy rate and current balance}
\label{supp:entropy-current-balance}

For the chosen thermodynamic record, the accessible entropy is
\begin{equation}
S_R
=
-k_{\rm B}
\sum_R
p_R
\ln
\left(
\frac{p_R}{\Omega_R}
\right).
\end{equation}
The expressions below apply at times where \(S_R(t)\) is differentiable. If a macroprobability vanishes, \(p_R\ln p_R\) is defined by continuity and the rate formulas are understood through the corresponding limit whenever that limit exists.

Since the multiplicities \(\Omega_R\) are fixed by the thermodynamic partition, differentiation gives
\begin{equation}
\dot S_R
=
-k_{\rm B}
\sum_R
\dot p_R
\left[
\ln
\left(
\frac{p_R}{\Omega_R}
\right)
+
1
\right].
\end{equation}
Normalization implies \(\sum_R\dot p_R=0\), so
\begin{equation}
\dot S_R
=
-k_{\rm B}
\sum_R
\dot p_R
\ln
\left(
\frac{p_R}{\Omega_R}
\right).
\end{equation}
Substituting the exact current representation yields
\begin{equation}
\dot S_R
=
-k_{\rm B}
\sum_{R,R'}
J_{RR'}
\ln
\left(
\frac{p_R}{\Omega_R}
\right).
\end{equation}
Using \(J_{RR'}=-J_{R'R}\), the ordered-pair sum can be antisymmetrized to give
\begin{equation}
\dot S_R
=
\frac{k_{\rm B}}{2}
\sum_{R,R'}
J_{RR'}
\ln
\left[
\frac{
p_{R'}/\Omega_{R'}
}{
p_R/\Omega_R
}
\right].
\end{equation}
This is an exact instantaneous identity. The logarithmic factor compares the probability per available microscopic direction in the two macrospaces. A current from a macrospace with larger \(p_{R'}/\Omega_{R'}\) toward one with smaller \(p_R/\Omega_R\) contributes positively to the accessible-entropy rate, whereas the opposite flow contributes negatively.

For each unordered pair of macrospaces, define
\begin{equation}
c_{RR'}
=
k_{\rm B}
J_{RR'}
\ln
\left[
\frac{
p_{R'}/\Omega_{R'}
}{
p_R/\Omega_R
}
\right].
\end{equation}
Both factors change sign under \(R\leftrightarrow R'\), so
\begin{equation}
c_{RR'}
=
c_{R'R}.
\end{equation}
The entropy rate may therefore be written as a sum over unordered pairs,
\begin{equation}
\dot S_R
=
\sum_{R<R'}
c_{RR'}.
\end{equation}
Separating the positive and negative pair contributions, with \([x]_+=\max(x,0)\), we define
\begin{equation}
\sigma_{\rm spread}
=
\sum_{R<R'}
[c_{RR'}]_+,
\quad
\sigma_{\rm ret}
=
\sum_{R<R'}
[-c_{RR'}]_+.
\end{equation}
Both quantities are nonnegative, and the exact instantaneous balance becomes
\begin{equation}
\dot S_R
=
\sigma_{\rm spread}
-
\sigma_{\rm ret}.
\end{equation}
The quantity \(\sigma_{\rm spread}\) collects those resolved macrospace currents that instantaneously increase the accessible entropy, while \(\sigma_{\rm ret}\) collects those that instantaneously decrease it. These labels refer only to the sign of each contribution to the coarse entropy balance. They do not represent two independent microscopic forces, two distinct physical processes, or forward and backward quantum trajectories: the underlying system follows one exact unitary evolution.

In particular, an entropy-increasing thermodynamic evolution does not require every resolved current to point in an entropy-increasing direction. One may have
\begin{equation}
\sigma_{\rm ret}
>
0
\end{equation}
while the total accessible entropy continues to increase whenever
\begin{equation}
\sigma_{\rm spread}
>
\sigma_{\rm ret},
\end{equation}
so that \(\dot S_R>0\). Conversely, organized thermodynamic return corresponds to intervals in which the entropy-lowering currents become sufficiently coordinated that \(\sigma_{\rm ret}>\sigma_{\rm spread}\), yielding \(\dot S_R<0\).

The current decomposition and the relative-majorization construction therefore resolve different aspects of the same closed dynamics. The former gives an exact instantaneous balance among macrospace currents, whereas the latter compares the complete accessible distributions over a finite interval. A nonzero \(\sigma_{\rm ret}\) does not by itself imply entropy decrease, just as a nonzero history-sensitive correction \(r(t,\tau)\) does not by itself imply finite-step thermodynamic return.

The current formula also yields an exact microscopic upper bound. For \(R\neq R'\), cyclicity gives
\begin{equation}
\Tr(P_RHP_{R'}\chi_t)
=
\Tr\!\left[
(P_RHP_{R'})(P_{R'}\chi_tP_R)
\right],
\end{equation}
so H\"older duality implies
\begin{equation}
|J_{RR'}(t)|
\leq
2
\left\|P_RHP_{R'}\right\|_\infty
\left\|P_{R'}\chi_tP_R\right\|_1.
\label{eq:supp-current-block-bound}
\end{equation}
With
\begin{equation}
A_{RR'}(t)
=
\ln\!\left[
\frac{p_{R'}(t)/\Omega_{R'}}{p_R(t)/\Omega_R}
\right],
\end{equation}
the entropy-lowering burden satisfies
\begin{equation}
\sigma_{\rm ret}(t)
\leq
2k_{\rm B}
\sum_{\substack{R<R'\\J_{RR'}A_{RR'}<0}}
|A_{RR'}|
\left\|P_RHP_{R'}\right\|_\infty
\left\|P_{R'}\chi_tP_R\right\|_1.
\label{eq:supp-sigma-return-block-bound}
\end{equation}
Equation~\eqref{eq:supp-sigma-return-block-bound} identifies the precise microscopic objects that carry instantaneous return: inter-macrospace hidden coherences whose currents point against the thermodynamic affinity. The following note resolves those exact macrocurrents in the energy-gap spectrum and separates the amount of available transition power from its coherent organization among equal-gap channels.

\suppnote{Exact spectral organization of thermodynamic macrocurrents}
\label{supp:note-spectral-return}

This note derives the exact state- and record-specific spectral representation underlying the current discussion in the main text. The Fourier decomposition of an observable under a time-independent Hamiltonian is standard. The additional structure used here is that the observables are the exact inter-macrospace current operators of Note~\ref{supp:note-macrocurrents}; their amplitudes therefore reconstruct \(\sigma_{\rm spread}\), \(\sigma_{\rm ret}\), and \(\dot S_R\) without replacing the thermodynamic dynamics by a surrogate model. The result distinguishes three logically different objects: the transition power available to the current operators, coherent addition among transitions with identical energy gaps, and thermodynamic alignment of the resulting currents with the affinities.

\suppsubsection{Gap-resolved current amplitudes}
\label{supp:spectral-current-operators}

Let
\begin{equation}
\mathcal E_{\rm mac}
=
\left\{
(R,R'):
R<R',\ P_RHP_{R'}\neq0
\right\}
\end{equation}
be the set of unordered coupled macrospace pairs. For \(e=(R,R')\in\mathcal E_{\rm mac}\), define
\begin{equation}
\widehat I_e
=
i\left(P_{R'}HP_R-P_RHP_{R'}\right),
\quad
J_e(t)
=
\Tr(\widehat I_e\rho_t).
\label{eq:supp-current-operator}
\end{equation}
The orientation of \(e\) fixes the sign of \(J_e\); all spectral powers below are unchanged if the orientation is reversed. Because the macro-uniform representative carries no inter-macrospace current, \(J_e(t)=\Tr(\widehat I_e\chi_t)\) as well as \(J_e(t)=\Tr(\widehat I_e\rho_t)\). The spectral analysis therefore resolves the same hidden currents that enter Eq.~\eqref{eq:instantaneous-current}.

Write the Hamiltonian in terms of projectors onto its \emph{distinct} energy eigenspaces,
\begin{equation}
H
=
\sum_{\alpha=1}^{d_E}
E_\alpha\Pi_\alpha.
\label{eq:supp-distinct-energy-decomp}
\end{equation}
The use of the complete projectors \(\Pi_\alpha\), rather than an arbitrary eigenbasis within degenerate eigenspaces, is essential. For a reference state \(\rho_0\), define
\begin{equation}
c_{e,\alpha\beta}
=
\Tr\!\left(
\widehat I_e\Pi_\alpha\rho_0\Pi_\beta
\right).
\label{eq:supp-transition-current-amplitude}
\end{equation}
Then
\begin{equation}
J_e(t)
=
\sum_{\alpha,\beta}
\exp[-i(E_\alpha-E_\beta)t]
\,c_{e,\alpha\beta}.
\label{eq:supp-current-energy-pairs}
\end{equation}
The diagonal part is the stationary current in the energy-dephased state,
\begin{equation}
J_e^\omega
=
\sum_\alpha c_{e,\alpha\alpha}
=
\Tr(\widehat I_e\omega),
\quad
\omega
=
\sum_\alpha\Pi_\alpha\rho_0\Pi_\alpha.
\label{eq:supp-current-stationary}
\end{equation}
For each nonzero distinct gap \(g\), let
\begin{equation}
\mathcal T_g
=
\left\{
(\alpha,\beta):
\alpha\neq\beta,\ E_\alpha-E_\beta=g
\right\},
\quad
d_g
=
|\mathcal T_g|,
\end{equation}
and define the coherent gap amplitude
\begin{equation}
C_{e,g}
=
\sum_{(\alpha,\beta)\in\mathcal T_g}
c_{e,\alpha\beta}.
\label{eq:supp-gap-current-amplitude}
\end{equation}
Equation~\eqref{eq:supp-current-energy-pairs} becomes
\begin{equation}
\boxed{
J_e(t)
=
J_e^\omega
+
\sum_{g\neq0}
C_{e,g}e^{-igt}.}
\label{eq:supp-exact-current-spectrum}
\end{equation}
Hermiticity of \(\rho_0\) and \(\widehat I_e\) gives \(C_{e,-g}=C_{e,g}^*\), so Eq.~\eqref{eq:supp-exact-current-spectrum} is real. A time shift \(\rho_0\mapsto\rho_{t_*}\) changes every member of one gap class by the same phase, \(c_{e,\alpha\beta}\mapsto e^{-igt_*}c_{e,\alpha\beta}\), and therefore \(C_{e,g}\mapsto e^{-igt_*}C_{e,g}\). The powers \(|c_{e,\alpha\beta}|^2\), \(|C_{e,g}|^2\), and all ratios constructed from them are consequently independent of the arbitrary time origin.

When the Hamiltonian, preparation, and macroprojectors are invariant under an antiunitary time-reversal operation \(\Theta\), the dephased state \(\omega\) is also invariant, while \(\Theta\widehat I_e\Theta^{-1}=-\widehat I_e\). Hence
\begin{equation}
J_e^\omega
=
0.
\label{eq:supp-stationary-current-zero}
\end{equation}
The static Hamiltonians and thermodynamic-class preparations used below are real in the computational basis, so complex conjugation provides this symmetry.

\suppsubsection{Exact finite-window current power}
\label{supp:spectral-finite-window}

Let \(\mathcal W=[t_0,t_0+T]\), and write \(t_c=t_0+T/2\). With
\begin{equation}
\operatorname{sinc}x
=
\begin{cases}
\sin x/x,&x\neq0,\\
1,&x=0,
\end{cases}
\end{equation}
the exact finite-window fluctuation power of one macrocurrent is
\begin{equation}
\begin{aligned}
\left\langle
|J_e-J_e^\omega|^2
\right\rangle_{\mathcal W}
={}&
\sum_{g,g'\neq0}
C_{e,g}C_{e,g'}^*
\\
&\times
\exp[-i(g-g')t_c]
\operatorname{sinc}\!\left[
\frac{(g-g')T}{2}
\right].
\end{aligned}
\label{eq:supp-finite-window-current-power}
\end{equation}
This identity is exact for the finite system. It retains not only exactly degenerate gaps but also the finite-time interference of nearby distinct gaps; no bin width, gap-concentration surrogate, or norm relaxation has been introduced. Summing Eq.~\eqref{eq:supp-finite-window-current-power} over macroedges gives the exact window-averaged squared Euclidean current norm.

For a finite discrete spectrum, the sinc factor removes all terms with \(g\neq g'\) in the infinite-time average. Thus
\begin{equation}
\overline{
\|\mathbf J-\mathbf J^\omega\|_2^2
}
=
\sum_e\sum_{g\neq0}|C_{e,g}|^2.
\label{eq:supp-long-time-current-power}
\end{equation}
This is the precise sense in which equal-gap amplitudes add coherently before the long-time current power is formed.

\suppsubsection{Equal-gap current-coherence theorem}
\label{supp:spectral-current-theorem}

Define
\begin{equation}
P_{\rm pair}
=
\sum_e
\sum_{\alpha\neq\beta}
|c_{e,\alpha\beta}|^2
\label{eq:supp-Pinc}
\end{equation}
and
\begin{equation}
P_{\rm spec}
=
\sum_e
\sum_{g\neq0}
|C_{e,g}|^2.
\label{eq:supp-Pspec}
\end{equation}
The first quantity is the distinct-energy-pair current power: squared amplitudes are added before different energy pairs sharing the same gap are grouped. The second is the exact long-time fluctuating current power after all amplitudes with the same gap have first been summed. Neither quantity, nor their ratio below, is an entropy or a resource-theory coherence monotone; they are state-, Hamiltonian-, record-, and current-operator-specific dynamical quantities.

For \(P_{\rm pair}>0\), define the dimensionless equal-gap current-coherence factor
\begin{equation}
\Gamma_{\rm spec}
=
\frac{P_{\rm spec}}{P_{\rm pair}}.
\label{eq:supp-Gamma-spec}
\end{equation}
Let
\begin{equation}
D_G
=
\max_{g\neq0}d_g
\label{eq:supp-DG-current}
\end{equation}
be the maximum multiplicity of one nonzero gap among pairs of \emph{distinct energy values}. For each edge and gap, Cauchy--Schwarz gives
\begin{equation}
|C_{e,g}|^2
=
\left|
\sum_{(\alpha,\beta)\in\mathcal T_g}
c_{e,\alpha\beta}
\right|^2
\leq
 d_g
\sum_{(\alpha,\beta)\in\mathcal T_g}
|c_{e,\alpha\beta}|^2.
\label{eq:supp-gap-cauchy}
\end{equation}
Summing over \(e\) and \(g\) yields
\begin{equation}
\boxed{
\begin{aligned}
P_{\rm spec}
&=
\Gamma_{\rm spec}P_{\rm pair},
\\
P_{\rm spec}
&\leq
D_GP_{\rm pair},
\quad
0\leq\Gamma_{\rm spec}\leq D_G.
\end{aligned}}
\label{eq:supp-Gamma-bound}
\end{equation}
Together with Eq.~\eqref{eq:supp-long-time-current-power}, this proves
\begin{equation}
\boxed{
\overline{
\|\mathbf J-\mathbf J^\omega\|_2^2
}
=
\Gamma_{\rm spec}P_{\rm pair}
\leq
D_GP_{\rm pair}.}
\label{eq:supp-spectral-current-theorem}
\end{equation}
If all nonzero gaps are nondegenerate, then \(D_G=1\), every set \(\mathcal T_g\) contains one transition, and therefore
\begin{equation}
\Gamma_{\rm spec}
=
1.
\label{eq:supp-Gamma-one}
\end{equation}
For \(D_G>1\), the factor can be larger or smaller than unity. Values \(\Gamma_{\rm spec}>1\) quantify coherent enhancement of the current power by equal-gap transition amplitudes, while \(\Gamma_{\rm spec}<1\) corresponds to net destructive addition. Gap degeneracy is therefore a capacity for coherent current organization, not a guarantee of it. The ratio \(\Gamma_{\rm spec}/D_G\) measures how much of the coarse global Cauchy bound in Eq.~\eqref{eq:supp-Gamma-bound} is realized by the actual state and current operators.

\suppsubsection{Thermodynamic return and the role of affinity}
\label{supp:spectral-return-thermodynamics}

For the orientation chosen above, the entropy-rate contribution of edge \(e\) is \(k_{\rm B}J_eA_e\). Hence
\begin{equation}
\sigma_{\rm ret}(t)
=
\sum_e
[-k_{\rm B}J_e(t)A_e(t)]_+
\leq
k_{\rm B}
\|\mathbf A(t)\|_2
\|\mathbf J(t)\|_2.
\label{eq:supp-return-current-envelope}
\end{equation}
On an interior observation window \(\mathcal W\), define
\begin{equation}
\mathcal A_2(\mathcal W)
=
\sup_{t\in\mathcal W}
\|\mathbf A(t)\|_2.
\end{equation}
Then the exact finite-window current power in Eq.~\eqref{eq:supp-finite-window-current-power} gives
\begin{equation}
\left\langle
\sigma_{\rm ret}^2
\right\rangle_{\mathcal W}^{1/2}
\leq
k_{\rm B}
\mathcal A_2(\mathcal W)
\left\langle
\|\mathbf J\|_2^2
\right\rangle_{\mathcal W}^{1/2}.
\label{eq:supp-exact-window-return-envelope}
\end{equation}
No approximation enters the current factor on the right-hand side; it can be evaluated directly from the \(C_{e,g}\) and the sinc kernel. If \(\mathbf J^\omega=0\) and the affinity norm remains bounded by a constant \(\mathcal A_*\) over the averaging sequence, Eqs.~\eqref{eq:supp-spectral-current-theorem} and \eqref{eq:supp-exact-window-return-envelope} imply the long-time sufficient bound
\begin{equation}
\overline{\sigma_{\rm ret}^2}^{\,1/2}
\leq
k_{\rm B}\mathcal A_*
\sqrt{\Gamma_{\rm spec}P_{\rm pair}}
\leq
k_{\rm B}\mathcal A_*
\sqrt{D_GP_{\rm pair}}.
\label{eq:supp-long-time-return-envelope}
\end{equation}
This is a current-specific sufficient condition, not a generic thermodynamic-limit theorem. Nondegenerate gaps prevent equal-gap enhancement because \(\Gamma_{\rm spec}=1\), but they do not force \(P_{\rm pair}\) to vanish. Conversely, a large \(\Gamma_{\rm spec}\) supplies enhanced current capacity but does not determine the sign of \(J_eA_e\). Thermodynamic return requires both current power and alignment against the thermodynamic affinity. A positive accessible-entropy rate additionally requires \(\sigma_{\rm spread}>\sigma_{\rm ret}\).

\suppsubsection{Exact numerical comparison: mixing, free, and engineered return}
\label{supp:spectral-model-comparison}

The central exact comparison uses \(N=16\), \(Q=2\), the same left-half thermodynamic preparation, and the same four-cell record. In addition to \(H_{\rm mix}\) and \(H_{\rm ret}\), we include the free nearest-neighbor Hamiltonian obtained from Eq.~\eqref{eq:mixing-H} by setting \(V_1=V_2=0\). This intermediate control is important: it separates the absence of interactions from the much stronger spectral synchronization deliberately built into \(H_{\rm ret}\).

All spectral groupings are performed on distinct-energy projectors, not on arbitrary eigenvectors inside degenerate eigenspaces. Numerically, eigenvalues are clustered with
\begin{equation}
\delta_E
=
\max\!\left[10^{-11}\max(J,\Delta_E),10^{-12}J\right],
\end{equation}
where \(\Delta_E\) is the spectral span, and gaps are clustered with
\begin{equation}
\delta_g
=
\max\!\left[5\delta_E,10^{-11}\max(J,g_{\max})\right],
\end{equation}
where \(g_{\max}=\max_{\alpha\neq\beta}|E_\alpha-E_\beta|\). For the central mixing, free, and engineered-return calculations, \(\delta_g/J=3.87\times10^{-10}\), \(3.83\times10^{-10}\), and \(3.50\times10^{-10}\), respectively. The smallest separation between distinct resolved gap values in the mixing case is \(2.50\times10^{-8}J\), more than sixty times the clustering tolerance. Statements of gap nondegeneracy below are therefore numerical-resolution statements, not analytic claims about the exact algebraic spectrum.

\begin{table}[t]
\centering
\caption{Exact gap-resolved macrocurrent diagnostics for \(N=16\), \(Q=2\), the left-half thermodynamic preparation, and the four-cell record. \(P_{\rm pair}\) and \(P_{\rm spec}\) are in units of \(J^2\). \(N_{\rm gap}^{(2)}\) is the participation number of the positive-gap current-power distribution; \(w_{\max}\) is its largest single-gap share.\justifying}
\label{tab:spectral-current-comparison}
\begin{tabular}{lccc}
\toprule
Quantity & Mixing & Free & Engineered return \\
\midrule
\(d_E\) & 119 & 113 & 29 \\
\(d_{\rm eff}\) & 102.958 & 79.967 & 22.366 \\
\(D_G\) & 1 & 26 & 28 \\
\(P_{\rm pair}/J^2\) & \(2.0869\!\times\!10^{-4}\) & \(1.9781\!\times\!10^{-4}\) & \(7.8923\!\times\!10^{-4}\) \\
\(P_{\rm spec}/J^2\) & \(2.0869\!\times\!10^{-4}\) & \(2.4829\!\times\!10^{-3}\) & \(1.5978\!\times\!10^{-2}\) \\
\(\Gamma_{\rm spec}\) & 1.000 & 12.552 & 20.245 \\
\(\Gamma_{\rm spec}/D_G\) & 1.000 & 0.483 & 0.723 \\
\(N_{\rm gap}^{(2)}\) & 152.38 & 11.37 & 2.30 \\
\(w_{\max}\) & 0.0331 & 0.1933 & 0.6302 \\
50\% power channels & 95 & 4 & 1 \\
90\% power channels & 817 & 24 & 5 \\
\bottomrule
\end{tabular}
\end{table}

\begin{figure}[t]
\centering
\includegraphics[width=0.5\linewidth]{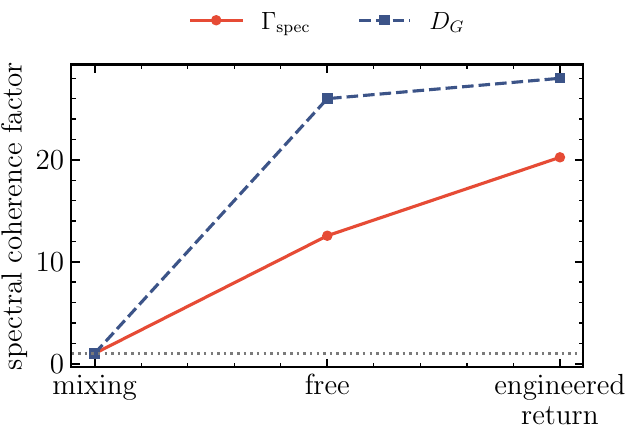}
\caption{Equal-gap current coherence for the exact \(N=16\), \(Q=2\) comparison. The solid curve shows the state- and current-specific factor \(\Gamma_{\rm spec}=P_{\rm spec}/P_{\rm pair}\); the dashed curve shows the absolute Cauchy upper limit \(D_G\). At the stated numerical spectral tolerance, no repeated nonzero gap is resolved for mixing, giving \(\Gamma_{\rm spec}=1\). The free and engineered controls possess increasing equal-gap enhancement, with engineered return realizing \(\Gamma_{\rm spec}=20.245\) out of \(D_G=28\).\justifying}
\label{fig:S20-spectral-coherence}
\end{figure}

The hierarchy is sharp. At the numerical spectral tolerance used in the exact diagonalization, the mixing Hamiltonian has no repeated nonzero gap among distinct energy values, so the resolved value is \(\Gamma_{\rm spec}=1\). The free chain possesses substantial equal-gap degeneracy and realizes a moderate coherent enhancement. The engineered spectrum realizes a still stronger enhancement, \(\Gamma_{\rm spec}=20.245\), corresponding to \(72.3\%\) of the global \(D_G\) Cauchy bound. Supplementary Fig.~\ref{fig:S20-spectral-coherence} displays this comparison.

\begin{figure}[t]
\centering
\includegraphics[width=0.5\linewidth]{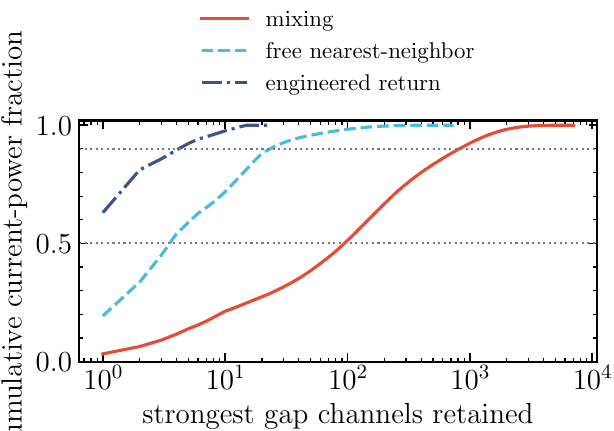}
\caption{Concentration of positive-gap macrocurrent power for \(N=16\), \(Q=2\). Gap channels are ordered by decreasing \(P_g=\sum_e|C_{e,g}|^2\), and the cumulative fraction is plotted against the number retained. Mixing distributes its current power over many gap channels, the free chain is intermediate, and the engineered return spectrum places more than half of its current power in a single positive-gap channel and 90\% in five.\justifying}
\label{fig:S22-gap-power}
\end{figure}

The distribution of current power among positive gaps is even more discriminating. Define
\begin{equation}
P_g
=
\sum_e|C_{e,g}|^2,
\quad
w_g
=
\frac{P_g}{\sum_{g>0}P_g},
\quad
N_{\rm gap}^{(2)}
=
\frac{1}{\sum_{g>0}w_g^2}.
\label{eq:supp-gap-participation}
\end{equation}
For mixing, the effective number of current-carrying positive-gap channels is \(152.38\), and the strongest one carries only \(3.31\%\) of the power. The free control has \(N_{\rm gap}^{(2)}=11.37\) and a largest share of \(19.33\%\). Engineered return has \(N_{\rm gap}^{(2)}=2.30\); one positive gap alone carries \(63.02\%\) of the current power, and five channels carry \(90\%\). Supplementary Fig.~\ref{fig:S22-gap-power} shows the cumulative current-power distributions.

\begin{figure}[t]
\centering
\includegraphics[width=0.5\linewidth]{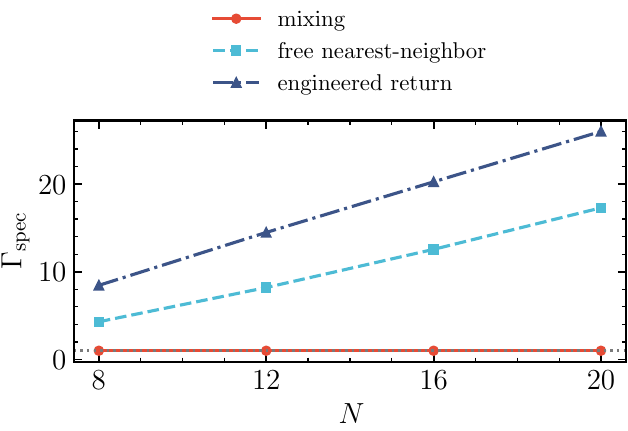}
\caption{Finite fixed-\(Q=2\) sequence of the equal-gap current-coherence factor. For the interacting mixing family, no repeated nonzero gap is resolved at any tested size \(N=8,12,16,20\), giving \(\Gamma_{\rm spec}=1\) throughout the sequence. The free and engineered families show progressively larger equal-gap enhancement. The four sizes are a finite-system diagnostic; no asymptotic scaling law is inferred.\justifying}
\label{fig:S21-spectral-size}
\end{figure}

The fixed-\(Q=2\) sequence \(N=8,12,16,20\) gives a complementary finite-size test. For mixing, no repeated nonzero gaps are resolved at any tested size, giving \(D_G=1\) and \(\Gamma_{\rm spec}=1\), while \(N_{\rm gap}^{(2)}\) grows from \(60.68\) to \(208.92\) and the largest positive-gap share falls from \(5.52\%\) to \(2.55\%\). For the free chain, \(\Gamma_{\rm spec}\) rises from \(4.30\) to \(17.28\). For engineered return it rises from \(8.45\) to \(25.99\), while \(\Gamma_{\rm spec}/D_G\) stays near \(0.72\) and \(N_{\rm gap}^{(2)}\) remains near \(2.3\). Supplementary Fig.~\ref{fig:S21-spectral-size} shows these values. The sequence is finite and fixed-particle-number; no thermodynamic-limit scaling law is inferred.

\begin{figure}[t]
\centering
\includegraphics[width=0.5\linewidth]{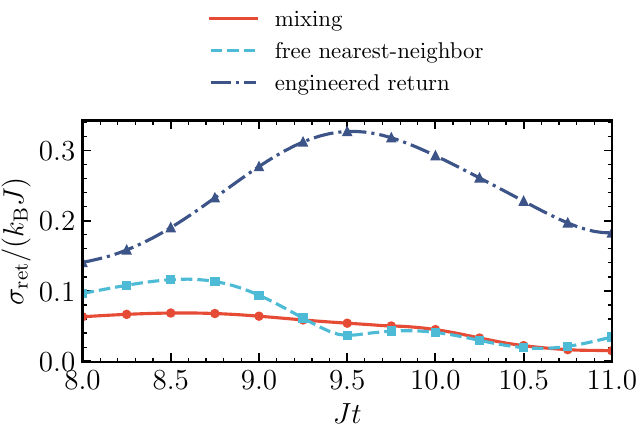}
\caption{End-to-end reconstruction of the return-oriented entropy-rate contribution on the common post-relaxation window \(8\leq Jt\leq11\). Curves are obtained from direct propagation; markers are reconstructed independently from the exact gap amplitudes in Eq.~\eqref{eq:supp-gap-current-amplitude} and the directly propagated affinities. The two calculations agree to machine precision.\justifying}
\label{fig:S23-spectral-reconstruction}
\end{figure}

Finally, we reconstruct the complete current trajectories from Eq.~\eqref{eq:supp-exact-current-spectrum} and combine them with the directly propagated affinities. On the common post-relaxation window \(8\leq Jt\leq11\), the maximum differences between the spectrally reconstructed and directly propagated \(\sigma_{\rm ret}\), \(\sigma_{\rm spread}\), and \(\dot S_R\) are below \(10^{-15}\). The corresponding root-mean-square \(\sigma_{\rm ret}/(k_{\rm B}J)\) values are
\begin{equation}
0.05229
\quad\text{(mixing)},
\quad
0.07220
\quad\text{(free)},
\quad
0.25312
\quad\text{(engineered return)}.
\end{equation}
Supplementary Fig.~\ref{fig:S23-spectral-reconstruction} displays this end-to-end reconstruction. These finite-window values are not determined by \(\Gamma_{\rm spec}\) alone: the affinities and total transition powers also differ. Their purpose is to verify that the gap-resolved amplitudes reconstruct the actual thermodynamic return current exactly.

\suppsubsection{Relation to generic equilibration bounds}
\label{supp:spectral-generic-bounds}

The exact current spectrum should be distinguished from generic observable-equilibration bounds. Reimann and Short--Farrelly bound temporal fluctuations using state-independent observable norms together with effective dimension and gap concentration~\cite{reimann2008foundation,short2012equilibration}. Those results remain important sufficient conditions for equilibration, but their worst-case relaxations need not be quantitatively informative for a particular thermodynamic current.

We evaluated the complete Short--Farrelly-based return bound for the same central models before introducing the current-specific decomposition. With the theorem-correct effective dimensions, \(d_{\rm eff}=102.958\) for mixing and \(22.366\) for engineered return. On the common window \(8\leq Jt\leq11\), the actual root-mean-square return burdens are \(0.05229\) and \(0.25312\,k_{\rm B}J\), whereas the corresponding generic upper bounds are \(124.7\) and \(449.1\,k_{\rm B}J\). The associated Chebyshev bounds on the fraction of nonpositive-entropy-rate times are therefore trivial. Along the fixed-\(Q=2\) mixing sequence, \(D_G/d_{\rm eff}\) decreases with \(N\), but the optimized finite-time gap-concentration factor on the fixed \(8\leq Jt\leq30\) window increases because the distinct gaps become increasingly dense. We consequently do not use the generic finite-time bound to establish the observed irreversible window.

This negative comparison is useful. The exact sinc kernel in Eq.~\eqref{eq:supp-finite-window-current-power} shows directly why a finite observation time can fail to resolve many nearby but nondegenerate gaps, while Eqs.~\eqref{eq:supp-Gamma-bound}--\eqref{eq:supp-spectral-current-theorem} isolate the different long-time question of coherent addition among exactly equal gaps. The two descriptions are complementary: generic equilibration bounds provide broad sufficient conditions, whereas the present gap amplitudes retain the actual preparation and current-operator matrix elements relevant to thermodynamic return.

\suppnote{Closed many-qubit models and thermodynamic records}
\label{supp:note-models}

\suppsubsection{Fixed-excitation Hilbert sector}
\label{supp:fixed-Q}

We consider \(N\) qubits on an open one-dimensional chain. Equivalently, the spin-up excitations are hard-core particles: each lattice site can contain at most one excitation. We define the local occupation operator
\begin{equation}
n_j
=
\sigma_j^+\sigma_j^-.
\end{equation}
The total excitation number is
\begin{equation}
Q
=
\sum_{j=1}^{N}
n_j.
\end{equation}
All qubit Hamiltonians used in the numerical tests conserve \(Q\). The dynamics can therefore be restricted from the outset to the fixed-\(Q\) invariant sector, whose dimension is
\begin{equation}
D_{N,Q}
=
\binom{N}{Q}.
\end{equation}
A computational-basis vector in this sector is specified uniquely by the ordered set of occupied sites,
\begin{equation}
|j_1,\ldots,j_Q\rangle,
\quad
1\leq j_1<\cdots<j_Q\leq N.
\end{equation}
All macroprojectors and thermodynamic records used for the hard-core calculations below are understood as acting within this conserved sector. Thus the thermodynamic coarse graining never mixes states with different values of the exactly conserved excitation number.

\suppsubsection{Mixing Hamiltonian}
\label{supp:mixing-H}

The principal autonomous Hamiltonian used to generate thermodynamic relaxation is
\begin{equation}
\begin{aligned}
H_{\rm mix}
={}&
-J
\sum_{j=1}^{N-1}
\left(
\sigma_j^+
\sigma_{j+1}^-
+
\sigma_j^-
\sigma_{j+1}^+
\right)
\\
&
+
V_1
\sum_{j=1}^{N-1}
n_jn_{j+1}
+
V_2
\sum_{j=1}^{N-2}
n_jn_{j+2}.
\end{aligned}
\end{equation}
The first term transfers an excitation between neighboring sites while conserving \(Q\). The remaining terms assign interaction energies to simultaneous occupation of nearest- and next-nearest-neighbor sites. Unless stated otherwise, we use
\begin{equation}
V_1
=
J,
\quad
V_2
=
0.7J.
\end{equation}
After preparation the Hamiltonian is time independent, so every fine initial state compatible with the same thermodynamic preparation evolves under the same autonomous unitary \(e^{-iH_{\rm mix}t}\). There is no stochastic drive, measurement, or external dynamical rule that imposes relaxation.

All times for the static many-body models are reported through the dimensionless combination \(Jt\).

\suppsubsection{Free nearest-neighbor control}
\label{supp:free-H}

As a noninteracting control, we use the nearest-neighbor hopping Hamiltonian
\begin{equation}
H_{\rm free}
=
-J
\sum_{j=1}^{N-1}
\left(
\sigma_j^+
\sigma_{j+1}^-
+
\sigma_j^-
\sigma_{j+1}^+
\right).
\end{equation}
This Hamiltonian preserves the same fixed-\(Q\) sector and the same lattice topology as the interacting model. It provides an intermediate control in which the microscopic dynamics remains structured and integrable while different single-particle modes can nevertheless accumulate distinct dynamical phases.

\suppsubsection{Engineered-return Hamiltonian}
\label{supp:return-H}

To construct a matched closed-system control with organized thermodynamic return, we use
\begin{equation}
H_{\rm ret}
=
-
\sum_{j=1}^{N-1}
J_j
\left(
\sigma_j^+
\sigma_{j+1}^-
+
\sigma_j^-
\sigma_{j+1}^+
\right),
\end{equation}
with position-dependent couplings
\begin{equation}
J_j
=
\frac{2J}{N}
\sqrt{
j(N-j)
}.
\end{equation}
For even \(N\),
\begin{equation}
J_j
\leq
J,
\end{equation}
so the engineered model operates with the same order of local hopping scale as \(H_{\rm mix}\).

The coupling profile produces a commensurate single-particle spectrum and perfect mirror transfer at
\begin{equation}
t_{\rm mir}
=
\frac{\pi N}{4J}.
\end{equation}
A fine occupation configuration initially supported in the left half is mapped, up to the corresponding unitary phase, to its spatially reflected configuration in the right half. This is not required to be a recurrence of the complete microscopic state to its initial configuration. Thermodynamically, however, the left- and right-confined configurations belong to sectors with equal multiplicity under the spatial records used below. The accessible entropy can therefore return to its initial value at the mirror time. The purpose of this model is to show that the same closed-system setting and the same thermodynamic coarse graining can support organized macroscopic return when the microscopic dynamics is deliberately structured to rephase.

\suppsubsection{Two-cell record}
\label{supp:two-cell}

For the coarser spatial description, the chain is divided into equal left and right halves and the thermodynamic record retains only the number of excitations in the left half,
\begin{equation}
R
=
N_L.
\end{equation}
If the observed value is \(N_L=m\), then \(m\) excitations occupy the \(N/2\) sites of the left half and \(Q-m\) occupy the \(N/2\) sites of the right half. The corresponding hard-core macrospace multiplicity is therefore
\begin{equation}
\Omega_m
=
\binom{N/2}{m}
\binom{N/2}{Q-m}.
\end{equation}
This record resolves only the total left-right imbalance and leaves all microscopic positions within each half unresolved.

\suppsubsection{Four-cell record}
\label{supp:four-cell}

For the finer spatial thermodynamic description, the chain is divided into four equal cells of length
\begin{equation}
\ell
=
\frac{N}{4}.
\end{equation}
The accessible record is
\begin{equation}
R
=
(N_1,N_2,N_3,N_4),
\end{equation}
where \(N_a\) is the number of excitations in cell \(a\), subject to
\begin{equation}
\sum_{a=1}^{4}
N_a
=
Q.
\end{equation}
For hard-core excitations, cell \(a\) contains \(\binom{\ell}{N_a}\) microscopic occupation patterns compatible with the same cell occupation. Hence the macrospace multiplicity is
\begin{equation}
\Omega_R
=
\prod_{a=1}^{4}
\binom{\ell}{N_a}.
\end{equation}
Summing over all admissible records gives, by repeated application of the Vandermonde identity,
\begin{equation}
\sum_R
\Omega_R
=
\binom{N}{Q}
=
D_{N,Q}.
\end{equation}
The multiplicity-defined reference distribution for this record is consequently
\begin{equation}
\pi_R
=
\frac{\Omega_R}{D_{N,Q}}
=
\frac{
\displaystyle
\prod_{a=1}^{4}
\binom{\ell}{N_a}
}{
\displaystyle
\binom{N}{Q}
}.
\end{equation}
Thus \(\pi_R\) is fixed entirely by the dimensions of the four-cell macrospaces within the conserved fixed-\(Q\) sector.

\suppsubsection{Interacting Bose--Hubbard realization}
\label{supp:bose-hubbard-model}

To test the framework with different particle statistics and a different local Hilbert structure, we consider a one-dimensional Bose--Hubbard chain,
\begin{equation}
H_{\rm BH}
=
-J
\sum_{j=1}^{L-1}
\left(
b_j^\dagger b_{j+1}
+
b_{j+1}^\dagger b_j
\right)
+
\frac{U}{2}
\sum_{j=1}^{L}
n_j(n_j-1),
\end{equation}
where \(n_j=b_j^\dagger b_j\). The total boson number
\begin{equation}
Q
=
\sum_{j=1}^{L}
n_j
\end{equation}
is conserved. The representative calculation uses
\begin{equation}
L
=
8,
\quad
Q
=
3,
\quad
U
=
1.3J.
\end{equation}
Unlike the hard-core realization, more than one boson may occupy the same lattice site.

The chain is divided into four equal cells of length \(\ell=L/4\), and the thermodynamic record again retains the cell occupations \(R=(N_1,N_2,N_3,N_4)\). For \(N_a\) indistinguishable bosons distributed over \(\ell\) sites with unrestricted on-site occupancy, the number of compatible Fock states is the stars-and-bars multiplicity \(\binom{\ell+N_a-1}{N_a}\). Hence
\begin{equation}
\Omega_R^{\rm B}
=
\prod_{a=1}^{4}
\binom{\ell+N_a-1}{N_a},
\end{equation}
while the dimension of the complete fixed-\(Q\) bosonic sector is
\begin{equation}
\Omega_{\rm tot}^{\rm B}
=
\binom{L+Q-1}{Q}.
\end{equation}
The initial thermodynamic class consists of all fixed-\(Q\) bosonic Fock states supported within the left half of the chain. Its multiplicity is
\begin{equation}
\Omega_i^{\rm B}
=
\binom{L/2+Q-1}{Q}.
\end{equation}
The Bose--Hubbard realization therefore changes both the microscopic statistics and the local occupation structure while preserving the thermodynamic logic of a fixed-particle-number preparation that is equilibrated over the initially available left-half volume.

\suppsubsection{Two-dimensional interacting hard-core realization}
\label{supp:2d-hardcore-model}

To test the role of spatial geometry and connectivity, we also consider hard-core excitations on an open \(L_x\times L_y\) square lattice,
\begin{equation}
H_{2{\rm D}}
=
-J
\sum_{\langle ij\rangle}
\left(
\sigma_i^+\sigma_j^-
+
\sigma_i^-\sigma_j^+
\right)
+
V
\sum_{\langle ij\rangle}
n_in_j,
\end{equation}
where \(\langle ij\rangle\) denotes nearest-neighbor lattice edges. The calculation reported in the main text uses
\begin{equation}
L_x
=
L_y
=
4,
\quad
Q
=
3,
\quad
V
=
J.
\end{equation}
The thermodynamic record partitions the square into four equal spatial quadrants and retains only the excitation numbers \(R=(N_1,N_2,N_3,N_4)\) in those quadrants. Each quadrant contains \(L_xL_y/4\) sites, so the corresponding hard-core multiplicity is
\begin{equation}
\Omega_R^{2{\rm D}}
=
\prod_{a=1}^{4}
\binom{L_xL_y/4}{N_a}.
\end{equation}
For the \(4\times4\) realization, each quadrant contains four sites. The initial thermodynamic class contains all fixed-\(Q\) product configurations supported in the left half of the square. The two-dimensional calculation therefore changes the connectivity and transport geometry while preserving the same meaning of the thermodynamic preparation and spatial record used in the one-dimensional tests.

\suppsubsection{Number-conserving Floquet realization}
\label{supp:floquet-model}

The exact coarse-propagation result requires only a unitary step and does not require evolution generated by a single time-independent Hamiltonian. To test this explicitly, we consider a number-conserving stroboscopic unitary on a hard-core chain. One Floquet period is
\begin{equation}
U_F
=
U_{\rm int}(\phi)
U_{\rm odd}(\theta_{\rm o})
U_{\rm even}(\theta_{\rm e}),
\end{equation}
where the even- and odd-bond layers are generated by number-conserving nearest-neighbor hopping on the corresponding disjoint bonds, and the diagonal interaction layer is
\begin{equation}
U_{\rm int}(\phi)
=
\exp
\left[
-i\phi
\sum_{j=1}^{N-1}
n_jn_{j+1}
\right].
\end{equation}
The representative realization uses
\begin{equation}
N
=
12,
\quad
Q
=
3,
\quad
\theta_{\rm e}
=
0.63,
\quad
\theta_{\rm o}
=
0.83,
\quad
\phi
=
0.77.
\end{equation}
We use the same four-cell record and the same left-half thermodynamic preparation as in the static hard-core calculations. For one Floquet period, the exact coarse matrix is obtained simply by replacing \(U_\tau\) with \(U_F\) in
\begin{equation}
K^{(U)}_{RR'}
=
\frac{1}{\Omega_{R'}}
\Tr
\left[
P_RUP_{R'}U^\dagger
\right].
\end{equation}
Its nonnegativity, stochasticity, and exact preservation of \(\pi\) therefore follow from the same unitary proof as for continuous-time evolution. This realization tests the structural result beyond dynamics generated by one static Hamiltonian.

\suppsubsection{Thermodynamic free-expansion preparation}
\label{supp:free-expansion-prep}

The free-expansion calculations use a preparation record that is deliberately coarser than the four-cell record used to monitor the subsequent dynamics. The thermodynamic preparation specifies only that all \(Q\) excitations lie somewhere within the left half of the chain, whose length is
\begin{equation}
L_i
=
\frac{N}{2}.
\end{equation}
Their exact microscopic positions inside that initially available volume are not resolved.

Let \(P_{L_i,Q}\) denote the projector onto the fixed-\(Q\) subspace spanned by all hard-core product configurations supported entirely within the first \(L_i\) sites. Its dimension is
\begin{equation}
\Omega_i
=
\Tr P_{L_i,Q}
=
\binom{L_i}{Q}.
\end{equation}
The state containing exactly the information specified by this thermodynamic preparation and no further distinction among its compatible microscopic configurations is
\begin{equation}
\rho_i^{\rm th}
=
\frac{
P_{L_i,Q}
}{
\Omega_i
}.
\end{equation}
This thermodynamic representative should not be interpreted as stochastic branching of the isolated microscopic dynamics or as a physical randomization operation. At the fine level, each compatible microscopic initial state follows one definite unitary history under the same Hamiltonian. By linearity, evolving \(\rho_i^{\rm th}\) gives exactly the equal-weight average of the macroprobabilities generated by those compatible orthogonal fine initial configurations.

For the four-cell observation record, the left half consists of cells \(1\) and \(2\). The initial four-cell macrodistribution is therefore not concentrated on one record. Instead,
\begin{equation}
p_R(0)
=
\begin{cases}
\Omega_R/\Omega_i,
&
N_3=N_4=0,
\quad
N_1+N_2=Q,
\\
0,
&
\text{otherwise}.
\end{cases}
\end{equation}
For every initially allowed record,
\begin{equation}
\frac{p_R(0)}{\Omega_R}
=
\frac{1}{\Omega_i}.
\end{equation}
Thus the probability per compatible microscopic configuration is uniform throughout the initially available volume. This is the discrete lattice analogue of a gas that is thermodynamically equilibrated within its initial volume before the expansion, rather than one prepared in a particular fine microscopic configuration.

The corresponding accessible entropy is
\begin{equation}
S_R(0)
=
k_{\rm B}\ln\Omega_i
=
k_{\rm B}
\ln
\binom{L_i}{Q}.
\end{equation}
After complete multiplicity-defined spreading over the full final volume \(L_f=N\), the total hard-core multiplicity is
\begin{equation}
\Omega_f
=
\binom{L_f}{Q},
\end{equation}
and the corresponding reference macrodistribution is
\begin{equation}
p_R^{\rm ref}
=
\frac{\Omega_R}{\Omega_f}.
\end{equation}
Its accessible entropy is therefore
\begin{equation}
S_R^{\rm ref}
=
k_{\rm B}\ln\Omega_f.
\end{equation}
The difference \(S_R^{\rm ref}-S_R(0)\) is consequently fixed entirely by the change in microscopic multiplicity between the initially accessible and final volumes. Its finite-lattice form and dilute classical limit are derived separately below.

\suppnote{Fine-history robustness and the thermodynamic compatibility class}
\label{supp:note-history-class}

This robustness test addresses a distinction that is essential to the interpretation of the thermodynamic preparation. A thermodynamic preparation fixes only a restricted record, whereas one exact microscopic initial state together with the fixed Hamiltonian determines one unique unitary history. We therefore distinguish explicitly between the label \(x\) of a fine initial configuration and the corresponding microscopic history \(\gamma_x\) generated from it.

For this test we choose
\begin{equation}
N
=
16,
\quad
Q
=
4,
\end{equation}
and, unlike the main free-expansion preparation, deliberately fix the finer four-cell initial record
\begin{equation}
R_0
=
(2,2,0,0).
\end{equation}
Each thermodynamic cell has length \(\ell=4\). The record specifies that two excitations occupy cell \(1\), two occupy cell \(2\), and none occupy cells \(3\) and \(4\), while leaving the exact occupied sites within the first two cells unresolved. The number of compatible computational-basis microstates is therefore
\begin{equation}
\Omega_{R_0}
=
\binom{4}{2}
\binom{4}{2}
=
36.
\end{equation}
Let
\begin{equation}
\mathcal X_{R_0}
=
\left\{
x:
|x\rangle
\text{ is a computational-basis state with record }R_0
\right\}.
\end{equation}
For each \(x\in\mathcal X_{R_0}\), the exact fine initial state is
\begin{equation}
\rho_{x,0}
=
|x\rangle\langle x|.
\end{equation}
All \(36\) compatible states evolve under the same autonomous Hamiltonian \(H_{\rm mix}\). The corresponding microscopic history is
\begin{equation}
\gamma_x
=
\left\{
|\psi_x(s)\rangle
=
e^{-iH_{\rm mix}s}|x\rangle:
0\leq s\leq T
\right\}.
\end{equation}
Thus the different histories in this calculation arise exclusively from different microscopic initial configurations compatible with the same thermodynamic preparation. There is no stochastic branching of a single exact initial state, no history-dependent Hamiltonian, and no distinct microscopic dynamical law assigned to different members of the compatibility class.

For a particular history \(\gamma_x\), the probability of observing macrorecord \(R\) at time \(t\) is
\begin{equation}
p_R(t|\gamma_x)
=
\langle
\psi_x(t)
|
P_R
|
\psi_x(t)
\rangle.
\end{equation}
One may associate with this individual fine history the history-conditioned coarse entropy
\begin{equation}
S_{R|\gamma_x}(t)
=
-k_{\rm B}
\sum_R
p_R(t|\gamma_x)
\ln
\left[
\frac{
p_R(t|\gamma_x)
}{
\Omega_R
}
\right].
\end{equation}
This quantity is useful as a diagnostic of the coarse behavior generated by a particular microscopic initial condition. It is not, however, the thermodynamic entropy assigned to the preparation, because the preparation does not specify which \(x\in\mathcal X_{R_0}\) was realized.

The thermodynamic representative of the complete preparation class is instead
\begin{equation}
\rho_{R_0}^{\rm th}
=
\frac{
P_{R_0}
}{
\Omega_{R_0}
}
=
\frac{1}{36}
\sum_{x\in\mathcal X_{R_0}}
|x\rangle\langle x|.
\end{equation}
Evolving this representative under the same Hamiltonian is exactly equivalent, by linearity, to averaging the macroprobabilities generated by the \(36\) compatible fine histories,
\begin{equation}
\begin{aligned}
p_R^{\rm th}(t)
&=
\Tr
\left[
P_R
e^{-iH_{\rm mix}t}
\rho_{R_0}^{\rm th}
e^{iH_{\rm mix}t}
\right]
\\
&=
\frac{1}{36}
\sum_{x\in\mathcal X_{R_0}}
p_R(t|\gamma_x).
\end{aligned}
\end{equation}
This equality is purely a consequence of the thermodynamic representative and the linearity of quantum evolution and measurement probabilities. It does not mean that one physical isolated system jumps randomly among the \(36\) histories. Each fine initial state determines one definite unitary trajectory; the average represents only the information specified by the unresolved thermodynamic preparation.

The corresponding class-level accessible entropy is
\begin{equation}
S_R^{\rm th}(t)
=
-k_{\rm B}
\sum_R
p_R^{\rm th}(t)
\ln
\left[
\frac{
p_R^{\rm th}(t)
}{
\Omega_R
}
\right].
\end{equation}
The distinction between \(S_R^{\rm th}(t)\) and the set of history-conditioned quantities \(S_{R|\gamma_x}(t)\) is important. The entropy of the thermodynamic class is obtained from the class-averaged macrodistribution \(p_R^{\rm th}(t)\); it is not the arithmetic average of the individual history-conditioned entropies. The nonlinearity of the entropy functional therefore preserves a genuine distinction between the thermodynamic class and its individual fine realizations.

\begin{figure}[t]
\centering
\includegraphics[width=0.8\linewidth]{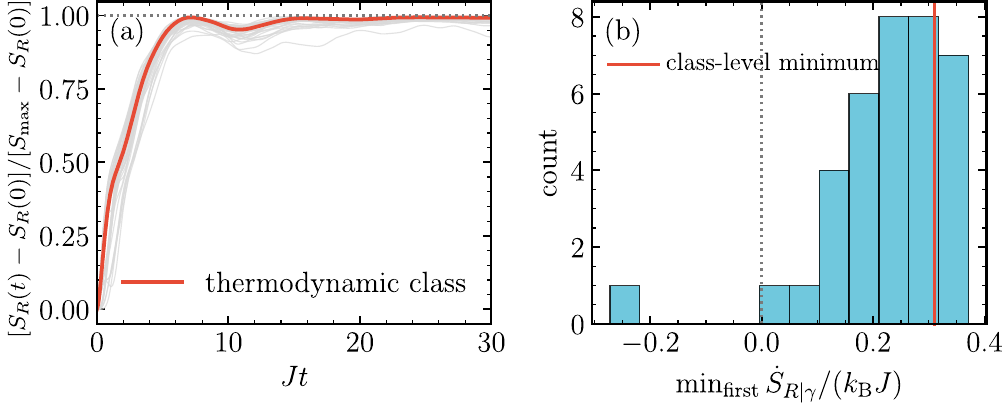}
\caption{Fine-history versus thermodynamic-class dynamics for \(N=16\), \(Q=4\), and the fixed four-cell preparation \(R_0=(2,2,0,0)\).
(a) History-conditioned coarse-entropy trajectories generated by all \(36\) compatible computational-basis initial microstates under the same Hamiltonian \(H_{\rm mix}\), together with the entropy of the complete thermodynamic class.
(b) Distribution of the minimum sampled history-conditioned accessible-entropy rates evaluated over the class-level first \(10\%\)--\(90\%\) relaxation window. The corresponding class-level minimum is indicated separately. Individual compatible histories can display entropy-lowering intervals even when the complete thermodynamic class has a positive accessible-entropy rate, demonstrating that thermodynamic monotonicity is a property of the accessible preparation rather than a condition imposed on every unresolved microscopic history.\justifying
}
\label{fig:S1}
\end{figure}

Figure~\ref{fig:S1} compares the individual fine-history diagnostics with the complete thermodynamic class. Panel~\figpanel{fig:S1}{a} shows all \(36\) history-conditioned coarse-entropy trajectories together with the class-level entropy. Panel~\figpanel{fig:S1}{b} shows the distribution of the minimum sampled history-conditioned accessible-entropy rates evaluated over the primary \(10\%\)--\(90\%\) relaxation window defined by the class-level trajectory, together with the corresponding class-level minimum. Individual compatible histories can exhibit substantially stronger entropy-lowering intervals even while the complete thermodynamic class retains a positive accessible-entropy rate over the same thermodynamic window.

The purpose of this calculation is therefore not to impose a Second Law separately on every microscopic history. It demonstrates precisely the opposite. A fixed thermodynamic preparation represents a class of fine states that are experimentally or operationally indistinguishable at the chosen level of description, and individual members of that class may display nonmonotonic coarse behavior. The thermodynamic Second-Law statement applies to the accessible preparation and its class-level macrodistribution, not to every microscopic trajectory that the preparation leaves unresolved.

\suppnote{Robustness under refinement of the thermodynamic record}
\label{supp:note-record-refinement}

The thermodynamic behavior should not be an artifact of choosing an exceptionally coarse spatial record. We therefore compare two descriptions of otherwise identical microscopic dynamics: the two-cell left-right record \(R=N_L\) and the finer four-cell density record \(R=(N_1,N_2,N_3,N_4)\). The microscopic Hamiltonian, preparation, and unitary evolution are unchanged; only the information retained by the thermodynamic description is refined.

For either record, the accessible entropy is defined by
\begin{equation}
S_R
=
-k_{\rm B}
\sum_R
p_R
\ln
\left(
\frac{p_R}{\Omega_R}
\right).
\end{equation}
The corresponding projectors \(P_R\), macrospace dimensions \(\Omega_R\), and probability distributions \(p_R\) differ because the two records resolve different amounts of spatial information. Refining the record partitions the coarse macrospaces into smaller ones and can therefore expose probability currents that are invisible at lower resolution.

To quantify the competition between entropy-increasing and entropy-lowering currents during the primary relaxation interval \(\mathcal T_{\rm first}\), we use the exact decomposition
\begin{equation}
\dot S_R
=
\sigma_{\rm spread}
-
\sigma_{\rm ret},
\end{equation}
and define the integrated return burden
\begin{equation}
\mathcal R
=
\frac{
\displaystyle
\int_{\mathcal T_{\rm first}}
dt\,
\sigma_{\rm ret}(t)
}{
\displaystyle
\int_{\mathcal T_{\rm first}}
dt\,
\sigma_{\rm spread}(t)
}.
\end{equation}
The quantity \(\mathcal R\) measures the cumulative entropy-lowering current relative to the cumulative entropy-increasing current over the same thermodynamic relaxation window. It does not require \(\sigma_{\rm ret}(t)\) to vanish instantaneously. Rather, a positive accessible-entropy rate throughout the interval requires only that the total spreading contribution dominate the total return contribution at each relevant time.

The four-cell description resolves spatial redistribution within each half of the chain and therefore exposes local entropy-lowering currents that are absent from the binary left-right description. Consequently,
\begin{equation}
\sigma_{\rm ret}(t)
>
0
\end{equation}
can occur under the finer record even while
\begin{equation}
\dot S_R(t)
>
0.
\end{equation}
This distinction is important: refinement need not eliminate thermodynamic monotonicity, but it can reveal additional internal competition that was hidden by a coarser record.

\begin{figure}[t]
\centering
\includegraphics[width=0.8\linewidth]{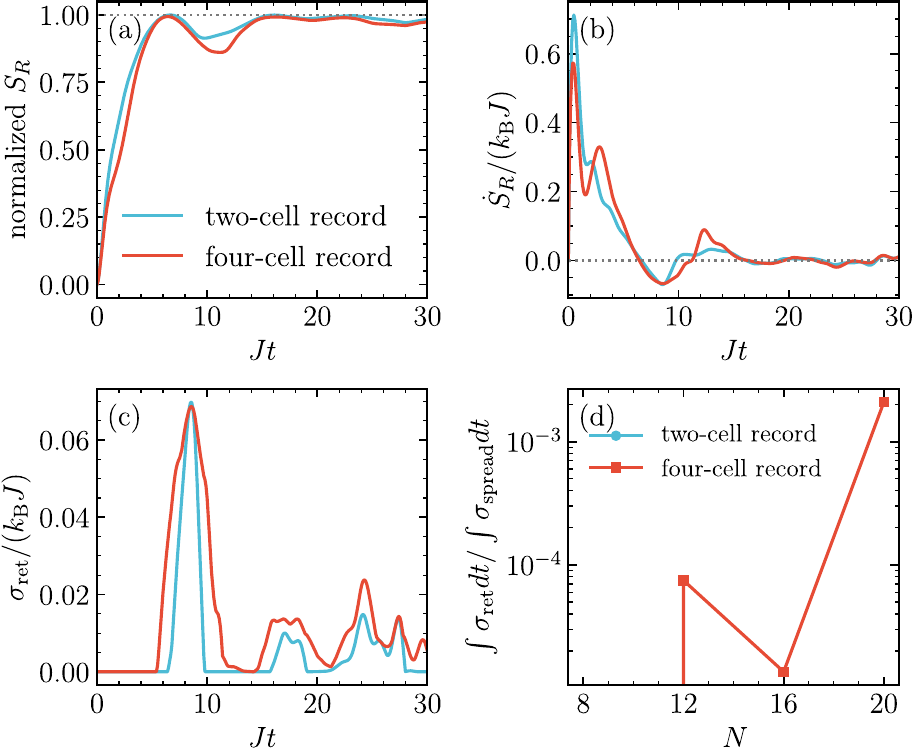}
\caption{Robustness under refinement of the accessible thermodynamic record.
(a) Normalized accessible entropy for the two-cell left--right record and the finer four-cell spatial record under otherwise identical microscopic dynamics.
(b) Corresponding exact instantaneous accessible-entropy rates.
(c) Entropy-lowering macrocurrent contribution \(\sigma_{\rm ret}\). The four-cell record resolves local return currents that are invisible to the coarser two-cell description, even while the total accessible-entropy rate remains positive during primary relaxation.
(d) Integrated return burden \(\mathcal R\) during the primary relaxation for
\((N,Q)=(8,2),(12,3),(16,4),(20,5)\).
This sequence increases the total chain size \(N\) at fixed excitation density \(Q/N=1/4\), equivalently at fixed initial filling \(Q/L_i=1/2\) with \(L_i=N/2\).
The persistence of a positive primary accessible-entropy rate under the finer record shows that the observed irreversible regime is not an artifact of an excessively coarse spatial description or of suppressing all locally entropy-lowering currents.\justifying
}
\label{fig:S2}
\end{figure}

Figure~\ref{fig:S2} compares the two resolutions directly. Panels~\figpanel{fig:S2}{a} and \figpanel{fig:S2}{b} show the normalized accessible entropy and its exact instantaneous rate. Panel~\figpanel{fig:S2}{c} displays the entropy-lowering contribution \(\sigma_{\rm ret}\) exposed by the finer four-cell record, while panel~\figpanel{fig:S2}{d} shows the integrated return burden \(\mathcal R\) over the investigated system sizes. The comparison demonstrates that the positive primary accessible-entropy-rate behavior survives a substantial refinement of the accessible thermodynamic record even though the finer description resolves additional entropy-lowering currents. For the size dependence in panel~\figpanel{fig:S2}{d}, we use
\((N,Q)=(8,2),(12,3),(16,4),(20,5)\), so that the total
system size increases at fixed excitation density \(Q/N=1/4\).

\suppnote{Finite-size recurrence and return suppression}
\label{supp:note-finite-size-return}

Exact finite-dimensional unitary dynamics retains the microscopic information required for recurrence, as emphasized by the Loschmidt reversal and Zermelo recurrence objections~\cite{loschmidt1876waermegleichgewicht,zermelo1896satz}. A strictly monotonic accessible entropy for all mathematical times is therefore neither assumed nor expected. The relevant thermodynamic question is instead how strongly finite-size return affects the accessible macrodistribution within and beyond the primary relaxation window.

For each trajectory, let \(t_{90}\) denote the first time at which the normalized accessible entropy
\begin{equation}
\Sigma_R(t)
=
\frac{
S_R(t)-S_R(0)
}{
S_{\max}-S_R(0)
}
\end{equation}
reaches \(90\%\) of its multiplicity-defined range,
\begin{equation}
\Sigma_R(t_{90})
\geq
0.9.
\end{equation}
Let \(T_{\rm obs}\) denote the end of the finite observation interval used for the numerical comparison. We quantify the largest subsequent normalized entropy return within that interval by
\begin{equation}
\Delta_{\rm ret}^{\max}(T_{\rm obs})
=
\max_{t_{90}\leq t_1<t_2\leq T_{\rm obs}}
\left[
\Sigma_R(t_1)
-
\Sigma_R(t_2)
\right]_+.
\end{equation}
Thus \(\Delta_{\rm ret}^{\max}(T_{\rm obs})\) measures the maximum downward excursion of the normalized accessible entropy after the first \(90\%\) crossing and before the stated observation cutoff. The cutoff is essential: over unrestricted mathematical time, finite-dimensional recurrence precludes interpreting the plotted quantity as a permanent suppression of return.

A complementary current-level diagnostic is the root-mean-square late-time entropy-lowering contribution,
\begin{equation}
A_{\rm ret}
=
\sqrt{
\left\langle
\sigma_{\rm ret}^2
\right\rangle_{\rm late}
}.
\end{equation}
Here the average is taken over the late-time window used in the numerical analysis. Whereas \(\Delta_{\rm ret}^{\max}(T_{\rm obs})\) measures the largest net macroscopic entropy return within the finite observation interval, \(A_{\rm ret}\) measures the typical magnitude of the resolved entropy-lowering current in the late-time dynamics.

These are finite-size diagnostics. They do not assume, and the available data do not establish, any particular asymptotic scaling law with system size. In particular, we do not infer exponential, power-law, or other functional suppression of recurrence from the finite sequence studied here.

\begin{figure}[t]
\centering
\includegraphics[width=0.8\linewidth]{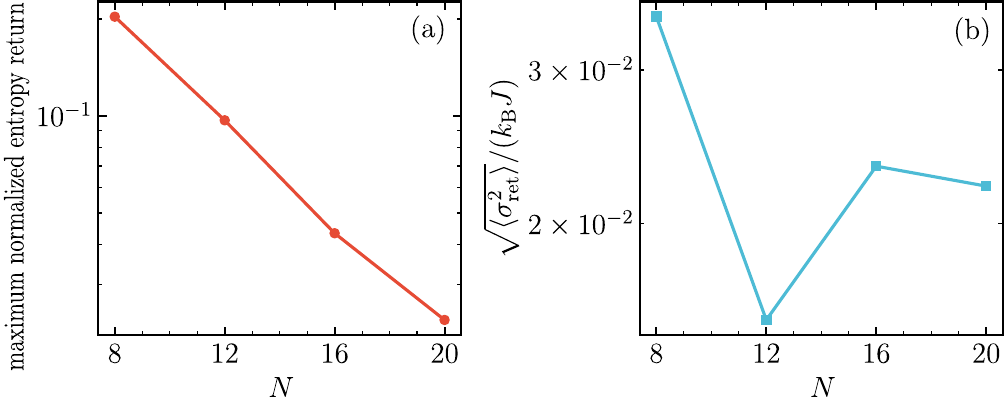}
\caption{Finite-size return diagnostics for the mixing Hamiltonian.
(a) Maximum normalized entropy-return depth \(\Delta_{\rm ret}^{\max}(T_{\rm obs})\) after the first \(90\%\) entropy crossing and within the finite observation interval.
(b) Late-time root-mean-square return-oriented entropy-rate contribution \(A_{\rm ret}\). Both quantities become substantially smaller over the investigated finite-size sequence. The data are used only to characterize the observed finite-size trend; no exponential, power-law, or other asymptotic scaling form is inferred.\justifying
}
\label{fig:S3}
\end{figure}

Figure~\ref{fig:S3} shows both quantities for the investigated mixing systems. Over this sequence, the finite-window maximum entropy-return depth and the late-time root-mean-square return-oriented entropy-rate contribution both become substantially smaller. The observed trend is consistent with increasingly weak thermodynamic return over the accessible finite-size sequence, while remaining fully compatible with exact microscopic reversibility and the possibility of recurrence at sufficiently long times.

\suppnote{Current-level comparison of mixing and engineered return}
\label{supp:note-current-control}

The exact macroprobability-current decomposition provides an instantaneous dynamical comparison between ordinary thermodynamic relaxation and deliberately organized return. For both Hamiltonians,
\begin{equation}
\dot S_R
=
\sigma_{\rm spread}
-
\sigma_{\rm ret},
\end{equation}
with \(\sigma_{\rm spread}\geq0\) and \(\sigma_{\rm ret}\geq0\).

For the mixing dynamics, the entropy-lowering contribution need not vanish. Instead, during the primary relaxation one finds that the entropy-increasing currents dominate,
\begin{equation}
\sigma_{\rm spread}
>
\sigma_{\rm ret},
\end{equation}
so that
\begin{equation}
\dot S_R
>
0.
\end{equation}
Thus the observed thermodynamic relaxation is not produced by eliminating every return-oriented macrospace current. Local entropy-lowering currents can remain dynamically active while being outweighed by the total entropy-increasing redistribution.

The engineered-return Hamiltonian realizes the complementary regime. It can initially satisfy the same inequality and therefore display an initial entropy-increasing spreading stage. As the microscopic phases subsequently rephase, however, the entropy-lowering currents become sufficiently coordinated that
\begin{equation}
\sigma_{\rm ret}
>
\sigma_{\rm spread},
\end{equation}
and hence
\begin{equation}
\dot S_R
<
0.
\end{equation}
This reversal is the instantaneous current-level signature of organized thermodynamic return.

\begin{figure}[t]
\centering
\includegraphics[width=0.8\linewidth]{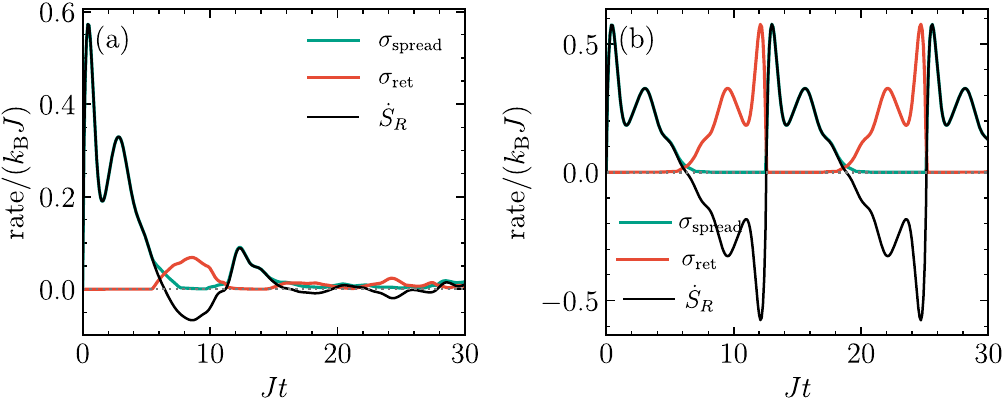}
\caption{Exact instantaneous entropy-current decomposition for the \(N=16\), \(Q=2\) closed chain.
(a) Mixing Hamiltonian.
(b) Engineered-return Hamiltonian.
The total accessible-entropy rate is \(\dot S_R=\sigma_{\rm spread}-\sigma_{\rm ret}\). Entropy-lowering currents are present in the mixing dynamics without overturning the positive primary relaxation, whereas the engineered dynamics develops intervals in which \(\sigma_{\rm ret}>\sigma_{\rm spread}\), producing macroscopic entropy decrease. The labels ``spread'' and ``return'' refer to the signs of contributions to the coarse entropy balance, not to distinct microscopic forces or trajectories.\justifying
}
\label{fig:S4}
\end{figure}

Figure~\ref{fig:S4} compares the complete balances for the \(N=16\), \(Q=2\) mixing and engineered-return systems. The key distinction is therefore not whether \(\sigma_{\rm ret}\) exists: a nonzero entropy-lowering contribution can already be present in the mixing dynamics. What distinguishes the engineered-return regime is that these contributions become sufficiently synchronized to overcome \(\sigma_{\rm spread}\) and reverse the macroscopic entropy flow.

\suppnote{Numerical verification of the \texorpdfstring{\(\pi\)-preserving}{pi-preserving} coarse map}
\label{supp:note-K-verification}

The stochasticity of \(K^{(U)}(\tau)\) and its exact preservation of the multiplicity-defined reference,
\begin{equation}
\sum_R
K^{(U)}_{RR'}(\tau)
=
1,
\quad
K^{(U)}(\tau)\pi
=
\pi,
\end{equation}
follow analytically from unitarity and completeness of the thermodynamic macroprojectors. We nevertheless verify both identities explicitly in the numerical implementation.

For the \(N=16\), \(Q=2\) mixing model with the four-cell record, we construct the complete matrix \(K^{(U)}(\tau)\) for
\begin{equation}
J\tau
=
0.02,
\,
0.05,
\,
0.10,
\,
0.20,
\,
0.50,
\,
1.00.
\end{equation}
We quantify deviations from column stochasticity by
\begin{equation}
\epsilon_{\rm stoch}
=
\max_{R'}
\left|
\sum_R
K^{(U)}_{RR'}
-
1
\right|,
\end{equation}
and deviations from exact preservation of the multiplicity-defined reference by
\begin{equation}
\epsilon_{\pi}
=
\left\|
K^{(U)}\pi-\pi
\right\|_\infty.
\end{equation}
Both residuals remain at floating-point precision over all tested finite time steps.

\begin{figure}[t]
\centering
\includegraphics[width=0.5\linewidth]{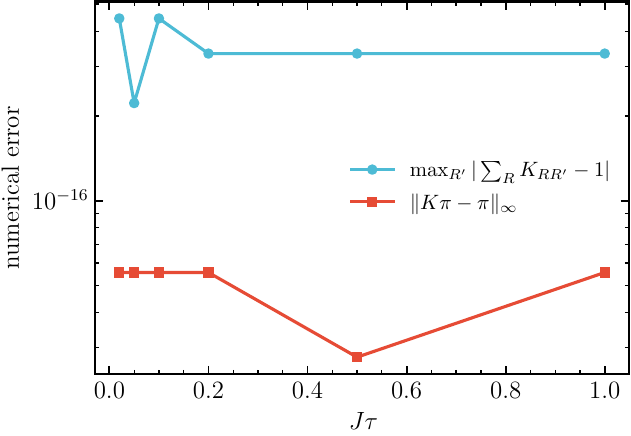}
\caption{
Numerical verification of the exact properties of \(K^{(U)}(\tau)\) for the \(N=16\), \(Q=2\) mixing model with the four-cell thermodynamic record. Shown are the maximum column-stochasticity residual \(\epsilon_{\rm stoch}\) and the maximum deviation \(\epsilon_\pi=\|K^{(U)}\pi-\pi\|_\infty\). Both remain at numerical roundoff over the tested finite propagation intervals, providing an implementation check of the analytical identities.\justifying
}
\label{fig:S5}
\end{figure}

Figure~\ref{fig:S5} reports these errors explicitly. Their magnitude is consistent with numerical roundoff and should therefore be interpreted as a verification of the implementation of the analytically exact identities, not as numerical evidence from which those identities are inferred.

\suppnote{Numerical verification of the history-defect bound}
\label{supp:note-bound-verification}

Supplementary Note~\ref{supp:note-history-bound} established the exact finite-step inequality
\begin{equation}
\Vpi
\left[
p(t)\rightarrow p(t+\tau)
\right]
\leq
\epsilon_{\rm hist}(t,\tau),
\end{equation}
where
\begin{equation}
\epsilon_{\rm hist}(t,\tau)
=
\frac{1}{2}
\left\|
r(t,\tau)
\right\|_1.
\end{equation}
The bound states that the history-sensitive correction provides an upper bound on the strongest possible violation of the relative-majorization ordering. We test this relation directly for the finite steps generated by both the mixing and engineered-return dynamics.

\begin{figure}[t]
\centering
\includegraphics[width=0.5\linewidth]{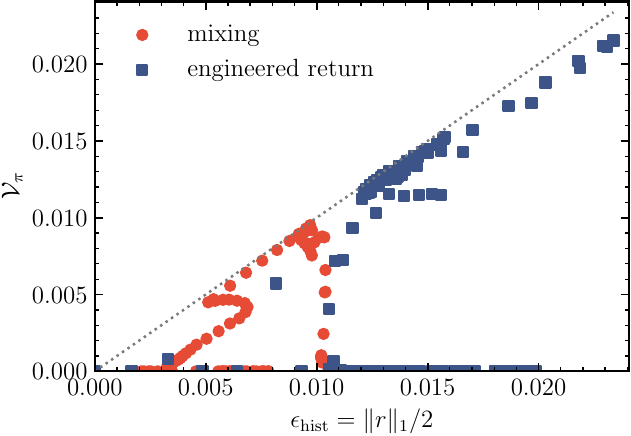}
\caption{
Direct numerical verification of the exact bound \(\Vpi\leq\epsilon_{\rm hist}\). Each point represents one finite step of either the mixing or engineered-return dynamics. The dotted diagonal is \(\Vpi=\epsilon_{\rm hist}\), and all numerical data lie on or below it. The points with \(\epsilon_{\rm hist}>0\) and \(\Vpi=0\) demonstrate directly that microscopic information can remain dynamically active without generating thermodynamically relevant return.\justifying
}
\label{fig:S6}
\end{figure}

For every sampled pair \(p(t)\rightarrow p(t+\tau)\), we evaluate independently the relative-majorization return defect \(\Vpi\) and the history-sensitive distance \(\epsilon_{\rm hist}\). Figure~\ref{fig:S6} plots the resulting values against one another. The diagonal \(\Vpi=\epsilon_{\rm hist}\) is the analytical upper bound, and every numerical point lies on or below it.

A particularly important part of the data is the set of points on the horizontal axis with
\begin{equation}
\epsilon_{\rm hist}
>
0,
\quad
\Vpi
=
0.
\end{equation}
These points directly realize the distinction central to the main text. The exact future macrodistribution depends on microscopic information absent from the present thermodynamic record, so the hidden information is dynamically active, yet its influence remains entirely compatible with the stronger \(\pi\)-preserving relative-majorization ordering.

Thus nonzero history sensitivity does not imply thermodynamic return. The analytical implication is only
\begin{equation}
\epsilon_{\rm hist}
=
0
\quad\Longrightarrow\quad
\Vpi
=
0,
\end{equation}
whereas the converse fails. Thermodynamically relevant return requires the history-sensitive contribution not merely to survive, but to become organized in a direction that violates the stronger coarse contraction.

\suppnote{Time-step convergence of local relative majorization}
\label{supp:note-majorization-dt}

The instantaneous Second-Law implication follows when the relative-majorization ordering holds for every sufficiently small positive time step. It is therefore important to verify that the ordering observed during the primary relaxation is not an artifact of choosing a particular coarse temporal resolution.

For each consecutive pair \(p(t)\) and \(p(t+\Delta t)\), we define the minimum generalized-Lorenz margin
\begin{equation}
\mathcal M_\pi(t,\Delta t)
=
\min_{x\in[0,1]}
\left[
L_{p(t)}^\pi(x)
-
L_{p(t+\Delta t)}^\pi(x)
\right].
\end{equation}
The finite-step relative-majorization condition
\begin{equation}
p(t)
\succ_\pi
p(t+\Delta t)
\end{equation}
is equivalent to
\begin{equation}
\mathcal M_\pi(t,\Delta t)
\geq
0.
\end{equation}
Because both Lorenz curves pass through \((0,0)\) and \((1,1)\), an exactly ordered pair has \(\mathcal M_\pi=0\); numerically, small negative values at floating-point scale are therefore interpreted as zero rather than as physical violations.

\begin{figure}[t]
\centering
\includegraphics[width=0.5\linewidth]{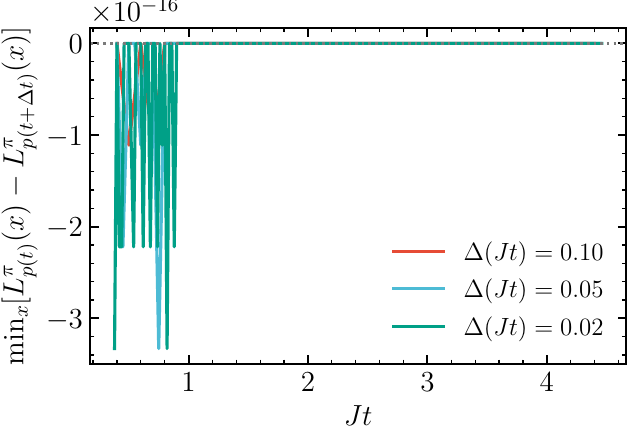}
\caption{Temporal-resolution test of local relative majorization for the exact \(N=16\), \(Q=2\) mixing trajectory. The minimum generalized-Lorenz margin \(\mathcal M_\pi(t,\Delta t)\) is shown throughout the primary \(10\%\)--\(90\%\) relaxation window for \(\Delta(Jt)=0.10\), \(0.05\), and \(0.02\). Values at the scale of negative floating-point roundoff are numerically equivalent to zero. The relative-majorization ordering persists as the observation interval is reduced to \(\Delta(Jt)=0.02\), for which all \(205\) consecutive comparisons in the primary window satisfy the ordering to numerical precision.\justifying
}
\label{fig:S7}
\end{figure}

We repeat the exact \(N=16\), \(Q=2\) mixing calculation using three temporal resolutions,
\begin{equation}
\Delta(Jt)
=
0.10,
\quad
0.05,
\quad
0.02.
\end{equation}
Figure~\ref{fig:S7} shows \(\mathcal M_\pi(t,\Delta t)\) throughout the primary \(10\%\)--\(90\%\) relaxation window. The Lorenz ordering persists, up to floating-point roundoff, as the finite observation step is successively reduced. The relative-majorization behavior observed at the coarser resolution is therefore not generated by undersampling the trajectory.

At the finest tested resolution, \(\Delta(Jt)=0.02\), the first-passage interval contains \(205\) consecutive finite-step comparisons, and every pair satisfies
\begin{equation}
p(t)
\succ_\pi
p(t+\Delta t)
\end{equation}
to numerical precision. This finite-resolution convergence test supports the local ordering over the investigated range; it does not replace the mathematical requirement of relative majorization for every sufficiently small positive \(\tau\) in the exact instantaneous statement.

\suppnote{Random-phase trace estimation for large thermodynamic classes}
\label{supp:note-trace-estimation}

For sufficiently large thermodynamic preparation classes, explicitly propagating every compatible orthogonal basis state becomes computationally expensive. We therefore evaluate the same class-level traces using random-phase trace estimation.

Let the initial thermodynamic macrospace have dimension \(D_0\), projector \(P_0\), and orthonormal basis \(\{|n\rangle\}_{n=1}^{D_0}\). For each trace vector \(a\), define
\begin{equation}
|\phi_a\rangle
=
\frac{1}{\sqrt{D_0}}
\sum_{n=1}^{D_0}
e^{i\theta_{an}}
|n\rangle,
\end{equation}
where the phases \(\theta_{an}\) are sampled independently and uniformly from \([0,2\pi)\). Averaging over the numerical phase ensemble removes the off-diagonal contributions in this basis, so for any operator \(A\),
\begin{equation}
\mathbb E_\theta
\left[
\langle\phi_a|A|\phi_a\rangle
\right]
=
\frac{1}{D_0}
\Tr
\left[
P_0A
\right].
\end{equation}
The finite-sample estimator is therefore
\begin{equation}
\frac{1}{N_{\rm tr}}
\sum_{a=1}^{N_{\rm tr}}
\langle\phi_a|A|\phi_a\rangle,
\end{equation}
which converges to the uniform trace over the initial thermodynamic macrospace as the estimator is refined.

The random phases appearing here have no physical role. They are introduced only to estimate a deterministic class trace efficiently and are not stochastic phases in the microscopic dynamics, a source of decoherence, or part of the mechanism of thermodynamic irreversibility.

For the largest dynamical free-expansion calculation,
\begin{equation}
N
=
40,
\quad
Q
=
4,
\end{equation}
the complete fixed-\(Q\) Hilbert sector has dimension
\begin{equation}
\binom{40}{4}
=
91390,
\end{equation}
whereas the initial left-half thermodynamic class contains
\begin{equation}
D_0
=
\binom{20}{4}
=
4845
\end{equation}
orthogonal fine configurations. Direct propagation of the entire class is therefore replaced by the random-phase estimator for this largest point.

\begin{figure}[t]
\centering
\includegraphics[width=0.8\linewidth]{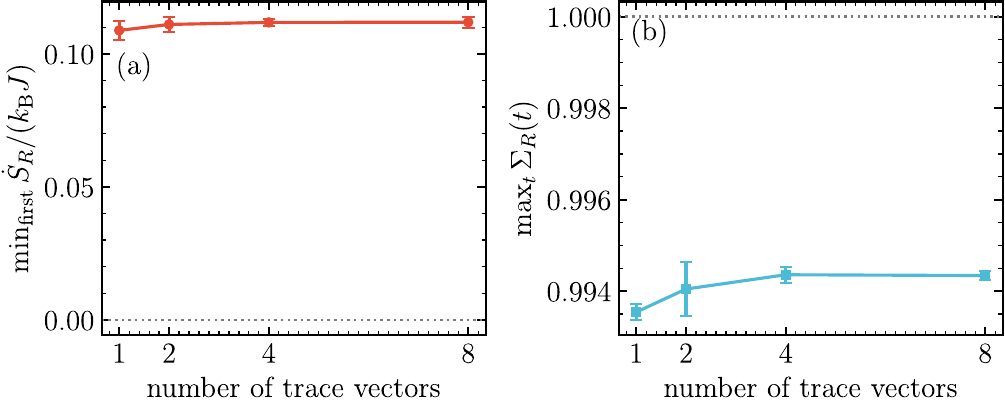}
\caption{Random-phase trace-estimator convergence for the \(N=40\), \(Q=4\) mixing calculation.
(a) Minimum sampled instantaneous accessible-entropy rate during the primary relaxation as the number \(N_{\rm tr}\) of random-phase trace vectors is increased.
(b) Maximum normalized accessible entropy reached during the simulated interval. Error bars show the spread among independent random-phase batches. The positive accessible-entropy-rate margin and the entropy trajectory remain stable as the numerical trace estimator is refined. The random phases are used only for trace estimation and are not part of the physical dynamics.\justifying
}
\label{fig:S8}
\end{figure}

Figure~\ref{fig:S8} tests convergence of the two principal large-system observables as the number \(N_{\rm tr}\) of trace vectors is increased. Multiple statistically independent random-phase batches are used to assess the residual estimator variability. The minimum sampled instantaneous accessible-entropy rate during the primary relaxation remains positive and stable under refinement, while the maximum normalized entropy reached during the simulated interval likewise converges.

\suppnote{Temporal sampling of the instantaneous entropy rate}
\label{supp:note-rate-sampling}

The instantaneous accessible-entropy rate used in the numerical tests is not obtained by finite-difference differentiation of a sampled entropy trajectory. It is evaluated directly from the Schr\"odinger dynamics. For the chosen thermodynamic record,
\begin{equation}
\dot S_R
=
-k_{\rm B}
\sum_R
\dot p_R
\ln
\left(
\frac{p_R}{\Omega_R}
\right),
\end{equation}
with \(\dot p_R\) calculated directly from the Hamiltonian action on the microscopic state. The rate at every sampled time is therefore an instantaneous quantity rather than a finite-difference estimate.

\begin{figure}[t]
\centering
\includegraphics[width=0.8\linewidth]{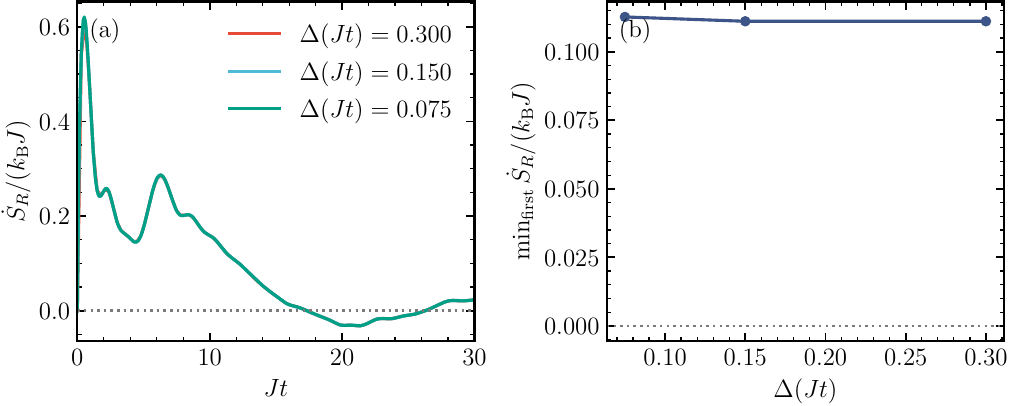}
\caption{
Temporal-grid convergence of the instantaneous accessible-entropy-rate test for the \(N=40\), \(Q=4\) mixing calculation.
(a) Directly evaluated \(\dot S_R(t)\) on progressively finer temporal grids.
(b) Minimum value of \(\dot S_R\) within the primary first-passage relaxation interval for each grid. The positive minimum remains stable under temporal refinement. This is a sampling test of the directly evaluated instantaneous rate, not a finite-difference differentiation of \(S_R(t)\).\justifying
}
\label{fig:S9}
\end{figure}

Nevertheless, the minimum value observed over a finite relaxation interval can depend on how densely the continuous trajectory is sampled. We therefore repeat the \(N=40\), \(Q=4\) calculation on the temporal grids
\begin{equation}
\Delta(Jt)
=
0.30,
\quad
0.15,
\quad
0.075.
\end{equation}
Figure~\ref{fig:S9} compares the directly evaluated rate trajectories and the corresponding minimum values within the primary first-passage window. The positive minimum remains stable as the temporal grid is refined.

This check should therefore be distinguished from a finite-difference convergence test. The expression for \(\dot S_R\) is evaluated directly at every time point; refining the grid tests only whether temporal sampling has missed a lower value of that already instantaneous rate between adjacent sampled times.

\suppnote{Exact combinatorial free-expansion entropy}
\label{supp:note-free-expansion}

\suppsubsection{Finite lattice}
\label{supp:finite-lattice}

Consider \(Q\) hard-core excitations thermodynamically equilibrated within an initial region containing \(L_i\) available lattice sites. The counting is the finite-lattice form of the Boltzmann--Gibbs multiplicity construction~\cite{boltzmann1877beziehung,gibbs1902elementary}. The number of compatible microscopic configurations is
\begin{equation}
\Omega_i
=
\binom{L_i}{Q},
\end{equation}
so the corresponding accessible entropy is \(S_i=k_{\rm B}\ln\Omega_i\). After expansion into a larger region containing \(L_f>L_i\) available sites, complete multiplicity-defined spreading gives
\begin{equation}
\Omega_f
=
\binom{L_f}{Q},
\end{equation}
and \(S_f=k_{\rm B}\ln\Omega_f\). The exact finite-lattice entropy difference is therefore
\begin{equation}
\Delta S_R
=
k_{\rm B}
\ln
\left[
\frac{
\binom{L_f}{Q}
}{
\binom{L_i}{Q}
}
\right].
\end{equation}
This result is purely combinatorial. It follows from the thermodynamic multiplicities before and after the expansion and does not require ETH, diffusion, Markovianity, a dynamical equilibration approximation, or any assumption concerning the detailed route by which the final thermodynamic distribution is reached.

\suppsubsection{Dilute expansion}
\label{supp:dilute-expansion}

To connect the exact finite-lattice result with the classical free-expansion entropy, write
\begin{equation}
\binom{L}{Q}
=
\frac{
L(L-1)\cdots(L-Q+1)
}{
Q!
}.
\end{equation}
Taking the logarithm gives
\begin{equation}
\ln
\binom{L}{Q}
=
Q\ln L
-
\ln Q!
+
\sum_{m=0}^{Q-1}
\ln
\left(
1-\frac{m}{L}
\right).
\end{equation}
For \(Q/L\ll1\),
\begin{equation}
\ln
\left(
1-\frac{m}{L}
\right)
=
-\frac{m}{L}
-
\frac{m^2}{2L^2}
+
\mathcal O
\left(
\frac{m^3}{L^3}
\right).
\end{equation}
Using
\begin{equation}
\sum_{m=0}^{Q-1}
m
=
\frac{Q(Q-1)}{2},
\end{equation}
the multiplicity becomes
\begin{equation}
\ln
\binom{L}{Q}
=
Q\ln L
-
\ln Q!
-
\frac{
Q(Q-1)
}{
2L
}
+
\mathcal O
\left(
\frac{Q^3}{L^2}
\right).
\end{equation}
Taking the difference between the final and initial volumes gives
\begin{equation}
\begin{aligned}
\frac{\Delta S_R}{k_{\rm B}}
={}&
Q
\ln
\left(
\frac{L_f}{L_i}
\right)
\\
&
+
\frac{
Q(Q-1)
}{
2
}
\left(
\frac{1}{L_i}
-
\frac{1}{L_f}
\right)
\\
&
+
\mathcal O
\left(
\frac{Q^3}{L_i^2}
\right),
\end{aligned}
\end{equation}
where the final-to-initial expansion ratio is held fixed. The leading finite-density correction is positive for \(L_f>L_i\), so the hard-core result approaches the dilute classical value from above.

\begin{figure}[t]
\centering
\includegraphics[width=0.5\linewidth]{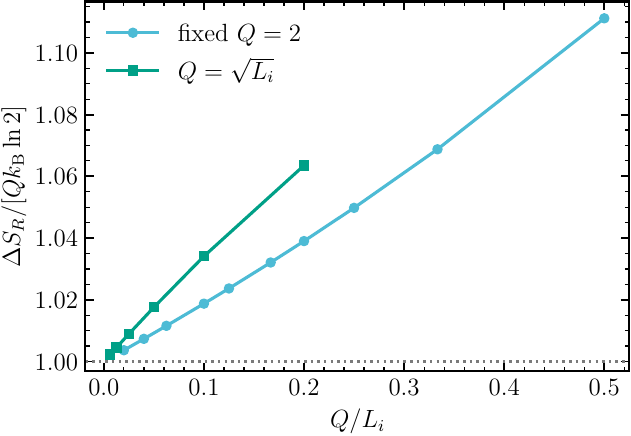}
\caption{
Exact combinatorial approach to the dilute free-expansion limit. Shown is \(\Delta S_R/[Qk_{\rm B}\ln2]\) versus the initial filling \(Q/L_i\) for a fixed-\(Q=2\) dilute sequence and for the thermodynamic sequence \(L_i=M^2\), \(Q=M\), for which \(Q\rightarrow\infty\) while \(Q/L_i\rightarrow0\). Both sequences converge to the classical volume-doubling value \(1\).\justifying
}
\label{fig:S10}
\end{figure}

Now take the simultaneous dilute thermodynamic limit
\begin{equation}
Q
\rightarrow
\infty,
\quad
L_i
\rightarrow
\infty,
\quad
\frac{Q}{L_i}
\rightarrow
0,
\end{equation}
at fixed \(L_f/L_i\). Dividing by \(Qk_{\rm B}\) gives
\begin{equation}
\frac{
\Delta S_R
}{
Qk_{\rm B}
}
\longrightarrow
\ln
\left(
\frac{L_f}{L_i}
\right).
\end{equation}
Since the physical volume is proportional to the number of available sites in the dilute lattice limit, the equivalent extensive asymptotic statement is
\begin{equation}
\Delta S_R
=
Qk_{\rm B}
\ln
\left(
\frac{V_f}{V_i}
\right)
+
o(Q).
\end{equation}
For volume doubling, \(L_f=2L_i\), this reduces to
\begin{equation}
\Delta S_R
=
Qk_{\rm B}\ln2
+
o(Q).
\end{equation}
Figure~\ref{fig:S10} displays the exact finite-lattice convergence using both a fixed-\(Q\) dilute sequence and a genuine thermodynamic sequence in which \(Q\) itself increases while the filling \(Q/L_i\) tends to zero.

\suppsubsection{Bosonic counting and the common classical limit}
\label{supp:bosonic-counting}

The same dilute classical endpoint follows from a microscopically different quantum statistics. For \(Q\) indistinguishable bosons distributed over \(L\) lattice sites with unrestricted on-site occupation, the number of fixed-particle-number Fock configurations is
\begin{equation}
\Omega^{\rm B}(L,Q)
=
\binom{L+Q-1}{Q}.
\end{equation}
The exact bosonic free-expansion entropy is therefore
\begin{equation}
\Delta S_R^{\rm B}
=
k_{\rm B}
\ln
\left[
\frac{
\binom{L_f+Q-1}{Q}
}{
\binom{L_i+Q-1}{Q}
}
\right].
\end{equation}
At low filling,
\begin{equation}
\ln
\binom{L+Q-1}{Q}
=
Q\ln L
-
\ln Q!
+
\frac{Q(Q-1)}{2L}
+
\mathcal O
\left(
\frac{Q^3}{L^2}
\right),
\end{equation}
and hence
\begin{equation}
\frac{\Delta S_R^{\rm B}}{k_{\rm B}}
=
Q
\ln
\left(
\frac{L_f}{L_i}
\right)
-
\frac{Q(Q-1)}{2}
\left(
\frac{1}{L_i}
-
\frac{1}{L_f}
\right)
+
\mathcal O
\left(
\frac{Q^3}{L_i^2}
\right).
\end{equation}
The leading finite-density correction therefore has the opposite sign from the hard-core result. For \(L_f>L_i\), hard-core counting approaches the classical entropy from above, whereas bosonic counting approaches it from below. Nevertheless, both statistics satisfy
\begin{equation}
\frac{\Delta S_R}{Qk_{\rm B}}
\longrightarrow
\ln
\left(
\frac{L_f}{L_i}
\right)
\end{equation}
when \(Q/L_i\rightarrow0\) at fixed expansion ratio. The common classical limit is consequently independent of these opposite finite-density quantum-statistical corrections. Main-text Fig.~\ref{fig:statistics-limit} displays this two-sided convergence for volume doubling and for several general expansion ratios.

\suppsubsection{General expansion ratio}
\label{supp:volume-ratio}

\begin{figure}[t]
\centering
\includegraphics[width=0.5\linewidth]{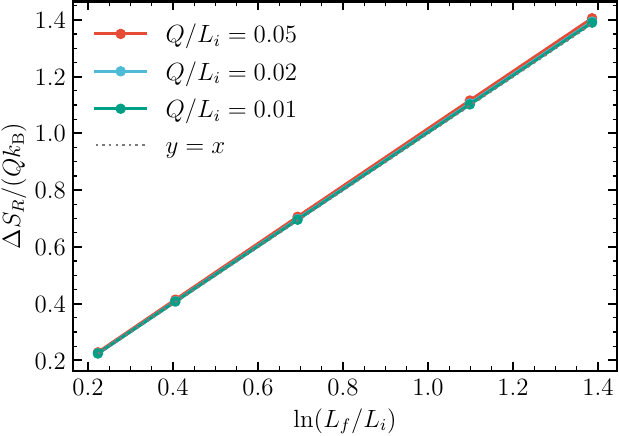}
\caption{
General volume-ratio test of the exact hard-core combinatorial entropy. The quantity \(\Delta S_R/(Qk_{\rm B})\) is plotted against \(\ln(L_f/L_i)\) for several small initial fillings. As the filling decreases, the exact finite-lattice data approach the classical line \(y=x\), corresponding in the dilute lattice limit to \(\Delta S_R=Qk_{\rm B}\ln(V_f/V_i)\).\justifying
}
\label{fig:S11}
\end{figure}

The dilute-limit result is not specific to doubling the available volume. Define the expansion ratio
\begin{equation}
\lambda
=
\frac{L_f}{L_i}.
\end{equation}
For the hard-core lattice, the exact finite-size entropy per particle is
\begin{equation}
\frac{
\Delta S_R
}{
Qk_{\rm B}
}
=
\frac{1}{Q}
\ln
\left[
\frac{
\binom{\lambda L_i}{Q}
}{
\binom{L_i}{Q}
}
\right].
\end{equation}
In the dilute limit,
\begin{equation}
\frac{
\Delta S_R
}{
Qk_{\rm B}
}
\longrightarrow
\ln\lambda
=
\ln
\left(
\frac{L_f}{L_i}
\right).
\end{equation}
Identifying the ratio of available lattice sites with the physical volume ratio in the dilute lattice limit gives
\begin{equation}
\Delta S_R
=
Qk_{\rm B}
\ln
\left(
\frac{V_f}{V_i}
\right)
+
o(Q).
\end{equation}
Figure~\ref{fig:S11} tests several expansion ratios and initial fillings directly. As the filling is reduced, the exact finite-lattice data approach the classical line \(\Delta S_R/(Qk_{\rm B})=\ln(L_f/L_i)\), confirming that the volume-doubling result is one member of the more general free-expansion relation.

\suppnote{Simultaneous increase of particle number and dilution}
\label{supp:note-simultaneous-dilution}

The exact combinatorial result establishes the dilute free-expansion endpoint independently of the dynamics. To test whether the closed-system relaxation remains entropy increasing while the particle number grows and the filling simultaneously decreases, we consider the sequence
\begin{equation}
(N,Q)
=
(8,2),
\quad
(20,3),
\quad
(40,4).
\end{equation}
Since the initially available region is the left half, \(L_i=N/2\), the corresponding initial fillings are
\begin{equation}
\rho_i
=
\frac{Q}{L_i}
=
\frac{2Q}{N}
=
0.50,
\quad
0.30,
\quad
0.20.
\end{equation}
Thus both the number of mobile excitations and the number of available microscopic configurations increase along the sequence, while the initial density moves systematically toward the dilute regime.

For volume doubling, the exact hard-core endpoint is
\begin{equation}
\Delta S_{\max}
=
k_{\rm B}
\ln
\left[
\frac{
\binom{N}{Q}
}{
\binom{N/2}{Q}
}
\right].
\end{equation}
The corresponding ratios to the classical dilute value are
\begin{equation}
\frac{
\Delta S_{\max}
}{
Qk_{\rm B}\ln2
}
=
1.1112,
\quad
1.0826,
\quad
1.0594.
\end{equation}
The exact finite-lattice endpoints therefore approach \(Qk_{\rm B}\ln2\) monotonically along this simultaneous increase of particle number and dilution.

The dynamical calculation uses the same interacting mixing Hamiltonian and four-cell thermodynamic record as in the main text. During the first \(10\%\)--\(90\%\) relaxation interval, the minimum sampled instantaneous accessible-entropy rates are
\begin{equation}
\frac{
\min_{\rm first}\dot S_R
}{
k_{\rm B}J
}
\simeq
0.278,
\quad
0.160,
\quad
0.113,
\end{equation}
for \((N,Q)=(8,2)\), \((20,3)\), and \((40,4)\), respectively. No negative sampled rate occurs within the primary relaxation window for any of the three systems.

The first two thermodynamic-class traces are evaluated exactly. For \(N=40\), \(Q=4\), the class trace is evaluated with the random-phase estimator described in Supplementary Note~\ref{supp:note-trace-estimation}. An independent random-phase batch and the denser temporal sampling described in Supplementary Note~\ref{supp:note-rate-sampling} both preserve a clearly positive primary minimum.

The decrease of the raw minimum rate with increasing \(N\) should not be interpreted as an asymptotic thermodynamic-limit rate. The microscopic spreading time also changes with system size, and no hydrodynamic time rescaling is imposed here. The robust dynamical statement supported by this sequence is the positive sign of the sampled instantaneous rate throughout the primary relaxation, while the exact combinatorial endpoint approaches the classical dilute value.

Table~\ref{tab:S1} collects the Hilbert-space dimensions, initial thermodynamic-class sizes, fillings, and exact endpoint ratios.
\begin{table}[t]
\caption{
Parameters of the simultaneous dilution sequence. \(D_{N,Q}=\binom{N}{Q}\) is the dimension of the complete fixed-\(Q\) Hilbert sector, and \(D_0=\binom{N/2}{Q}\) is the number of orthogonal fine product configurations compatible with the initial left-half thermodynamic preparation.\justifying}
\label{tab:S1}
\centering
\begin{tabular}{c c c c c c}
\hline\hline
\(N\) &
\(Q\) &
\(Q/L_i\) &
\(D_{N,Q}\) &
\(D_0\) &
\(\Delta S_{\max}/(Qk_{\rm B}\ln2)\)
\\
\hline
8  & 2 & 0.50 & 28    & 6    & 1.1112 \\
20 & 3 & 0.30 & 1140  & 120  & 1.0826 \\
40 & 4 & 0.20 & 91390 & 4845 & 1.0594 \\
\hline\hline
\end{tabular}
\end{table}

\suppnote{Relative-majorization diagnostics for the exact primary relaxation}
\label{supp:note-majorization-diagnostics}

We next collect the finite-step relative-majorization diagnostics for the exact four-cell thermodynamic-class trajectories. The primary relaxation interval is defined independently for each system by the first passage of the normalized accessible entropy from \(0.1\) to \(0.9\).

At temporal resolution
\begin{equation}
\Delta(Jt)
=
0.10,
\end{equation}
every tested consecutive pair satisfies
\begin{equation}
p(t)
\succ_\pi
p(t+\Delta t)
\end{equation}
throughout the primary intervals for
\begin{equation}
(N,Q)
=
(8,2),
\quad
(16,2),
\quad
(20,3).
\end{equation}
Thus the complete accessible distributions, rather than only their scalar entropies, follow the stronger \(\pi\)-preserving ordering at every tested primary step for these exact class trajectories.

For the representative \(N=16\), \(Q=2\) system, the calculation is additionally repeated at the finer resolution
\begin{equation}
\Delta(Jt)
=
0.02.
\end{equation}
The resulting primary interval contains \(205\) consecutive finite-step comparisons, all of which satisfy the relative-majorization ordering to floating-point precision. The temporal-resolution analysis is developed further in Supplementary Note~\ref{supp:note-majorization-dt}.

Table~\ref{tab:S2} summarizes the primary intervals and pass fractions.
\begin{table}[t]
\caption{
Exact relative-majorization tests during the primary free-expansion relaxation. The pass fraction is the fraction of consecutive finite steps satisfying \(p(t)\succ_\pi p(t+\Delta t)\). The final row gives the denser temporal-resolution test for the representative \(N=16\), \(Q=2\) trajectory.\justifying}
\label{tab:S2}
\centering
\begin{tabular}{c c c c c}
\hline\hline
\(N\) &
\(Q\) &
\(\Delta(Jt)\) &
primary interval \(Jt\) &
pass fraction
\\
\hline
8  & 2 & 0.10 & \(0.30\)--\(2.30\) & 1.000 \\
16 & 2 & 0.10 & \(0.40\)--\(4.50\) & 1.000 \\
20 & 3 & 0.10 & \(0.50\)--\(5.90\) & 1.000 \\
16 & 2 & 0.02 & \(0.38\)--\(4.48\) & 1.000 \\
\hline\hline
\end{tabular}
\end{table}

The same intervals retain measurable sensitivity to microscopic information absent from the current thermodynamic record. For \(\tau=0.1/J\), the representative maximum history-sensitive distances are
\begin{equation}
\max\epsilon_{\rm hist}
\simeq
3.94\times10^{-2},
\quad
1.98\times10^{-2},
\quad
2.35\times10^{-2},
\end{equation}
for \((N,Q)=(8,2)\), \((16,2)\), and \((20,3)\), respectively. Nevertheless, in the same primary windows,
\begin{equation}
\max\Vpi
\simeq
0
\end{equation}
to numerical precision. These exact class calculations therefore realize the regime \(\epsilon_{\rm hist}>0\) with \(\Vpi=0\): microscopic information excluded from the present record continues to modify future macroprobabilities without producing a detectable violation of the stronger \(\pi\)-preserving ordering.

The later finite-size evolution of the \(N=16\), \(Q=2\) trajectory further separates the onset of the different return diagnostics. The relative-majorization defect first becomes measurably nonzero near
\begin{equation}
Jt
\simeq
5.9,
\end{equation}
after the primary relaxation. The history-sensitive entropy correction becomes negative near
\begin{equation}
Jt
\simeq
6.4,
\end{equation}
and an actual negative finite-step entropy change appears at approximately the same time. Along this particular trajectory, the observed onset sequence is therefore
\begin{equation}
\Vpi>0
\quad\longrightarrow\quad
\Delta_\tau S_{\rm corr}<0
\quad\longrightarrow\quad
\Delta_\tau S_R<0.
\end{equation}
These arrows record the ordering of the observed onset times and are not universal logical implications. In particular, \(\Vpi>0\) need not make \(\Delta_\tau S_{\rm corr}\) negative, and a negative correction need not overcome the nonnegative mixing contribution sufficiently to make the total entropy decrease.

\suppnote{Intermediate integrable control}
\label{supp:note-integrable-control}

To test whether the distinction between mixing and organized return can be reduced to a simple integrable-versus-nonintegrable classification, we introduce the uniform free nearest-neighbor chain as an intermediate control between the interacting mixing Hamiltonian and the engineered mirror-transfer Hamiltonian.

The free control is
\begin{equation}
H_{\rm free}
=
-J
\sum_{j=1}^{N-1}
\left(
\sigma_j^+\sigma_{j+1}^-
+
\sigma_j^-\sigma_{j+1}^+
\right).
\end{equation}
It is integrable and preserves the same fixed-\(Q\) sector and chain topology. Its normal-mode frequencies, however, are not deliberately synchronized in the manner imposed by \(H_{\rm ret}\). The microscopic evolution can therefore exhibit substantial spreading and dephasing even though the model is integrable.

\begin{figure}[t]
\centering
\includegraphics[width=0.8\linewidth]{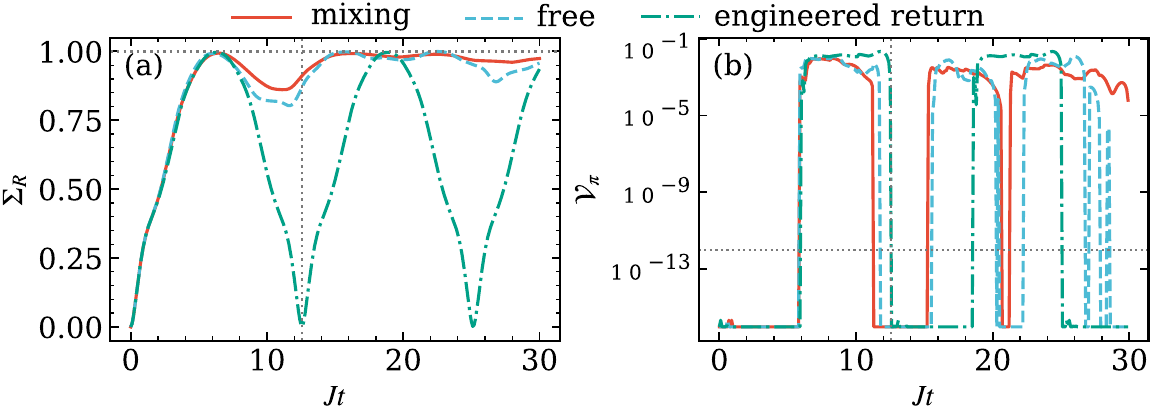}
\caption{Matched control family for \(N=16\), \(Q=2\), using the same fixed-\(Q\) Hilbert sector, left-half thermodynamic preparation, and four-cell record.
(a) Normalized accessible entropy for the interacting mixing chain, uniform free nearest-neighbor chain, and engineered mirror-transfer chain. The vertical line marks the mirror time of the engineered model.
(b) Corresponding relative-majorization return defect. The free model is integrable but can still undergo substantial internal dephasing without the synchronized macroscopic return generated by the engineered chain. Integrability alone is therefore not the relevant classification for thermodynamic return.\justifying
}
\label{fig:S12}
\end{figure}

Figure~\ref{fig:S12} compares the interacting mixing model, the free integrable chain, and the engineered mirror-transfer chain using the same \(N=16\), \(Q=2\) Hilbert sector, left-half thermodynamic preparation, and four-cell record. The comparison is not intended to classify all integrable systems. Its purpose is narrower: integrability alone is neither equivalent to synchronized thermodynamic return nor sufficient to explain the contrast between the mixing and engineered-return dynamics.

\suppnote{Microscopic recurrence fidelity}
\label{supp:note-microscopic-fidelity}

Complete recurrence of the microscopic quantum state is a stronger requirement than thermodynamic return. Spin-echo experiments provide the canonical physical example of macroscopic rephasing without microscopic information destruction~\cite{hahn1950echo,ridderbos1998spin,anastopoulos2011spin}. For a pure fine initial state \(|\psi_0\rangle\), we define the microscopic recurrence fidelity
\begin{equation}
F(t)
=
\left|
\langle
\psi_0
|
e^{-iHt}
|
\psi_0
\rangle
\right|^2.
\end{equation}
A full microscopic recurrence, \(F(t)=1\), restores the initial pure state up to a global phase and therefore restores every observable and every thermodynamic record. The converse is not true: a chosen thermodynamic macrostate can return even when the exact wavefunction is different from its initial state.

The engineered mirror-transfer Hamiltonian provides an explicit example. At \(t=t_{\rm mir}\), a fine occupation configuration initially confined to the left half is transferred, up to a phase, to its spatially reflected configuration in the right half. The left- and right-confined thermodynamic sectors have equal multiplicity, so the accessible entropy can return to its initial value even though the final microscopic state is not the original left-confined state and its fidelity with that state can remain small.

\begin{figure}[t]
\centering
\includegraphics[width=0.5\linewidth]{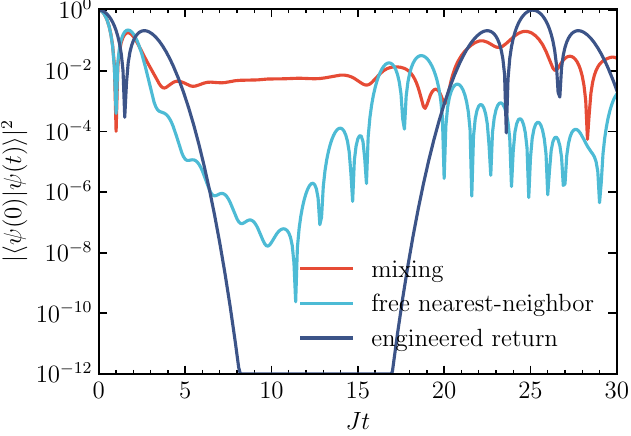}
\caption{Microscopic recurrence fidelity for representative \(N=16\), \(Q=2\) fine initial states under the mixing, free nearest-neighbor, and engineered-return Hamiltonians. Complete microscopic recurrence is stronger than thermodynamic return: the accessible thermodynamic state can return even when the exact wavefunction does not coincide with its initial state. The fidelity is therefore not used as the primary irreversibility criterion.\justifying
}
\label{fig:S13}
\end{figure}

Figure~\ref{fig:S13} compares representative microscopic fidelities for the interacting mixing, free nearest-neighbor, and engineered-return controls. The fidelity is included only to distinguish microscopic state recurrence from thermodynamic return; it is not used as the irreversibility diagnostic anywhere in the analysis.

\suppnote{Generality across microscopic realizations}
\label{supp:note-generality}

In the quantum setting, the exact coarse-propagation theorem requires only a unitary map and a projective thermodynamic partition; its finite-measure classical counterpart is proved in Supplementary Note~\ref{supp:note-projection-classical}. The additional dynamical question is whether the complete accessible trajectory lies in the strong relative-majorization sector during its primary relaxation. To test whether the behavior found in the baseline hard-core chain is tied to a particular microscopic realization, we apply the same thermodynamic-class construction and the same return diagnostics to the Bose--Hubbard, two-dimensional hard-core, and Floquet systems defined in Supplementary Note~\ref{supp:note-models}.

For the one-dimensional interacting hard-core chain with \((N,Q)=(16,2)\), the primary first-passage window is
\begin{equation}
Jt
=
0.4\text{--}4.5,
\end{equation}
with
\begin{equation}
\frac{
\min_{\rm first}\dot S_R
}{
k_{\rm B}J
}
=
0.143119,
\quad
\max_{\rm first}\Vpi
=
2.22\times10^{-16},
\quad
\max_{\rm first}\epsilon_{\rm hist}
=
1.9786\times10^{-2}.
\end{equation}
For the interacting Bose--Hubbard chain with
\begin{equation}
(L,Q,U/J)
=
(8,3,1.3),
\end{equation}
the primary window is
\begin{equation}
Jt
=
0.3\text{--}2.6,
\end{equation}
and
\begin{equation}
\frac{
\min_{\rm first}\dot S_R
}{
k_{\rm B}J
}
=
0.230861,
\quad
\max_{\rm first}\Vpi
=
0,
\quad
\max_{\rm first}\epsilon_{\rm hist}
=
4.0956\times10^{-2}
\end{equation}
at numerical resolution.

For the \(4\times4\) interacting hard-core lattice with \(Q=3\) and \(V=J\), the primary window is
\begin{equation}
Jt
=
0.2\text{--}1.2,
\end{equation}
with
\begin{equation}
\frac{
\min_{\rm first}\dot S_R
}{
k_{\rm B}J
}
=
0.592311,
\quad
\max_{\rm first}\Vpi
=
0,
\quad
\max_{\rm first}\epsilon_{\rm hist}
=
7.7845\times10^{-2}
\end{equation}
at numerical resolution. The relative-majorization pass fraction is unity throughout all three primary windows.

These values exhibit the same separation across different particle statistics, local Hilbert structures, and spatial connectivity: the future thermodynamic distribution retains a measurable dependence on microscopic information absent from the current record, while that influence produces no detectable violation of the stronger relative-majorization order during the tested primary relaxation.

\begin{figure}[t]
\centering
\includegraphics[width=0.8\linewidth]{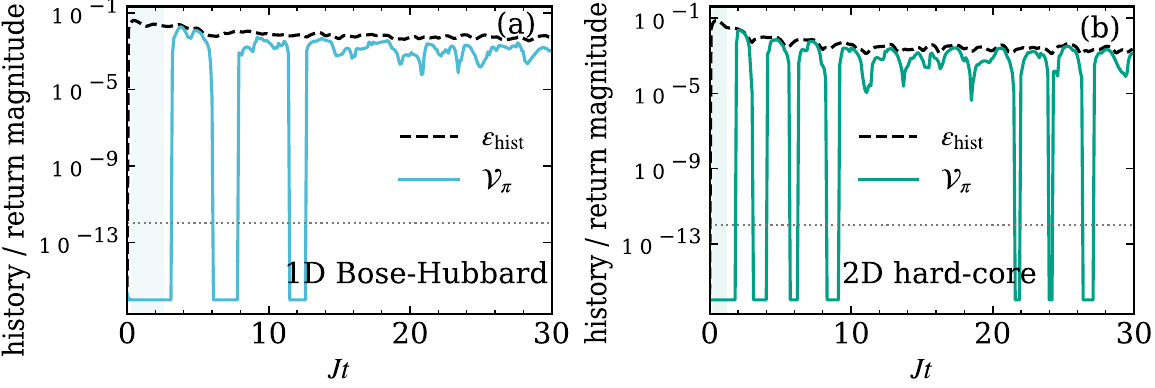}
\caption{
History activity and thermodynamic return in two microscopically distinct closed systems.
(a) Interacting Bose--Hubbard chain.
(b) Interacting two-dimensional hard-core lattice.
The dashed black curve is the history-sensitive distance \(\epsilon_{\rm hist}\), the solid curve is the relative-majorization return defect \(\Vpi\), and shading marks the primary \(10\%\)--\(90\%\) relaxation interval. In both systems, microscopic information absent from the current record remains dynamically active while the stronger return defect is zero at the tested numerical resolution. Later growth of \(\Vpi\) reflects finite-size return outside the primary relaxation window.\justifying
}
\label{fig:S14-generality-history}
\end{figure}

Figure~\ref{fig:S14-generality-history} isolates the history-sensitive and return-defect curves for the Bose--Hubbard and two-dimensional hard-core realizations. In both cases, \(\epsilon_{\rm hist}\) remains clearly nonzero in the shaded primary interval while \(\Vpi\) stays at the numerical floor. The later growth of the return defect is a finite-size return effect and is not used as evidence for the primary irreversible regime.

\suppsubsection{Floquet generality}
\label{supp:floquet-generality}

The structural result does not require a time-independent Hamiltonian. For the number-conserving Floquet realization defined in Supplementary Note~\ref{supp:floquet-model}, the thermodynamic evolution is sampled stroboscopically once per Floquet period.

The normalized accessible entropy rapidly enters its high-entropy region. During the corresponding primary first-passage interval,
\begin{equation}
\max_{\rm first}\Vpi
=
2.22\times10^{-16},
\end{equation}
at numerical resolution, while
\begin{equation}
\max_{\rm first}\epsilon_{\rm hist}
=
1.1740\times10^{-1}.
\end{equation}
The history-sensitive modification is therefore larger than in the representative static models while the tested return defect remains at numerical zero. This case makes the logical distinction particularly transparent: substantial dependence of future macroprobabilities on inaccessible microscopic information can coexist with the strong \(\pi\)-preserving ordering.

\begin{figure}[t]
\centering
\includegraphics[width=0.8\linewidth]{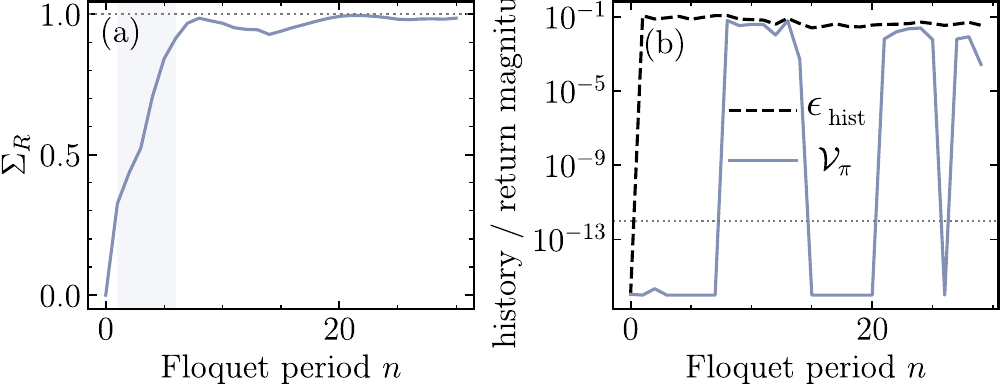}
\caption{
Closed number-conserving Floquet realization.
(a) Normalized accessible entropy versus Floquet period. The shaded region marks the primary first-passage interval.
(b) History-sensitive distance \(\epsilon_{\rm hist}\) and relative-majorization return defect \(\Vpi\) over the same stroboscopic evolution. The primary defect remains at numerical zero while the hidden-history influence is appreciable, demonstrating the same separation beyond dynamics generated by a time-independent Hamiltonian.\justifying
}
\label{fig:S15-floquet}
\end{figure}

Figure~\ref{fig:S15-floquet} shows the normalized entropy together with the two finite-step diagnostics over the same stroboscopic evolution. The Floquet calculation demonstrates that the distinction between history sensitivity and organized return is not tied to evolution generated by a single static Hamiltonian.

\suppnote{Hamiltonian-parameter robustness and the stronger relative-majorization sector}
\label{supp:note-strong-order}

The representative interaction point \((V_1/J,V_2/J)=(1,0.7)\) is not selected from an isolated fine-tuned region. We scan the \(9\times7\) parameter grid
\begin{equation}
\frac{V_1}{J}
\in
\{0,0.25,\ldots,2\},
\quad
\frac{V_2}{J}
\in
\{0,0.25,0.5,0.7,0.9,1.2,1.5\},
\end{equation}
using \(N=12\), \(Q=3\), and finite-step resolution
\begin{equation}
\Delta(Jt)
=
0.15.
\end{equation}
At every parameter point, the primary window is determined independently from the first \(10\%\)--\(90\%\) passage of the normalized accessible entropy.

\begin{figure}[t]
\centering
\includegraphics[width=0.8\linewidth]{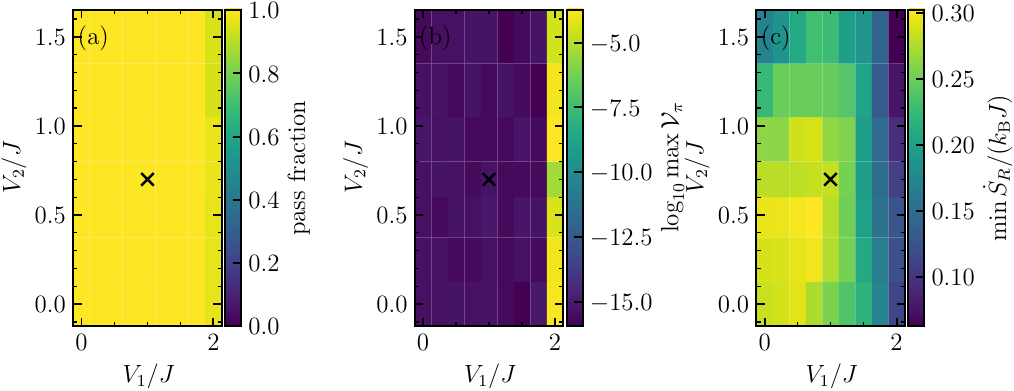}
\caption{
Interaction-parameter robustness for the \(N=12\), \(Q=3\) hard-core chain.
(a) Fraction of primary finite steps satisfying the numerical criterion \(\Vpi\leq10^{-11}\).
(b) Maximum primary relative-majorization return defect.
(c) Minimum sampled value of the exact instantaneous accessible-entropy rate within the primary window.
The cross marks the baseline interaction point \((V_1/J,V_2/J)=(1,0.7)\). Small violations of the stronger relative-majorization ordering appear only on the sampled \(V_1/J=2\) boundary, whereas the accessible-entropy rate remains positive throughout the complete parameter grid.\justifying
}
\label{fig:S16-parameter-scan}
\end{figure}

Figure~\ref{fig:S16-parameter-scan} shows the relative-majorization pass fraction, the maximum primary return defect, and the minimum sampled value of the exact instantaneous accessible-entropy rate across the complete scan. Of the \(63\) sampled parameter points, all \(56\) points with \(V_1/J<2\) have pass fraction one throughout the primary window. Small Lorenz-order violations occur only at the seven sampled points on the boundary \(V_1/J=2\).

Across the complete grid, the median pass fraction remains one, the largest primary return defect is
\begin{equation}
\max\Vpi
=
1.81\times10^{-4},
\end{equation}
and the minimum sampled primary accessible-entropy rate remains positive,
\begin{equation}
\min
\left[
\frac{\dot S_R}{k_{\rm B}J}
\right]_{\rm first}
=
0.062874.
\end{equation}
The baseline point \((1,0.7)\) therefore lies well inside the broad region displaying the stronger finite-step ordering rather than on its sampled boundary.

Equally importantly, the boundary points establish that relative majorization is sufficient but not necessary for a positive accessible-entropy rate. Small violations of the generalized Lorenz ordering coexist with a strictly positive instantaneous accessible-entropy rate.

\suppsubsection{Boundary time-step refinement}
\label{supp:boundary-refinement}

To test whether the small defects observed on the \(V_1/J=2\) boundary are merely artifacts of the original finite observation step, we repeat all seven boundary points using
\begin{equation}
J\tau
=
0.15,
\quad
0.075,
\quad
0.0375,
\end{equation}
without changing the Hamiltonian, system size, thermodynamic record, or numerical majorization tolerance.

The absolute maximum defects generally decrease as the finite step is refined, while the minimum sampled primary accessible-entropy rates remain positive for every tested boundary point and resolution. The pass fraction need not increase monotonically with decreasing \(\tau\), because reducing the time step changes both the number and the physical locations of the consecutive state pairs included in the primary window.

\begin{figure}[t]
\centering
\includegraphics[width=0.8\linewidth]{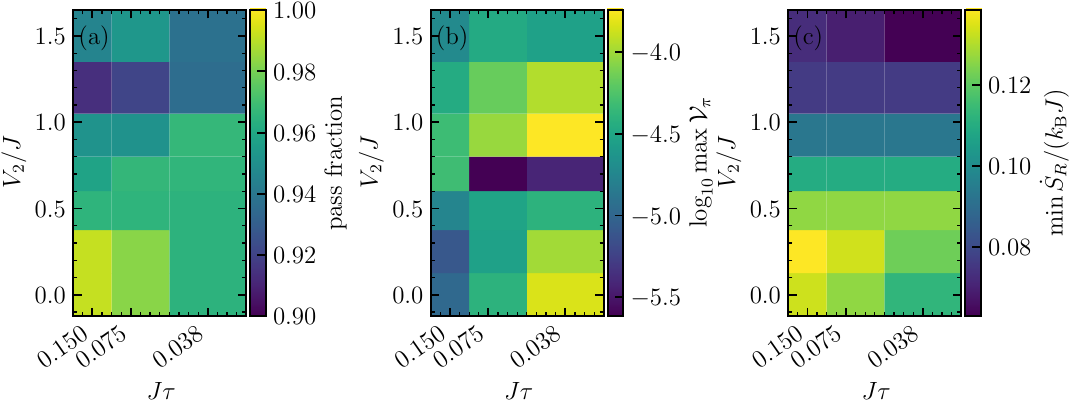}
\caption{
Time-step refinement along the sampled strong-coupling boundary \(V_1/J=2\).
(a) Relative-majorization pass fraction.
(b) Maximum primary return defect.
(c) Minimum sampled value of the exact instantaneous accessible-entropy rate.
The absolute defect becomes small as \(J\tau\) is reduced, while the accessible-entropy rate remains positive throughout all tested boundary points. Vanishing absolute defect under finite-step refinement is not by itself evidence for restoration of exact local relative majorization.\justifying
}
\label{fig:S17-boundary}
\end{figure}

Figure~\ref{fig:S17-boundary} displays the pass fraction, maximum defect, and minimum sampled accessible-entropy rate over the complete refined boundary. The decreasing absolute defect under refinement is useful numerically, but by itself it does not establish exact local relative majorization: continuity already guarantees that \(p(t+\tau)\rightarrow p(t)\) and hence that any finite-step defect must vanish in absolute magnitude as \(\tau\rightarrow0\).

\suppsubsection{Fixed-time local defect-rate test}
\label{supp:local-defect-rate}

Because
\begin{equation}
\Vpi
\left[
p(t)\rightarrow p(t+\tau)
\right]
\longrightarrow
0
\end{equation}
for any continuous trajectory as \(\tau\rightarrow0^+\), the absolute finite-step defect alone cannot determine the local strength of a Lorenz-order violation. We therefore introduce the finite-step defect-rate proxy
\begin{equation}
\nu_\pi(t,\tau)
=
\frac{
\Vpi[p(t)\rightarrow p(t+\tau)]
}{
\tau
},
\end{equation}
and evaluate it at a fixed physical reference time \(t_{\rm ref}\) while varying \(\tau\). Holding \(t_{\rm ref}\) fixed avoids comparing defects evaluated at different points of the trajectory.

For \(V_1/J=2\), \(V_2/J=0.9\), and
\begin{equation}
Jt_{\rm ref}
=
4.8,
\end{equation}
the dimensionless defect-rate proxy \(\nu_\pi/J\) takes the values
\begin{equation}
1.207\times10^{-3},
\quad
1.242\times10^{-3},
\quad
1.248\times10^{-3},
\quad
1.248\times10^{-3},
\quad
1.247\times10^{-3},
\end{equation}
for
\begin{equation}
J\tau
=
0.15,
\quad
0.075,
\quad
0.0375,
\quad
0.01875,
\quad
0.009375,
\end{equation}
respectively. Over the same refinement, the finite-step entropy slope approaches the exact instantaneous value
\begin{equation}
\frac{
\dot S_R
}{
k_{\rm B}J
}
=
0.106688
>
0.
\end{equation}
For \(V_2/J=1.2\), the defect-rate proxy approaches approximately
\begin{equation}
\frac{\nu_\pi}{J}
\simeq
8.60\times10^{-4},
\end{equation}
over the tested refinement range, while the exact instantaneous accessible-entropy rate at the corresponding fixed reference time is
\begin{equation}
\frac{
\dot S_R
}{
k_{\rm B}J
}
=
0.104180
>
0.
\end{equation}
The \(V_2/J=0.7\) case has a much smaller defect-rate scale, with \(\nu_\pi/J\) of order \(10^{-5}\), while again retaining a clearly positive instantaneous accessible-entropy rate.

\begin{figure}[t]
\centering
\includegraphics[width=0.8\linewidth]{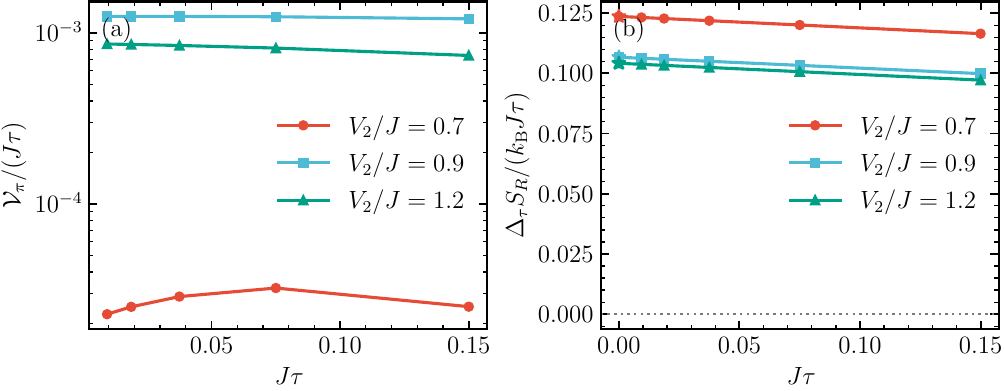}
\caption{
Fixed-time local refinement on the \(V_1/J=2\) boundary.
(a) Finite-step relative-majorization defect-rate proxy \(\Vpi/(J\tau)\) for three values of \(V_2/J\), evaluated while holding the physical reference time fixed.
(b) Corresponding finite-step entropy slope \(\Delta_\tau S_R/(k_{\rm B}J\tau)\). Stars at \(J\tau=0\) denote the exact instantaneous accessible-entropy rate at the same physical reference time. The selected strong-coupling points display finite-step violations of the stronger Lorenz ordering while retaining a positive accessible-entropy rate.\justifying
}
\label{fig:S18-local-rate}
\end{figure}

Figure~\ref{fig:S18-local-rate} presents these fixed-time refinements. Over the accessible range, the strongest boundary cases are consistent with a nonzero contribution to \(\Vpi\) that is approximately linear in \(\tau\), despite the positive accessible-entropy rate. These finite calculations establish that violations of the stronger relative-majorization order can coexist with a robustly positive accessible-entropy rate over the tested range. They do not constitute a proof of the mathematical limit
\begin{equation}
\nu_\pi(t)
=
\limsup_{\tau\rightarrow0^+}
\nu_\pi(t,\tau),
\end{equation}
nor do we extrapolate an exact limiting value from the five finite resolutions.

\begin{figure}[t]
\centering
\includegraphics[width=0.5\linewidth]{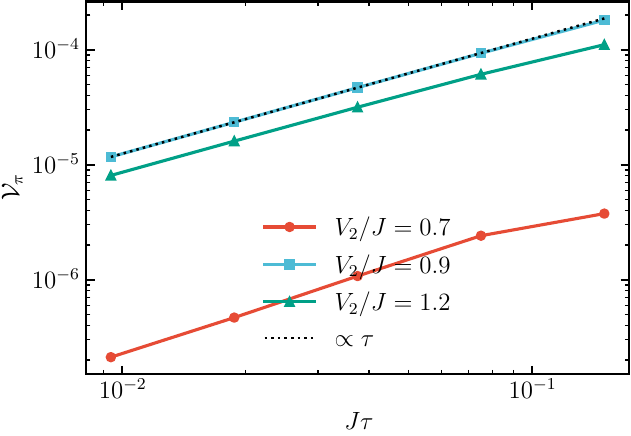}
\caption{Absolute relative-majorization defect at the fixed physical reference times used in Fig.~\ref{fig:S18-local-rate}. The dotted guide is proportional to \(\tau\). For the strongest sampled cases, the defect decreases approximately linearly with the finite step over the tested range. Consequently, a vanishing absolute defect under temporal refinement does not by itself imply restoration of exact local relative majorization.\justifying
}
\label{fig:S19-local-absolute}
\end{figure}

Figure~\ref{fig:S19-local-absolute} shows the same fixed-time return defects without division by \(\tau\). For the strongest cases, the approximately slope-one behavior on the tested range explains why the absolute defect decreases as the time step is refined even while the defect-rate proxy remains approximately finite. This explicitly illustrates why a shrinking absolute \(\Vpi\) cannot, by itself, be interpreted as restoration of exact local relative majorization.

\suppnote{Equilibration on average versus finite-step thermodynamic return}
\label{supp:note-meier-comparison}

The Second Law on average developed by Meier \textit{et al.}~\cite{meier2025secondlaw} and the finite-step formulation developed here begin from closely related coarse-grained entropy functionals but address different dynamical questions. This note makes their common entropy structure explicit, identifies the additional ingredients entering the equilibration-on-average theorem, and compares the two descriptions for the same microscopic models. The purpose is not to replace one construction by the other, but to separate long-time equilibration around an orbit-average reference from the exact finite-step question of which microscopic contribution can generate thermodynamic return.

\suppsubsection{Common entropy structure}
\label{supp:meier-common-entropy}

Meier \textit{et al.} consider a projective observable
\begin{equation}
O
=
\sum_i
O_i\Pi_i,
\end{equation}
with outcome probabilities
\begin{equation}
p_i
=
\Tr(\Pi_i\rho)
\end{equation}
and eigenspace dimensions
\begin{equation}
V_i
=
\Tr\Pi_i.
\end{equation}
They distinguish the Shannon entropy of the observable outcomes,
\begin{equation}
S_{\rm Sh}^{O}[\rho]
=
-\sum_i
p_i\ln p_i,
\end{equation}
from the observational entropy
\begin{equation}
S_{\rm obs}^{O}[\rho]
=
-\sum_i
p_i
\ln
\left(
\frac{p_i}{V_i}
\right).
\end{equation}
The latter is the von Neumann entropy of the corresponding coarse-grained state
\begin{equation}
\rho_{\rm cg}
=
\sum_i
\frac{p_i}{V_i}
\Pi_i.
\end{equation}
Under the direct identification
\begin{equation}
\Pi_i
\longleftrightarrow
P_R,
\quad
V_i
\longleftrightarrow
\Omega_R,
\end{equation}
one has
\begin{equation}
\rho_{\rm cg}
=
\cG[\rho],
\quad
S_{\rm obs}^{O}[\rho]
=
\frac{S_R(\rho)}{k_{\rm B}}.
\end{equation}
The observational-entropy functional itself is therefore not the distinction between the two approaches. The difference lies in how the closed unitary dynamics is organized, which reference distribution is relevant to the dynamical statement, and what type of Second-Law behavior is being established.

\suppsubsection{Structural economy of the finite-step decomposition}
\label{supp:meier-structural-economy}

The present construction begins directly from the exact decomposition
\begin{equation}
\rho(t)
=
\cG[\rho(t)]
+
\chi(t)
\end{equation}
and propagates both contributions through an arbitrary finite unitary interval \(U_\tau\). The accessible probabilities then satisfy
\begin{equation}
p(t+\tau)
=
K^{(U)}(\tau)p(t)
+
r(t,\tau),
\end{equation}
where
\begin{equation}
K^{(U)}_{RR'}(\tau)
=
\frac{1}{\Omega_{R'}}
\Tr
\left[
P_R
U_\tau
P_{R'}
U_\tau^\dagger
\right],
\end{equation}
and
\begin{equation}
r_R(t,\tau)
=
\Tr
\left[
P_R
U_\tau
\chi(t)
U_\tau^\dagger
\right].
\end{equation}
Positivity, completeness of the macroprojectors, and unitarity imply exactly
\begin{equation}
K^{(U)}_{RR'}
\geq
0,
\quad
\sum_R
K^{(U)}_{RR'}
=
1,
\quad
K^{(U)}\pi
=
\pi,
\end{equation}
with
\begin{equation}
\pi_R
=
\frac{\Omega_R}{\Omega_{\rm tot}}.
\end{equation}
Since
\begin{equation}
S_R(p)
=
k_{\rm B}
\left[
\ln\Omega_{\rm tot}
-
D(p\Vert\pi)
\right],
\end{equation}
the classical data-processing inequality gives
\begin{equation}
D
\left(
K^{(U)}p
\Vert
\pi
\right)
=
D
\left(
K^{(U)}p
\Vert
K^{(U)}\pi
\right)
\leq
D(p\Vert\pi),
\end{equation}
and therefore
\begin{equation}
S_R
\left(
K^{(U)}p
\right)
\geq
S_R(p).
\end{equation}
Thus the contribution predicted from the current thermodynamic representative alone is entropy nondecreasing for every finite unitary step, recovering the established observational-entropy H theorem for a macro-uniform initial state~\cite{strasberg2021firstsecond,nagasawa2024generic}. Its role here is to provide the exact baseline inside the decomposition of an arbitrary current state, without introducing an infinite-time orbit average, an effective dimension, an energy-gap concentration function, or an entropy-continuity estimate.

The exact entropy change is correspondingly decomposed as
\begin{equation}
\Delta_\tau S_R
=
\Delta_\tau S_{\rm mix}
+
\Delta_\tau S_{\rm corr},
\quad
\Delta_\tau S_{\rm mix}
\geq
0,
\end{equation}
where \(\Delta_\tau S_{\rm corr}\) contains the complete effect of the history-sensitive correction \(r(t,\tau)\). The relative-majorization defect is likewise bounded by
\begin{equation}
\Vpi
\left[
p(t)\rightarrow p(t+\tau)
\right]
\leq
\frac{1}{2}
\left\|
r(t,\tau)
\right\|_1.
\end{equation}
The principal structural simplification is therefore precise: for the fixed record, macro-uniform representative, and finite step, every accessible-entropy-lowering departure from the contraction baseline is isolated in the single history-sensitive object \(r(t,\tau)\).

This economy comes with a correspondingly precise limitation. The exact identities do not prove that \(r\) is small, that \(\Vpi=0\), or that entropy increases for a generic many-body Hamiltonian. They instead reduce the dynamical problem to determining when the history-sensitive contribution fails to organize a macroscopic return. By contrast, the additional spectral machinery of Ref.~\cite{meier2025secondlaw} supplies rigorous sufficient conditions for equilibration for most times together with quantitative bounds on fluctuations. The two constructions therefore control different parts of the problem.

\suppsubsection{Three distinct reference distributions}
\label{supp:meier-three-references}

For a time-independent Hamiltonian, Ref.~\cite{meier2025secondlaw} introduces the infinite-time averaged state
\begin{equation}
\omega
=
\lim_{T\rightarrow\infty}
\frac{1}{T}
\int_0^T
\rho(t)\,dt,
\end{equation}
together with the corresponding observable population vector
\begin{equation}
p_{\omega,R}
=
\Tr
\left(
P_R\omega
\right).
\end{equation}
The state \(\omega\) serves as the equilibrium reference in the specific sense of equilibration on average. It is a dynamical object determined jointly by the Hamiltonian and the initial microscopic state.

The present finite-step construction instead singles out the multiplicity-defined distribution
\begin{equation}
\pi_R
=
\frac{\Omega_R}{\Omega_{\rm tot}},
\end{equation}
which depends only on the thermodynamic partition and the selected invariant Hilbert sector. Its defining property for the present dynamics is the exact identity
\begin{equation}
K^{(U)}(\tau)\pi
=
\pi
\end{equation}
for every unitary interval.

A third distribution enters when one asks for an equilibrium state constrained to have the same mean energy. The energy-matched Gibbs state used in the main text defines
\begin{equation}
\pi_R^{(E)}
=
\Tr
\left[
P_R
\rho_{\beta_E,Q}^{\rm eq}
\right].
\end{equation}
Thus
\begin{equation}
p_\omega,
\quad
\pi,
\quad
\pi^{(E)}
\end{equation}
have different definitions and answer different physical questions. They should not be identified without additional assumptions. Equality \(p_\omega=\pi\) expresses an equidistribution property of the particular unitary orbit relative to the chosen macrospaces. Equality \(\pi^{(E)}=\pi\) instead concerns the energetic constraint and holds, for example, when \(\beta_E=0\) in the selected sector.

\suppsubsection{What the Second Law on average controls}
\label{supp:meier-average-bound}

For the same set of \(n_{\rm mac}\) projective outcomes, define the total-variation distance from the orbit-average populations by
\begin{equation}
d_{\rm TV}
\left[
p(t),p_\omega
\right]
=
\frac{1}{2}
\left\|
p(t)-p_\omega
\right\|_1.
\end{equation}
The population-equilibration result of Ref.~\cite{meier2025secondlaw} bounds its finite-time average as
\begin{equation}
\left\langle
d_{\rm TV}
\left[
p(t),p_\omega
\right]
\right\rangle_T
\leq
\eta_{\epsilon,T},
\end{equation}
where
\begin{equation}
\eta_{\epsilon,T}
=
\frac{1}{2}
\sqrt{
\frac{
n_{\rm mac}
}{
d_{\rm eff}
}
f(\epsilon,T)
},
\end{equation}
with
\begin{equation}
f(\epsilon,T)
=
N(\epsilon)
\left[
1
+
\frac{
8\log_2|\sigma(H)|
}{
\epsilon T
}
\right].
\end{equation}
Here
\begin{equation}
d_{\rm eff}
=
\frac{1}{\Tr(\omega^2)},
\end{equation}
\(|\sigma(H)|\) is the number of distinct energy eigenvalues, and \(N(\epsilon)\) is the maximum number of energy gaps contained within an interval of width \(\epsilon\).

The entropy theorem then combines this population-equilibration bound with a continuity inequality for the Shannon entropy. In the notation above,
\begin{equation}
\left\langle
\left|
S_{\rm Sh}^{O}[\rho(t)]
-
S_{\rm Sh}^{O}[\omega]
\right|
\right\rangle_T
\leq
\log(n_{\rm mac}-1)
\eta_{\epsilon,T}
+
H_2
\left(
\eta_{\epsilon,T}
\right)
\end{equation}
on the small-fluctuation branch \(\eta_{\epsilon,T}<1/2\), with a corresponding observational-entropy result derived in Ref.~\cite{meier2025secondlaw}. Combined with a past hypothesis in which the initial observable entropy lies sufficiently below its equilibrium value, this yields their Second Law on average.

The logical contrast is therefore direct. Ref.~\cite{meier2025secondlaw} asks under what conditions
\begin{equation}
p(t)
\simeq
p_\omega
\end{equation}
for most times and derives quantitative sufficient conditions for that behavior. The present construction asks instead, for one finite unitary interval, which part of the exact evolution is guaranteed not to lower the accessible entropy and which part retains the capacity to generate thermodynamic return.

Two qualifications are essential. First, Ref.~\cite{meier2025secondlaw} does not replace the instantaneous physical state by \(\omega\) when defining the observable entropy. The orbit average enters as the equilibrium reference around which the instantaneous observable entropy is controlled. Second, the equilibration-on-average construction explicitly permits finite fluctuations and recurrences. A deliberately synchronized return trajectory therefore does not contradict the theorem; it may instead lie in a regime where its small-fluctuation bound is not restrictive.

\suppsubsection{Exact matched-model comparison}
\label{supp:meier-matched-models}

We now evaluate the three reference distributions for the same \(N=16\), \(Q=2\) hard-core Hilbert sector, the same four-cell thermodynamic record, and the same left-half preparation used for the mixing and engineered-return controls. The initial thermodynamic representative is
\begin{equation}
\rho_0
=
\frac{
P_{L_i,Q}
}{
\Omega_i
},
\quad
\Omega_i
=
28.
\end{equation}
For either time-independent Hamiltonian, the exact infinite-time average is obtained by dephasing only between distinct energy eigenspaces,
\begin{equation}
\omega
=
\sum_\lambda
\Pi_\lambda
\rho_0
\Pi_\lambda,
\end{equation}
where \(\Pi_\lambda\) projects onto the complete degenerate eigenspace with energy \(\lambda\). The dynamical figures in the main text are displayed through \(Jt=30\), whereas all quantities involving \(\omega\) below are exact infinite-time quantities and are independent of that plotting cutoff.

To characterize the spectral synchronization relevant to the equilibration bound, define
\begin{equation}
D_G
=
\max_{g\neq0}
\#
\left\{
(\lambda,\lambda'):
\lambda-\lambda'
=
g
\right\},
\end{equation}
where \(\lambda\) and \(\lambda'\) run over distinct elements of \(\sigma(H)\). For sufficiently small \(\epsilon\) that distinct gap values are resolved, \(D_G\) is the corresponding resolved-gap limit of the concentration factor \(N(\epsilon)\). The matched comparison is summarized in Table~\ref{tab:meier-comparison}.
\begin{table}[t]
\centering
\caption{
Exact \(N=16\), \(Q=2\) comparison for the same left-half thermodynamic preparation and four-cell record. Infinite-time quantities are obtained by exact spectral dephasing. \(D_G\) is the maximum multiplicity of an equal nonzero gap among distinct energy values. Entropies are reported in units of \(k_{\rm B}\).\justifying}
\label{tab:meier-comparison}
\begin{tabular}{lcc}
\toprule
Quantity & Mixing & Engineered return \\
\midrule
\(\dim\mathcal H_Q\) & 120 & 120 \\
\(n_{\rm mac}\) & 10 & 10 \\
\(\Omega_i\) & 28 & 28 \\
\(|\sigma(H)|\) & 119 & 29 \\
\(d_{\rm eff}\) & 102.958 & 22.366 \\
\(D_G\) & 1 & 28 \\
\(\tfrac12\|p_\omega-\pi\|_1\) & 0.06296 & 0.16355 \\
\(D(p_\omega\Vert\pi)\) & 0.01110 & 0.06978 \\
\(S_R(p_\omega)/k_{\rm B}\) & 4.77639 & 4.71772 \\
\(\beta_EJ\) & -0.05336 & 0 \\
\(\tfrac12\|\pi^{(E)}-\pi\|_1\) & 0.00547 & \(<10^{-14}\) \\
\(\tfrac12\|\pi^{(E)}-p_\omega\|_1\) & 0.05825 & 0.16355 \\
\bottomrule
\end{tabular}
\end{table}

The theorem-correct effective dimension is defined with projectors onto complete distinct-energy eigenspaces. With that definition, the engineered-return spectrum has a substantially smaller effective dimension than the mixing spectrum because its large energy degeneracies concentrate the preparation weight into fewer eigenspaces. The gap organization is simultaneously much more degenerate. At the numerical resolution used for this diagnostic, the mixing Hamiltonian has \(119\) distinct energies and no repeated nonzero gap among the distinct energy values. The engineered-return Hamiltonian instead has only \(29\) distinct energies,
\begin{equation}
\frac{E}{J}
=
-3.5,
-3.25,
\ldots,
3.25,
3.5,
\end{equation}
with uniform spacing \(J/4\) and
\begin{equation}
D_G
=
28.
\end{equation}
When distinct gap values are resolved and the finite-time correction becomes negligible, the spectral part of the Meier equilibration parameter approaches
\begin{equation}
\eta_{\rm gap}
=
\frac{1}{2}
\sqrt{
\frac{
n_{\rm mac}D_G
}{
d_{\rm eff}
}
}.
\end{equation}
Numerically,
\begin{equation}
\eta_{\rm gap}^{\rm mix}
\simeq
0.156,
\quad
\eta_{\rm gap}^{\rm ret}
\simeq
1.769.
\end{equation}
The engineered-return value strongly exceeds \(1/2\) before the finite-time correction in \(f(\epsilon,T)\) is included. This does not violate the theorem; rather, its small-fluctuation branch is not restrictive for this deliberately synchronized spectrum. At the present finite sizes the rigorous finite-time bound is also quantitatively conservative for the mixing Hamiltonian. We therefore use the exact gap-resolved current amplitudes of Note~\ref{supp:note-spectral-return}, rather than the generic norm bound, to diagnose how the actual thermodynamic currents are spectrally organized.

The distinction between orbit-average and thermodynamic reference distributions is particularly transparent for the engineered-return system. Its conserved mean energy equals the infinite-temperature mean energy in the fixed-\(Q\) sector, giving
\begin{equation}
\beta_E
=
0,
\quad
\rho_{\beta_E,Q}^{\rm eq}
=
\frac{
P_Q
}{
120
},
\quad
\pi^{(E)}
=
\pi.
\end{equation}
Nevertheless,
\begin{equation}
\frac{1}{2}
\left\|
p_\omega-\pi
\right\|_1
\simeq
0.16355,
\end{equation}
so that
\begin{equation}
p_\omega
\neq
\pi
=
\pi^{(E)}.
\end{equation}
The infinite-time orbit average therefore exists but does not coincide with the energy-matched thermodynamic macrodistribution. This is fully compatible with Ref.~\cite{meier2025secondlaw}, which distinguishes equilibration from thermalization.

\suppsubsection{A complete recurrence period}
\label{supp:meier-recurrence-period}

The distinct engineered energies are uniformly separated by \(J/4\). All relative phases therefore recur after
\begin{equation}
T_{\rm rec}
=
\frac{8\pi}{J}
\simeq
\frac{25.133}{J}.
\end{equation}
The displayed interval \(0\leq Jt\leq30\) thus contains one complete recurrence period. Averaging the exact macrotrajectory over one complete period gives
\begin{equation}
\overline{
d_{\rm TV}
\left[
p(t),p_\omega
\right]
}
\simeq
0.3734.
\end{equation}
With temporal sampling \(\Delta(Jt)=0.02\), none of the sampled times satisfies
\begin{equation}
d_{\rm TV}
\left[
p(t),p_\omega
\right]
<
0.1.
\end{equation}
Thus the orbit average is perfectly well defined, but the instantaneous engineered trajectory does not spend most of its recurrence period close to that average.

The corresponding entropy values further illustrate why the entropy of an orbit-average distribution should not be conflated with the time average of the instantaneous thermodynamic entropy. For the engineered control,
\begin{equation}
\frac{
S_R(p_\omega)
}{
k_{\rm B}
}
\simeq
4.718,
\end{equation}
whereas
\begin{equation}
\frac{
\overline{
S_R[p(t)]
}
}{
k_{\rm B}
}
\simeq
4.260
\end{equation}
over one complete recurrence period. Concavity requires only
\begin{equation}
\overline{
S_R[p(t)]
}
\leq
S_R
\left[
\overline{p(t)}
\right]
=
S_R(p_\omega),
\end{equation}
so the two values are entirely consistent. This inequality is not a criticism of Ref.~\cite{meier2025secondlaw}, whose theorem controls the instantaneous observable entropy when its equilibration conditions are effective. The relevant point here is that the commensurate engineered spectrum prevents the instantaneous trajectory from remaining close to \(p_\omega\), thereby allowing the coherent macroscopic return for which the control was constructed.

\suppsubsection{Exact bridge between the two formulations}
\label{supp:meier-bridge}

The orbit-average and finite-step descriptions can be connected exactly. For any fixed interval \(\tau\),
\begin{equation}
p(t+\tau)
=
K^{(U)}(\tau)p(t)
+
r(t,\tau).
\end{equation}
Taking the infinite-time average and using time-translation invariance of that average gives
\begin{equation}
p_\omega
=
K^{(U)}(\tau)p_\omega
+
\overline{r}_\tau,
\end{equation}
where
\begin{equation}
\overline{r}_\tau
=
\lim_{T\rightarrow\infty}
\frac{1}{T}
\int_0^T
r(t,\tau)\,dt.
\end{equation}
Since
\begin{equation}
K^{(U)}(\tau)\pi
=
\pi,
\end{equation}
we obtain
\begin{equation}
\left[
I-K^{(U)}(\tau)
\right]
\left(
p_\omega-\pi
\right)
=
\overline{r}_\tau.
\end{equation}
This identity provides an exact bridge between the two viewpoints. If
\begin{equation}
p_\omega
=
\pi,
\end{equation}
then necessarily
\begin{equation}
\overline{r}_\tau
=
0.
\end{equation}
The converse does not follow without additional information. If \(K^{(U)}(\tau)\) possesses a nontrivial fixed-point subspace, then a nonzero component of \(p_\omega-\pi\) can lie in that subspace while still giving \(\overline{r}_\tau=0\). The orbit average therefore retains only the time-averaged imprint of the history-sensitive contribution, whereas the finite-step decomposition resolves how that contribution is organized at each particular stage of the trajectory.

The two routes to Second-Law behavior may consequently be summarized as
\begin{equation}
\begin{aligned}
H,\rho_0
&\longrightarrow
\omega
\\
&\longrightarrow
d_{\rm eff},
\,
N(\epsilon),
\,
f(\epsilon,T)
\\
&\longrightarrow
\left\langle
d_{\rm TV}
[
p(t),p_\omega
]
\right\rangle_T
\\
&\longrightarrow
\text{entropy-continuity bound}
\\
&\longrightarrow
S[p(t)]
\simeq
S[p_\omega]
\quad
\text{for most times},
\end{aligned}
\end{equation}
for the equilibration-on-average construction, whereas the present route is
\begin{equation}
\begin{aligned}
\rho_t
&=
\cG[\rho_t]
+
\chi_t
\\
&\longrightarrow
p(t+\tau)
=
K^{(U)}p(t)
+
r
\\
&\longrightarrow
\left.
\partial_\tau K^{(U)}
\right|_0
=
0,
\quad
\frac{r}{\tau}
\longrightarrow
\dot p
\\
&\longrightarrow
\dot S_R
=
\sigma_{\rm spread}
-
\sigma_{\rm ret}
\\
&\longrightarrow
J_e(t)
=
J_e^\omega
+
\sum_{g\neq0}C_{e,g}e^{-igt}
\\
&\longrightarrow
\overline{\|\mathbf J-\mathbf J^\omega\|_2^2}
=
\Gamma_{\rm spec}P_{\rm pair}
\leq
D_GP_{\rm pair}.
\end{aligned}
\end{equation}
The two spectral viewpoints are therefore complementary rather than successive steps of one theorem. The equilibration-on-average route uses effective dimension and gap concentration to bound distance from the orbit-average distribution for most times. The present route instead keeps the exact state- and current-operator amplitudes and resolves how equal-gap channels contribute to thermodynamic current power. It does not by itself prove that this power is small for a generic many-body Hamiltonian, and the generic finite-time norm bounds are quantitatively conservative for the models studied here. Neither construction establishes all-time positivity of the accessible-entropy rate.

\suppnote{Relation to recent coarse-grained Second Laws and hidden microscopic resources}
\label{supp:note-recent-literature}

The finite-step construction sits inside a mature literature on coarse states, hidden resources, and closed-system Second Laws. This note separates that established background from the additional structure used in the present work.

\suppsubsection{Established coarse state and macrostate entropy increase}
\label{supp:recent-established-baseline}

For a projective record, \(\cG[\rho]=\sum_Rp_RP_R/\Omega_R\) is the established observational coarse state. It is the Bayesian/Petz-recovered estimate generated from the macrostatistics and a uniform microscopic prior~\cite{buscemi2023coarse,bai2024priors}; its fixed points and relative-entropy deficit have also been developed as a resource theory of macroscopicity~\cite{nagasawa2025macroscopicity}. Maximum-entropy and measurement-based constructions can be placed in one prior-dependent framework~\cite{schindler2025unification}.

If a unitary step begins from a state satisfying \(\rho=\cG[\rho]\), observational entropy cannot decrease, and the increase is generic for the unitary ensembles studied in Ref.~\cite{nagasawa2024generic}; the microscopic Second-Law construction of Strasberg and Winter contains the corresponding monotonicity statement~\cite{strasberg2021firstsecond}. In the present notation this antecedent is exactly \(\Delta_\tau S_{\rm mix}\geq0\). The contribution of the current work is not that inequality in isolation, but its embedding in the exact identity
\begin{equation}
p(t+\tau)=K^{(U)}(\tau)p(t)+r(t,\tau)
\end{equation}
for an arbitrary current state, with the vector response of the hidden component retained explicitly.

\suppsubsection{Hidden resources, scale divisibility, and the residual}
\label{supp:recent-rignon}

Rignon-Bret and Elouard identify the scalar mismatch \(D(\rho\Vert\rho_{\rm cg})\) as an internal nonequilibrium resource and formulate a scale-divisibility condition that excludes information feedback from ignored microscopic variables~\cite{rignonbret2026algebraic}. For the projective record used here, their scalar resource coincides with \(I_R(\rho)=D(\rho\Vert\cG[\rho])\).

The residual resolves a different aspect of the same hidden structure:
\begin{equation}
r(t,\tau)=p(t+\tau)-K^{(U)}(\tau)p(t).
\end{equation}
It is state and interval specific. Thus \(r=0\) says that the exact and macro-uniform current states produce the same endpoint record over one chosen step, which is weaker than a map-level divisibility property. When \(r\neq0\), the pair \((\epsilon_{\rm hist},\Vpi)\) further distinguishes feedback that remains inside the full \(\pi\)-preserving contraction order from feedback organized into thermodynamic return. The scalar deficit measures how much information is hidden; the residual and its Lorenz defect measure how that hidden information is dynamically directed.

\suppsubsection{Complementary restricted-information and closed-system laws}
\label{supp:recent-other-laws}

Observation-dependent thermodynamics, geometric reduction by restricted observability, most-time equilibration laws, and thermodynamic-limit macroscopic operation theorems answer complementary questions~\cite{rubino2026coarse,xsqg-xvgc,meier2025secondlaw,chiba2026secondlaw}. Maximum-entropy assignment maps and exact memory representations likewise show that coarse dynamics can be nonlinear or non-Markovian without any loss of microscopic information~\cite{castillo2025coarse,aristoff2023coarse,espanol2026memory}. These works delimit the novelty claim: non-Markovianity, hidden resources, and coarse-state monotonicity are established. The additional structure here is the exact present-state residual, its instantaneous current limit, and the separation of hidden activity from thermodynamic return at both the entropy-current and full-distribution levels.

\suppsubsection{Thermodynamic boundary and companion application}
\label{supp:recent-work-scope}

The autonomous relaxation studied here does not, by itself, assign heat or work. Those terms refer to energy transfer across a specified boundary and therefore require an explicit control mechanism and comparator. A companion study supplies that additional structure for matched finite-particle quantum and classical Hamiltonian dynamics: it embeds the thermally isolated expansion stroke in a complete cycle and relates the resulting energy-and-record irreversibility difference to work and efficiency differences~\cite{ahmadi2026finiteparticle}. Separating that application from the present derivation keeps the logical boundary clear: this paper identifies the origin and organization of thermodynamic return, while the companion specifies how the same restricted-information geometry enters energy-transfer bookkeeping.

\suppsubsection{Defensible novelty boundary}
\label{supp:recent-novelty-boundary}

The antecedents include projective observational entropy, the macro-uniform coarse state and its Jaynes/Petz interpretations, its entropy nondecrease when it is the actual initial state, hidden microscopic mismatch as a thermodynamic resource, relative Lorenz curves, approximate-majorization geometry, projection memory, and the Ulam transition matrix. The finite-step contribution established here is their state-resolved synthesis:
\begin{equation}
\rho_t=\cG[\rho_t]+\chi_t
\quad\Longrightarrow\quad
p(t+\tau)=K^{(U)}p(t)+r(t,\tau),
\end{equation}
with \(K^{(U)}\pi=\pi\), the complete vector residual retained, and hidden activity separated from organized return through
\begin{equation}
\Vpi\leq\frac12\|r\|_1.
\end{equation}
The short-time theorem adds that the macro-uniform baseline has zero first-order entropy rate and that \(r/\tau\to\dot p\). The current formulation then identifies the exact competition \(\dot S_R=\sigma_{\rm spread}-\sigma_{\rm ret}\), while Note~\ref{supp:note-spectral-return} resolves the actual current-carrying amplitudes and separates pair transition power from coherent equal-gap organization. The engineered control and exact classical residual counterpart demonstrate the physical and structural reach of this synthesis. What remains open is a broad many-body theorem deriving both a positive spreading margin and asymptotic suppression of return from locality, transport, and record resolution.

\suppnote{Scope of the analytical results and numerical evidence}
\label{supp:note-scope}

The paper combines exact identities, finite-system observations, a classical counterpart, and combinatorial limits. Their logical status is different, and the conclusions should be read with those distinctions in view.

The finite-step relations
\begin{equation}
K^{(U)}\pi=\pi,
\quad
\Delta_\tau S_{\rm mix}\geq0,
\quad
\Vpi\leq\frac12\|r\|_1
\end{equation}
follow exactly from unitarity, completeness of the projective record, and data processing. The middle inequality is an established observational-entropy baseline rather than the central irreversibility result. Its first-order rate vanishes,
\begin{equation}
\left.\partial_\tau K^{(U)}(\tau)\right|_{\tau=0}=0,
\quad
\dot S_{\rm mix}=0,
\quad
\lim_{\tau\rightarrow0}\frac{r(t,\tau)}{\tau}=\dot p(t),
\end{equation}
so the complete instantaneous record motion is generated by the hidden component. These identities are independent of the specific many-body models and do not assume chaos, ETH, Markovianity, or equilibration. Although each finite-time \(K^{(U)}\) is stochastic, the family is not generally a semigroup; Eq.~\eqref{eq:supp-K-composition-defect} displays the exact memory-bearing composition defect.

Note~\ref{supp:note-projection-classical} proves the corresponding finite-step structure for invertible measure-preserving classical dynamics on a finite invariant measure space. The Ulam matrix itself is established; the additional element is its exact completion by the within-cell residual and the resulting invariant-reference thermodynamic bounds. The theorem does not normalize an unbounded Liouville measure and does not prove that a generic classical Hamiltonian suppresses the residual.

Relative majorization is a sufficient finite-step Second-Law condition, not a necessary condition for \(\dot S_R>0\). The strong-coupling boundary calculations explicitly show small Lorenz-order violations coexisting with a positive exact entropy rate. Conversely, the representative hard-core, Bose--Hubbard, two-dimensional, and Floquet calculations exhibit the stronger regime
\begin{equation}
\epsilon_{\rm hist}>0,
\quad
\Vpi\simeq0,
\quad
\dot S_R>0
\end{equation}
throughout their reported primary windows. These are finite-system numerical observations at stated resolutions, not a theorem for arbitrary Hamiltonians or records.

The engineered Hamiltonian establishes a constructive counterpoint. With the same preparation and record, deliberately synchronized dynamics reorganizes hidden microscopic structure into macroscopic return. The free chain shows that the distinction is not exhausted by an integrable-versus-nonintegrable label. The exact spectral theorem sharpens this comparison by resolving every thermodynamic macrocurrent into state- and record-specific gap amplitudes:
\begin{equation}
\overline{\|\mathbf J-\mathbf J^\omega\|_2^2}
=
\Gamma_{\rm spec}P_{\rm pair}
\leq
D_GP_{\rm pair}.
\end{equation}
This identifies coherent equal-gap organization of the actual current network. It does not make \(\Gamma_{\rm spec}\) a universal irreversibility criterion: \(D_G=1\) does not force \(P_{\rm pair}\) to vanish, and \(\Gamma_{\rm spec}>1\) does not determine the sign of the current--affinity product. A positive entropy rate still requires \(\sigma_{\rm spread}>\sigma_{\rm ret}\).

The hard-core and bosonic free-expansion endpoints are exact multiplicity statements. Both satisfy
\begin{equation}
\frac{\Delta S_R}{Qk_{\rm B}}
\longrightarrow
\ln\!\left(\frac{V_f}{V_i}\right)
\end{equation}
in the dilute limit, with opposite finite-density corrections. The accompanying simulations provide separate finite-size evidence for positive primary relaxation along a sequence approaching that endpoint; the combinatorial limit does not depend on those dynamics.

Heat and work are not assigned within the autonomous analysis because they require a specified boundary, transfer mechanism, and control class. No work statement is used to establish any result in this paper.

Taken together, the results establish exact finite-step and instantaneous structures, an exact state-resolved spectral organization of the thermodynamic current network, several finite-system irreversible windows, an exact classical residual counterpart, and the classical free-expansion endpoint. They do not prove
\begin{equation}
\Vpi\rightarrow0,
\quad
\nu_\pi\rightarrow0,
\quad
\text{or}
\quad
\dot S_R\geq0
\end{equation}
for arbitrary Hamiltonians, records, all times, or a thermodynamic limit. The remaining theoretical challenge is to derive broad local many-body conditions that simultaneously suppress return-oriented current power and maintain a positive spreading margin while keeping observation time and record resolution explicit.

\end{document}